\documentclass[aps, pra, reprint, superscriptaddress]{revtex4-2}

\pdfoutput=1
\usepackage{adjustbox}
\usepackage[linesnumbered,ruled,vlined]{algorithm2e}
\usepackage{algpseudocode}
\usepackage{amsmath}
\usepackage{amssymb}
\usepackage{amsthm}
\usepackage[page,header]{appendix}
\usepackage[T1]{fontenc}
\usepackage{graphicx}
\usepackage{hyperref}
\usepackage[capitalise]{cleveref}
\usepackage{makecell}
\usepackage{mathtools}
\usepackage{microtype}
\usepackage{multirow}
\usepackage{physics}
\usepackage{siunitx}
\usepackage{subcaption}
\usepackage{natbib}

\usepackage[all]{nowidow}
\usepackage{quantikz}
\usepackage{tabularx} 
\usepackage{tikz}
\usepackage{titletoc}
\usepackage{thmtools}
\hypersetup{
    colorlinks,
    allcolors=blue
}

\newtheorem*{corollary*}{Corollary}

\theoremstyle{definition}

\theoremstyle{remark}

\renewcommand{\selectlanguage}[1]{}
\DeclareMathOperator*{\argmin}{argmin}
\AtBeginDocument{\RenewCommandCopy\qty\SI}

\makeatletter
\def\l@subsubsection#1#2{}
\makeatother

\graphicspath{{Figures/}}

\begin{document}

\title{Quantum-Computing for Corrosion Simulation and Prevention: Workflow and Resource Analysis}

\author{Nam Nguyen}
\email{nam.h.nguyen5@boeing.com}
\affiliation{Applied Mathematics, Boeing Research \& Technology, USA}
\author{Thomas W. Watts}
\affiliation{HRL Laboratories, LLC, Malibu, CA, USA}
\author{Benjamin Link}
\affiliation{Applied Mathematics, Boeing Research \& Technology, USA}
\author{Kristen S. Williams}
\affiliation{Applied Mathematics, Boeing Research \& Technology, USA}
\author{Yuval R. Sanders}
\affiliation{Centre for Quantum Software and Information, School of Computer Science, Faculty of Engineering \& Information Technology,
University of Technology Sydney, NSW 2007, Australia}
\author{Samuel J. Elman}
\email{samuel.elman@uts.edu.au}
\affiliation{Centre for Quantum Software and Information, School of Computer Science, Faculty of Engineering \& Information Technology,
University of Technology Sydney, NSW 2007, Australia}
\author{Maria Kieferova}
\affiliation{Centre for Quantum Software and Information, School of Computer Science, Faculty of Engineering \& Information Technology,
University of Technology Sydney, NSW 2007, Australia}
\author{Michael J. Bremner}
\affiliation{Centre for Quantum Software and Information, School of Computer Science, Faculty of Engineering \& Information Technology,
University of Technology Sydney, NSW 2007, Australia}
\affiliation{Centre for Quantum Computation and Communication Technology, University of Technology Sydney, NSW 2007, Australia}
\author{Kaitlyn J. Morrell}
\affiliation{MIT Lincoln Laboratory, Lexington, MA, USA}
\author{Justin Elenewski}
\affiliation{MIT Lincoln Laboratory, Lexington, MA, USA}
\author{Eric B. Isaacs}
\affiliation{HRL Laboratories, LLC, Malibu, CA, USA}
\author{Samuel D. Johnson}
\affiliation{HRL Laboratories, LLC, Malibu, CA, USA}
\author{Luke Mathieson}
\affiliation{Centre for Quantum Software and Information, School of Computer Science, Faculty of Engineering \& Information Technology,
University of Technology Sydney, NSW 2007, Australia}
\author{Kevin M. Obenland}
\affiliation{MIT Lincoln Laboratory, Lexington, MA, USA}
\author{Matthew Otten}
\email{mjotten@wisc.edu}
\affiliation{Department of Physics, University of Wisconsin – Madison, Madison, WI, USA}

\author{Rashmi Sundareswara}
\affiliation{HRL Laboratories, LLC, Malibu, CA, USA}
\author{Adam Holmes}
\email{aholmes@hrl.com}
\affiliation{HRL Laboratories, LLC, Malibu, CA, USA}

\begin{abstract}
Corrosion is a pervasive issue that impacts the structural integrity and performance of materials across various industries, imposing a significant economic impact globally. In fields like aerospace and defense, developing corrosion-resistant materials is critical, but progress is often hindered by the complexities of material-environment interactions. While computational methods have advanced in designing corrosion inhibitors and corrosion-resistant materials, they fall short in understanding the fundamental corrosion mechanisms due to the highly correlated nature of the systems involved. This paper explores the potential of leveraging quantum computing to accelerate the design of corrosion inhibitors and corrosion-resistant materials, with a particular focus on magnesium and niobium alloys. We investigate the quantum computing resources required for high-fidelity electronic ground-state energy estimation (GSEE), which will be used in our hybrid classical-quantum workflow. Representative computational models for magnesium and niobium alloys show that 2292 to 38598 logical qubits and $(1.04$ to $1962) \times 10^{13}$ T-gates are required for simulating the ground-state energy of these systems under the first quantization encoding using plane waves basis.
\end{abstract}

\maketitle

Corrosion is a natural process that degrades materials through chemical or electrochemical reactions with their environment, impacting both the structural integrity and performance of engineering materials~\cite{Roberge2018corrosion}. It imposes a significant economic burden, with global costs exceeding \$2.5 trillion annually~\cite{koch2016international}.  
In aerospace and defense, corrosion-resistant materials reduce maintenance costs and enhance sustainability~\cite{Williams2019}. The development of corrosion-resistant materials relies on a combination of experimental and computational methods. Computational models can aid in the designing of corrosion inhibitors  and the development of corrosion-resistant materials~\cite{Barca2020Recent,kothe2018exascale}, with experiments validating these materials~\cite{ozkan2024laying}.
However, no single mathematical framework currently exists that can fully capture the complexity of the physical and chemical processes driving corrosion~\cite{Rodrguez2020,Ke2019}. As a result, different computational models need to be investigated for various corrosion applications. This work introduces a complete workflow from first principles for simulating corrosion processes. The novel quantum-classical hybrid approach unifies previously unrelated classical and quantum algorithms. We present computational models for aqueous corrosion in magnesium and magnesium-aluminum alloy surfaces, which may be extendable to other metallic systems; and for high-temperature oxidation in niobium alloys, which are critical for extreme environments but highly susceptible to degradation. Our models for magnesium corrosion are based on nudged elastic band modeling, while our niobium algorithm involves a coupled-cluster approximation to map the problem to a classical Ising model. In both cases, ground-state energy estimation is a key computational bottleneck.

Traditional modeling techniques, like finite element analysis (FEA), are poorly suited to corrosion due to the highly localized, non-uniform environments in which it occurs~\cite{Liu2014}. Corrosion processes often span multiple length and time scales and can evolve over time, transitioning from localized pitting to more catastrophic modes such as stress corrosion cracking or corrosion fatigue~\cite{Ke2019}. Factors such as material properties, corrosive environments, and component assembly further complicate these interactions~\cite{taylor2018integrated,smith2016development}. The lack of an integrated framework that combines chemistry, microstructure, and mechanical behavior into a single model necessitates significant simplifying assumptions~\cite{Liu2019}.

\begin{figure}[ht!]
    \centering
    \includegraphics[width=\columnwidth]{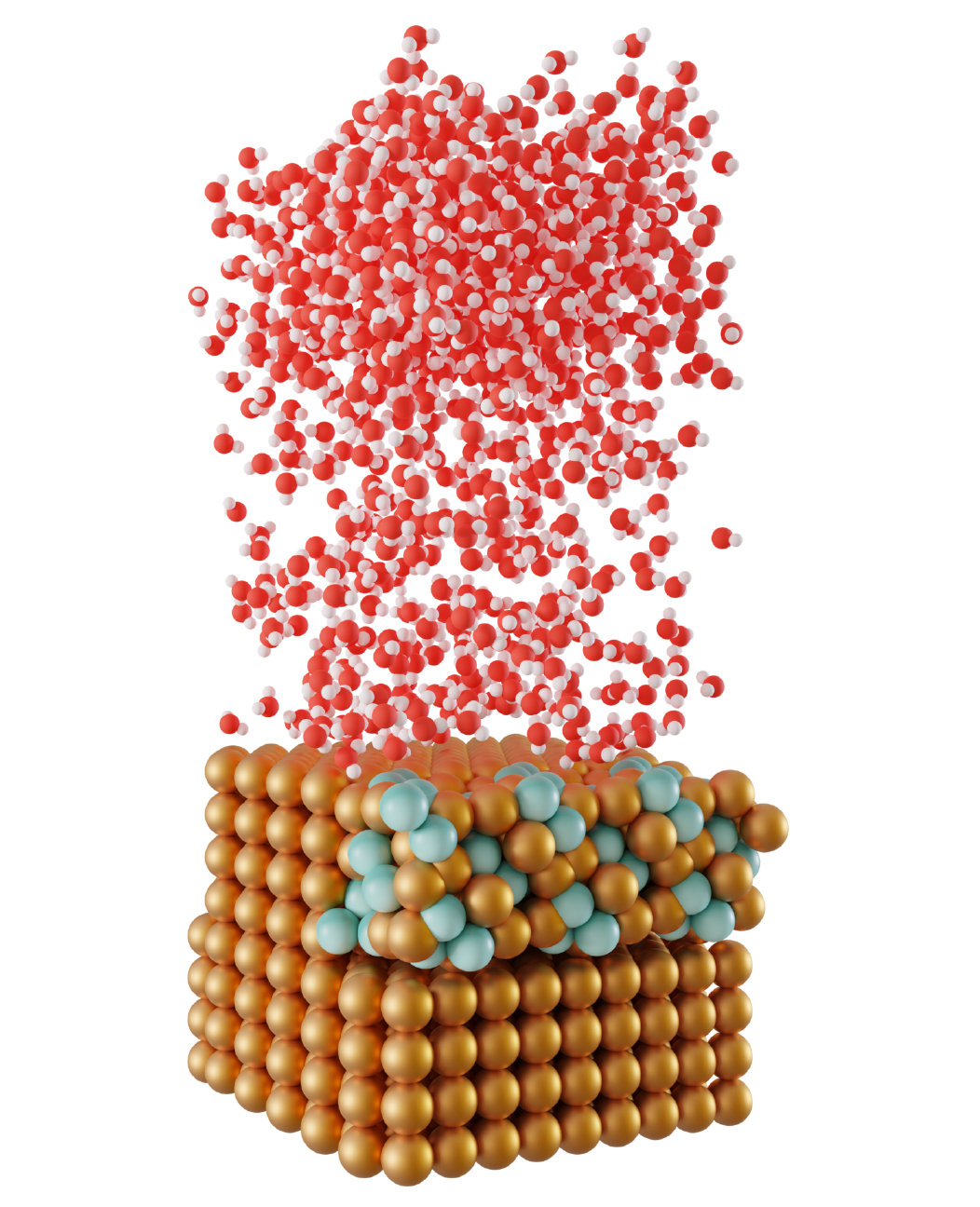}
    \caption{1000 Water Model of Mg Grain Boundary with Mg-Al secondary phase. The golden spheres represent Mg atoms, red oxygen, white hydrogen, and light blue aluminum.}
    \label{fig: huge_alloy_model}
\end{figure}

At the atomic scale, the chemical reactions driving corrosion are governed by the electronic structure of materials, which can be described by the time-independent Schrödinger equation. Accurate solutions to the Schrödinger equation are crucial for predicting properties such as corrosion rates but are computationally prohibitive due to the superpolynomial scaling of the wavefunction size with the number of orbitals ($N$) and electrons ($N_e$)~\cite{Sherrill1999TheCI}. Many of these processes involve highly correlated electronic states~\cite{Ke2019}, making exact diagonalization infeasible when $N$ and $N_e$ exceed 25~\cite{Guo2015Nelectron}. For example, Figure~\ref{fig: huge_alloy_model} shows a magnesium alloy model that incorporates 1000 water molecules and eight atomic layers, highlighting the scale and complexity of these systems. Simulating such a model with high accuracy is far beyond the capabilities of classical methods.

Quantum computers offer a promising path forward, as they are uniquely suited for simulating strongly correlated chemical systems~\cite{Bauer2020,Reiher2017}. Recent advances in quantum algorithms for simulating chemical models~\cite{babbush2018encoding,Berry2024DoublingEO}, along with new quantum software libraries~\cite{McClean2017OpenFermionTE,*Javadi2024qiskit}, make quantum computing increasingly practical. While quantum computers are not expected to completely negate high-performance computer costs, it will potentially enable high-fidelity computations on larger atomistic models and more accurate simulation of corrosion processes that are highly correlated. We conduct a comprehensive resource assessment for quantum modeling of aqueous and high-temperature corrosion processes, marking the first application of the pyLIQTR computational tool to a real-world materials science problem. 
Our work develops and implements a hybrid quantum-classical workflow that incorporates Ground-State Energy Estimation (GSEE) and details the required logical-level quantum hardware resources for high-fidelity simulations of magnesium and niobium alloy corrosion. In particular, we cost utility-scale instances and establish the feasibility of quantum-enhanced approaches in materials degradation studies by estimating the total number of logical qubits and $T$ gates necessary to implement the quantum phase estimation subroutine. An overview of these resources is provided in \cref{tab:Summary}.

\begin{table}[]
    \centering
    \begin{ruledtabular}
        \begin{tabular}{l c c c}
        \textbf{Model} & \makecell{\textbf{Cell size} \\ ({\AA})} & {$\boldsymbol{N_{\text{Log}}}$} & \makecell{\textbf{$T$ count}\\ ($\times 10^{13}$)} \\
        \hline
        \multicolumn{4}{c}{\textbf{Magnesium alloys}}\\
        \hline
        Mg dimer & \makecell{[12.7, 12.7, 19.9]} & 2292 & $1.04$\\
        Mg monolayer & \makecell{[19.8, 19.8, 32.3]} & 10358 & $68.2$\\
        Mg cluster & \makecell{[19.8, 19.8, 32.3]} & 26746 & $68.7$\\
        \makecell[l]{$\textrm{Mg}_{17}\textrm{Al}_{12}$\\2nd phase supercell} &\makecell{[33.0, 31.3, 50.3]} & \makecell{38598} & \makecell{1962}
        \\
        \hline
        \multicolumn{4}{c}{\textbf{Niobium-rich alloys}}\\
        \hline
        Nb$_{65}$Zr$_6$Hf$_7$Ti$_4$W$_3$ & \makecell{[8.6, 9.8, 14.0]} & 6305 & $21.2$\\
        Nb$_{97}$Hf$_{3}$Ti$_{22}$Zr$_6$ & \makecell{[13.3, 13.3, 13.3]} & 9370 & $62.0$\\
        Nb$_{42}$Ti$_3$Hf$_3$Ta$_3$Zr$_{3}$ & [10.0, 10.0, 10.1]&4142&7.17\\
        Nb$_{97}$Ta$_{22}$Zr$_3$W$_{6}$ & [13.3, 13.3, 13.3]& 9880& 65.3
        \end{tabular}
        \end{ruledtabular}
    \caption{Summary of logical resource estimate on a single shot of qubitized QPE for selected computational models, not including initial state preparation procedure. Here, $N_{\text{Log}}$ denotes the number of logical qubits, and $T$ count represents the number of non-Clifford $T$ gates.}
    \label{tab:Summary}
\end{table}

\section*{Results}
\begin{figure*}[ht]
    \centering
    \includegraphics[width=\linewidth]{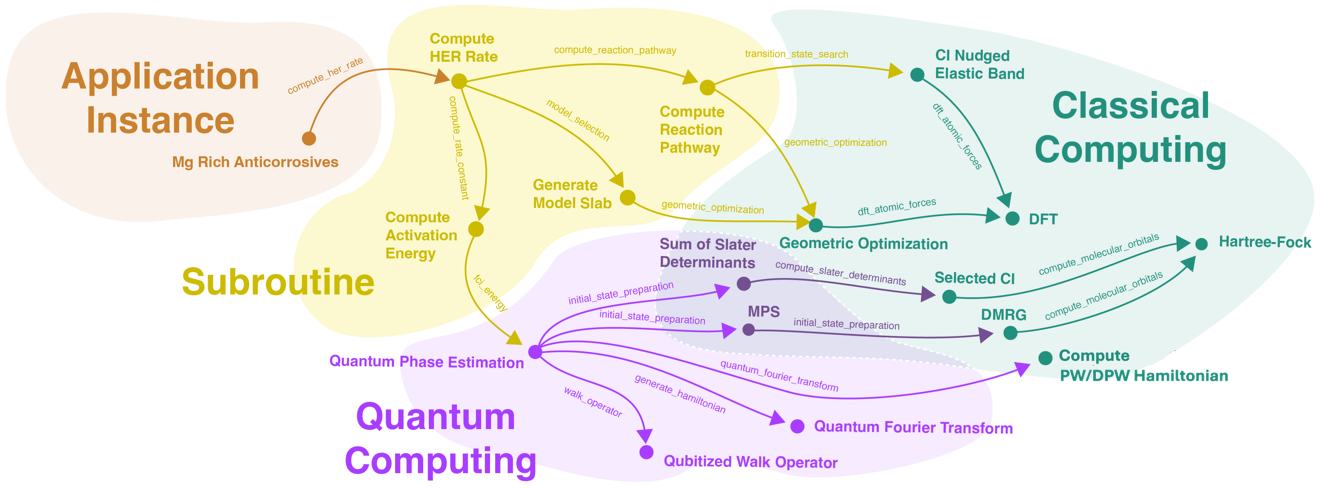}
    \caption{Magnesium QBG: the directed acyclic graph decomposes the magnesium workflow into individual computational subroutines along with its required inputs and actions. To compute the HER rate, we first need to determine the reaction pathways, generate the model slabs and compute the activation energies. Each node is then broken down further to visualize the workflow.}
    \label{fig:mg_qbg}
\end{figure*}

We develop a hybrid classical-quantum workflow to investigate the corrosion mechanisms at the atomic scale in two classes of alloys, under aqueous and high-temperature environments. These workflows are expressed using the Quantum Benchmarking Graph (QBG) formalism.
QBG is a technique intended to graphically and systematically decompose an application instance into foundational subroutines and core enabling computations. A QBG is realized as an attributed, directed acyclic graph (DAG). In these graphs, both nodes and edges have attributes that convey computational capabilities, requirements, performance measures, data requirements, and additional metadata. The QBG framework serves as a critical component for assessing the resources required for a given application, by composing the resource usage of the algorithms/modules upon which the application relies. We have developed a detailed QBG framework with cost modeling for both application instances, depicted in~\cref{fig:mg_qbg,fig:nb_qbg}.

In this section we present the workflows for both application instances: specifically, we study interface between water and magnesium (or magnesium-rich alloys), as well as the diffusivity of oxygen at high temperatures in niobium-rich alloys.
Magnesium-rich sacrificial coatings disrupt the electro-chemical processes that driving corrosion in aqueous environments, offering a promising avenue for effective corrosion mitigation and potential replacements for chromate-based coatings in aerospace aluminum alloys~\cite{johnson2007magnesium}. In contrast, Niobium alloys, with their high-temperature resistance, have great potential in the development of more efficient jet engines~\cite{Zhao2003UltrahighTemperatureMF,Prasad2017NiobiumAO}.

For both workflows, a major component is estimating the resources required for the quantum phase estimation (QPE) subroutine for GSEE. Resource estimations for QPE in the provided examples were calculated using the pyLIQTR software package~\cite{Obenland2024pyliqtr}. A summary of the resource estimates found is provided in \cref{tab:Summary}.

\subsection*{Magnesium and Mg-rich Alloys}
Here we give an overview of the full classical-quantum workflow, a more detailed exposition is provided in the supplementary information section \cref{sec: Mg corrosion}. In aqueous environment, the corrosion of Mg or Mg alloy is dominated by the highly-exothermic hydrogen evolution reaction (HER), which constitutes the cathodic part of the electrochemical corrosion reaction.
HER is further complicated by the highly-reactive, and short-lived corrosion intermediates formed during cathodic polarization, presenting a significant challenge in both computational and experimental investigations of the reaction mechanism. While experimental methods like mass loss, electrochemical characterization, and in situ monitoring of hydrogen gas evolution can provide corrosion rate measurements, these techniques are time-consuming, resource-intensive, and they do not offer detailed insights into the atomic-scale mechanisms of HER. As such, developing new materials that effectively mitigate corrosion in aqueous environments requires a deeper mechanistic understanding, and computational approaches, though challenging, are critical.

\subsubsection*{Workflow}

This workflow is designed to develop a comprehensive understanding of HER. The workflow is depicted schematically in the quantum benchmark graph (QBG) shown in \cref{fig:mg_qbg}, with dependencies of subroutines marked by arrows. A theoretical model for HER was proposed by~\citet{Williams2016Modeling}, based on the Volmer-Tafel reaction mechanisms, which decomposes HER into three simpler steps:
\begin{subequations}
    \begin{gather}
        \text{Mg}+2\text{H}_2\text{O}\rightarrow \text{Mg}(\text{OH}_{\text{ads}})(\text{H}_{\text{ads}})+\text{H}_2\text{O}
        \\
        \text{Mg}(\text{OH}_{\text{ads}})(\text{H}_{\text{ads}})+\text{H}_2\text{O}\rightarrow\text{Mg}(\text{OH}_{\text{ads}})_2(\text{H}_{\text{ads}})_2
        \\
        \text{Mg}(\text{OH}_{\text{ads}})_2(\text{H}_{\text{ads}})_2\rightarrow \text{Mg}(\text{OH}_{\text{ads}})_2+\text{H}_2(\text{g})
    \end{gather}
    \label{eq:pathways}%
\end{subequations}
\noindent where (ads) designates a species adsorbed on the surface and (g) designates a gaseous species.

The first stage of the computational workflow identifies the relevant reaction pathways associated with the reactions in \cref{eq:pathways}, using the nudged elastic band (NEB) method, a state-of-the-art approach for transition-state searches~\cite{Henkelman2000climbing}. NEB constructs a series of intermediate atomic configurations ("images") that represent intermediate configurations during the reaction. The resulting path is relaxed to the minimum energy path (MEP) between optimized reactant and product states, or roaming path, depending on the model for the reaction~\cite{Li2024Roaming}.
NEB simultaneously optimizes the geometries of all intermediate images, either serially or in parallel, but this process is computationally demanding due to the repeated ground-state energy calculations required.

In the second stage of the workflow, rate constants for reaction pathways are calculated using the Arrhenius equation:
\begin{equation}
    k=A\exp\left(-\frac{E_a}{RT}\right)
\end{equation}
where $A$ is the Arrhenius factor, $R$ the gas constant, and $T$ the temperature. Here, $E_a$ represents the energy difference between an initial state and a transition state. Accurate determination of $E_a$, often requiring a precision threshold of $10^{-3}$ Hartree, is crucial due to its exponential effect on $k$. If the MEP from NEB using DFT is sufficiently accurate, or if the discrepancy between DFT and high-fidelity methods is negligible, high-fidelity solvers can focus solely on refining $E_a$ for the reactant and transition state. However, the size of these models renders high-fidelity classical solvers intractable. At this stage, quantum computers, specifically through quantum phase estimation, can be strategically integrated to efficiently calculate $E_a$. While quantum computers could be used throughout the entire process, doing so might needlessly consume quantum resources, so strategically integrating quantum processing ensures improved efficiency and resource usage, as shown in the QBG in \cref{fig:mg_qbg}.

In this workflow, the practitioner (corrosion scientist or engineer) specifies the magnesium-rich alloy and solvent, the quantum chemist constructs the computational model (defining supercell size, slab geometry, accuracy thresholds, suitable basis sets, and performing transition state searches), and the quantum computing scientist uses these parameters—along with tools like PEST in pyLIQTR—to generate the electronic Hamiltonian for quantum algorithm execution~\cite{Obenland2024pyliqtr}.

By uncovering the most energy-efficient pathway for the HER on magnesium-alloy surfaces—using methods like comparing Gibbs free energies at various transition states or ab-initio molecular dynamics to directly track atomic movement, alongside analyses of chemical properties such as band structures and charge transfer—we gain insight into the reaction mechanism, and in this work, ground-state energies along the reaction coordinate are computed on a quantum computer.

To estimate the resource requirements for the quantum subroutines in our computational 
workflow on the magnesium--water interface, we selected three models of magnesium, 
labelled ``Mg Dimer,'' ``Mg Monolayer,'' and ``Mg Cluster,'' as well as the 
magnesium--aluminum alloy $\mathrm{Mg}_{17}\mathrm{Al}_{12}$, which is a secondary phase prevalent in alloys such as the AZ91 alloy. These four systems 
span a range of increasing size and electronic complexity, which in turn 
escalates their computational demands. The final resource estimates for these three 
models and the alloy are summarized in \cref{tab:Summary}. 
To reduce space--time volume in the quantum algorithms, we formulated the problem 
in first quantization.

\subsection*{Niobium-rich Alloys}

\begin{figure}[ht]
    \centering
	\includegraphics[scale=0.9]{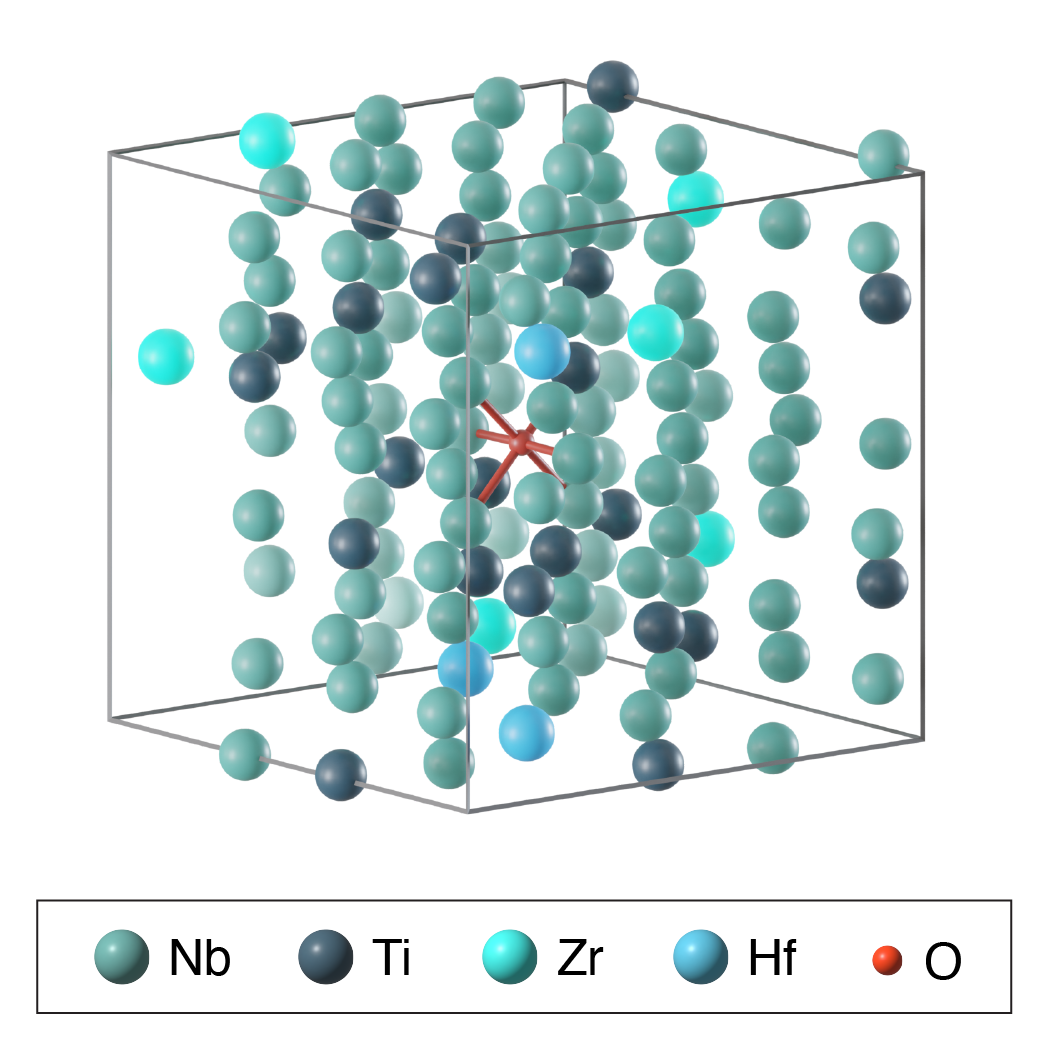}
	\caption{ Visualization of a simulation cell of a Nb alloy,   Nb$_{97}$Hf$_3$Ti$_{22}$Zr$_6$O, with interstitial oxygen.}
	\label{fig:nb_alloy_model}
\end{figure}

Niobium-rich refractory alloys are ideal candidates for aerospace, defense, and aviation applications due to their ductility, high melting point, and superior strength-to-weight ratios compared to Ni-based superalloys~\cite{Prasad2017NiobiumAO,Zhao2003UltrahighTemperatureMF}. Poor oxidation resistance in niobium necessitates identifying alloying elements that improve oxidation behavior, a challenge amplified by the vast search space of possible multi-component alloys, motivating the need for computational approaches that allows for efficient screening~\cite{Widom2016}. While modeling oxygen diffusivity within niobium alloys provides insight into their oxidation resistance, the computational cost dramatically increases with the addition of more alloying elements. 

Modeling oxygen diffusion using kinetic Monte Carlo (KMC) methods can help to identify promising alloys. While previous studies of binary Nb alloys  have had some success using cluster expansion methods~\cite{Samin2019ab}, the larger design space of multicomponent Nb-rich alloys, incorporating elements like Ti, Ta, Zr, W, and Hf, adds complexity to the search for optimal compositions.

\begin{figure*}[ht]
    \centering
    \includegraphics[width=\textwidth]{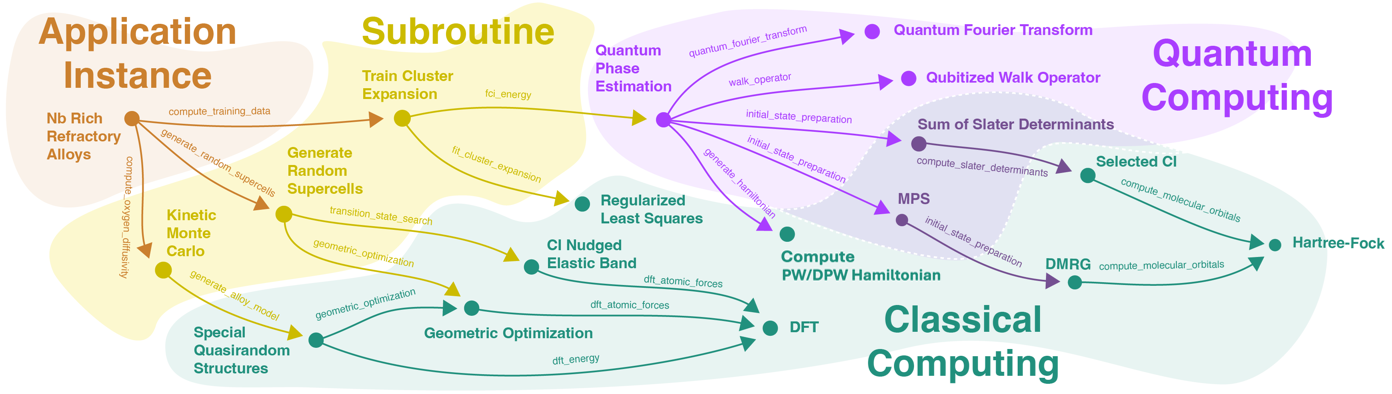}
    \caption{Quantum benchmark workflow for the niobium-rich refractory alloys. }
    \label{fig:nb_qbg}
\end{figure*}

\subsubsection*{Workflow}

In order to evaluate chemical formulas for different niobium-rich alloys, a set of alloying elements is selected. In general, the alloy's chemical formula is of the form $\text{Nb}_{(1-\sum_i x_i)} \text{A}_{x_1} \text{B}_{x_2}... \text{Z}_{x_n}$ where A through Z denote the $n$ alloying elements. This work focuses on compositions containing three or four alloying elements. The computational resource requirements grow with the size of the supercell, which must be large enough to accurately capture the proportions of each alloying element. Consequently, four-element alloys demand more computation than those with three elements (ternary), or simpler binary systems.

Once a set of alloying elements is selected, numerous alloy structures with varying amounts of each element are randomly generated and relaxed using density functional theory (DFT). These structures include an oxygen atom positioned at octahedral sites within the lattice, as well as transition state geometries between octahedral sites, obtained using the climbing-image nudged elastic band (CI-NEB) method~\cite{Henkelman2000climbing}. For each geometry, the electronic ground state energy is computed with as high accuracy as possible, serving as training data for cluster expansions (CEs). Fitting conventional CEs to multi-component alloys is difficult, however recently developed techniques provide \textit{embedded} CEs (eCEs) that can reproduce the energetics of quinary alloys using around 500 data points~\cite{Muller2024}. The objective is to compute high fidelity energy estimates for points for two separate cluster expansions: one for oxygen at octahedral sites and another for transition state geometries.

To determine the oxygen diffusivity $D(T)$ at high temperatures ($\ge1600$C(1873.15K)), one can simulate the movement of oxygen between interstitial sites in an atomic lattice. Using kinetic Monte Carlo (KMC) methods, oxygen atoms are propagated between these sites with probabilities proportional to the rate constants that govern each site-to-site jump. Determining this rate constant involves computing the energy barriers by solving the Hamiltonian DFT or other first-principles methods~\cite{Nakata2020Large}. Given the size of the search space, cluster expansions are incorporated into this workflow to obtain energy barriers~\cite{Samin2019ab}. Specifically, for a given choice of alloying elements, cluster models are trained on energies obtained from random configurations of the alloying atoms in various proportions. Once trained, the cluster model allows for the estimation of energy barriers of alloy models that sweep over many stoichiometries beyond those considered in the training dataset. This approach not only allows for an efficient computation of the energy barriers, but the cluster model can also provide entire phase diagrams~\cite{Walle2002Automating}.

At a high level, for a given set of alloying elements the objective is to estimate the energy barriers associated with oxygen transitioning from one octahedral interstitial site to another with high accuracy. During these transitions, the oxygen atom will pass through a transition state. For pure niobium, the transition state is an interstitial site with tetrahedral symmetry, and thus are amenable to classical computational methods. When considering disordered niobium alloys, the transition states are no longer retain perfect tetrahedral symmetry and require a transition state search algorithm to determine the geometry of the transition state, we consider the application of CI-NEB. 

The classical preprocessing in this workflow begins by generating a set of supercells, as shown in Fig.~\ref{fig:nb_alloy_model}, where alloying elements are embedded into a Niobium lattice for various ratios. Next, oxygen atoms are randomly placed at octahedral interstitial sites, and the entire supercell is relaxed. In addition, another set of transition-state geometries is produced by running a transition-state search algorithm. Both the octahedral and transition-state geometries are stored in a database~\cite{Samin2019ab}.
Special quasi-random structures are designed to statistically mimic the random distribution of atoms or elements in an alloy or compound while using a finite, periodic unit cell, making them computationally manageable.~\cite{Gao2016sqs}

We focus on modeling how oxygen diffuses in multi-component, Nb-rich alloys by using cluster expansions—a technique that approximates how different atomic arrangements affect the energy of an oxygen atom in the lattice~\cite{Muller2024}. First, we generate various alloy structures with different atomic configurations and compute their energies using high-accuracy electronic structure methods. These data points are then used to train the cluster expansions so that they can predict the solution energy of oxygen in either its usual (octahedral) site or a neighboring (tetrahedral) site that represents the transition state for migration.

Once the cluster expansions are fitted, we extract the energy barriers for oxygen hopping between adjacent octahedral sites by comparing these predicted energies. Finally, a KMC simulation uses those barriers to stochastically move oxygen atoms through the alloy and track the resulting diffusion over time. The KMC calculations yield estimates of the overall diffusivity of oxygen as a function of temperature. By identifying alloy compositions with lower diffusivities, we can highlight candidates that are likely to be more resistant to high-temperature oxidation, although other factors also play a role in selecting the best alloy.

\begin{figure*}[ht!]
        \centering
        \includegraphics[width=1 \linewidth]{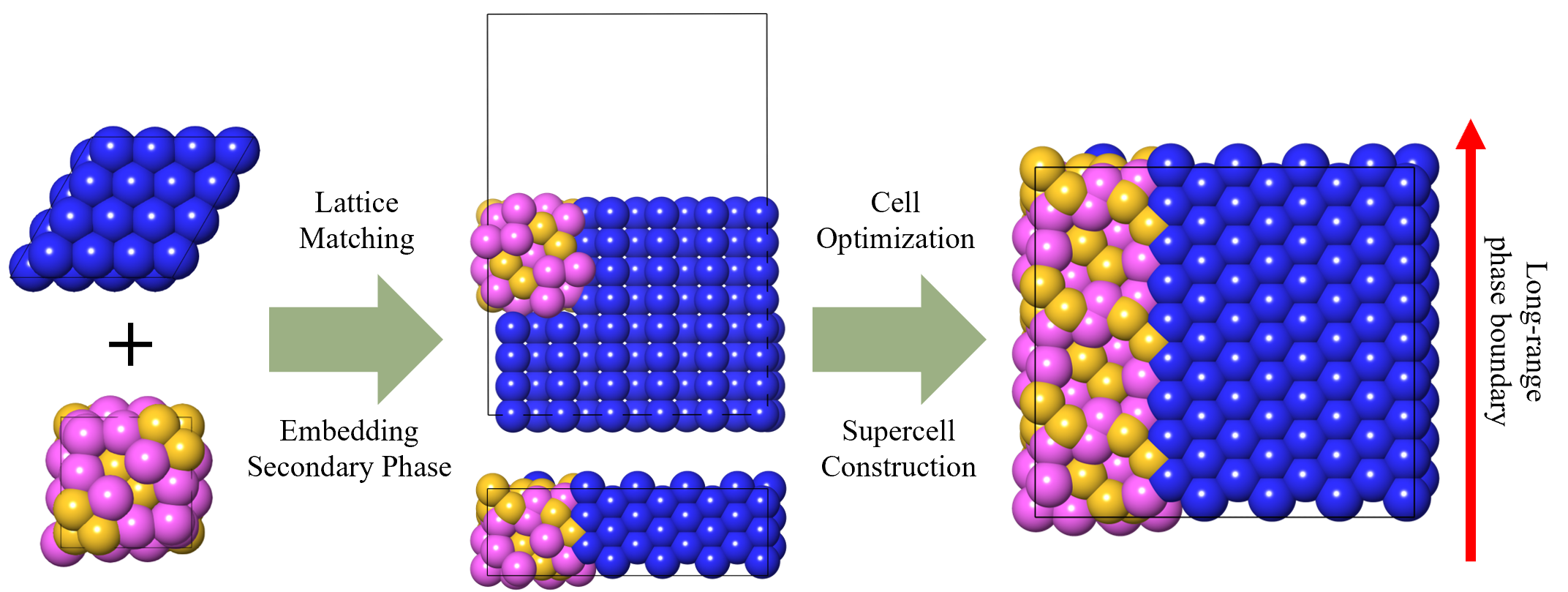}
        \caption{Schematic of approach for generating supercells of composite structures in magnesium alloys. Beginning with two independent primary and secondary phase structures, a guess-geometry is established by embedding Mg$_{17}$Al$_{12}$ within the primary structure and define a cubic periodic boundary that still maintains \textit{hcp} magnesium. Secondary phase aluminum is in gold, secondary phase magnesium is in pink, and primary phase magnesium is in blue.}
        \label{fig: alloy_supercell}
\end{figure*}

The inputs to this computational workflow are a set of alloying elements  $\{\mathrm{A},\mathrm{B},\mathrm{C}, \dots\}$ and specified ranges for their fractional
stoichiometries. For example, when considering alloys of the form $\mathrm{Nb}_{1-(x+y+z)}\,\mathrm{A}_x\,\mathrm{B}_y\,\mathrm{C}_z$, we focus on niobium-rich compositions by restricting $(x,y,z)$ to  $[0,a]\times[0,b]\times[0,c]$, where $a, b, c \ll 1$. The number of alloying elements and these composition ranges then determine the required size of the computational supercell, ensuring that there are enough atoms to accurately represent each element's proportion.

The primary output of this computational workflow is the oxygen diffusivity at various temperatures for different alloy compositions. If a cluster expansion is also trained on the alloy's formation energy, an additional phase diagram can be constructed to assess the alloy's stability and thermodynamics
across the composition range $(x,y,z)$. Because oxygen diffusivity $D(T)$ at elevated temperatures $T$ directly influences the rate of oxidation in niobium alloys, it is the key quantity of interest. Ideally, a protective oxide film forms---akin to the behavior in aluminum alloys---to prevent further oxidation ~\cite{Perkins1990Oxidation}.
\Cref{tab:Summary} presents the quantum resource estimates for four Nb-rich alloys that were selected using this computational workflow process.

\section*{Discussion}

CPC remains a significant challenge across various industries, resulting in enormous economic costs and impacts. Developing computational models to explain and/or predict corrosion typically focuses on empirical modeling using data from laboratory experiments, field tests or in-service findings. However, the application of physics-based techniques to model and understand fundamental corrosion mechanisms, including first principles approaches, has increased within the past decade as researchers have gained access to high performance computers and commercial software. 

Despite having access to supercomputers and highly-optimized classical codes, researchers are still limited in the sizes of physical models they can study. For the majority of first principles codes, a few hundred atoms is considered a large, computationally-demanding model: running an \textit{ab initio} calculation of a few thousand atoms is practically impossible without approximations (e.g., DFT), significant processing power and time~\cite{Nakata2020Large}. Furthermore, the DFT approximations that are typically used to reduce computational costs are known to give inaccurate results for highly-correlated states. This makes it challenging to develop atomistic models of corrosion that have both sufficient scale to mimic realistic systems and high fidelity to produce accurate, reliable predictions.  

This paper presents computational workflows associated with developing advanced corrosion-resistant materials and anti-corrosion coatings, focusing on magnesium-rich sacrificial coatings, corrosion resistant magnesium alloys, and niobium-rich refractory alloys. We assess the potential of quantum computing and its applicability to the described workflows. This assessment focuses on obtaining accurate ground state energy estimations for each computational model.

For the quantum algorithmic framework used in the present study, resource estimates indicate that the smallest model related to magnesium alloy design would require over 2,300 qubits, with the associated T-gate count for the QPE circuit being on the order of $10^{13}$. The largest computational model studied in this paper would require approximately 40,000 qubits and a T-gate count on the order of $10^{16}$. For computational models associated with the Nb-rich refractory alloys use case, the approximate number of qubits needed ranges from roughly 6,000 to 10,000, and the T-gate count for the targeted QPE algorithm ranges from $\sim10^{13}$ to $\sim10^{15}$. We believe these estimates are far from optimal, and that several research directions that could significantly reduce these resources while maintaining utility. For example, combining techniques like BLISS~\cite{Loaiza2023,*loaiza2023bliss} or tensor factorizations with the first quantized representation could further reduce the Hamiltonian 1-norm.

Further research into refining these algorithms is especially important when considering that the reported resource estimates are for a single instance of QPE. This is separate consideration from  the total quantum resources required to perform ground state energy estimation to high accuracy – which could potentially require many repetitions of this subroutine. The actual number of QPE shots needed depends on the overlap between the initial state generated by an ansatz and the true ground state  (see \cref{sec: classical vs quantum} for a more detailed discussion of this).

In summary, this work evaluates the quantum computing resources needed to address realistic, industrial-scale problems with high utility. By quantifying the resource requirements to study corrosion reaction mechanisms, this paper lays the groundwork for understanding the potential impact of quantum computing in the development of advanced corrosion-resistant materials.

\section*{Methods}

This section describes the methodology for constructing classical-quantum hybrid workflows using the QBG framework and the methods used to estimate resources presented in the main paper. QBG decomposes each workflow into individual nodes where each node represents a specific computational kernel along with its required inputs and actions. The key nodes are presented in \cref{tab: QBG Nodes} with their inputs, outputs and runtime complexity.

\begin{table*}
    \centering
    \begin{ruledtabular}
    \begin{tabular}{lcccc}
        \makecell{\textbf{Node}\\\textbf{Name}} & \makecell{\textbf{Quantum}/\\\textbf{Classical}} & \textbf{Input} & \textbf{Output} & \textbf{Runtime Complexity} \\
        \hline
        \makecell[l]{Geometry\\Optimisation} & \makecell{Classical} & \makecell{Initial Geo, Basis Set\\Functional} & Optimized Geo. & $O(N_{\text{iter}} N^3)$ \\
        NEB & \makecell{Classical} & \makecell{Product, Reactant,\\Image Geo, Basis Set,\\Functional} & Transition State Geo. & $O(N_{\text{iter}} N_{\text{image}} N^3)$ \\
        DFT & \makecell{Classical} & \makecell{Geometry, Basis Set,\\Functional} & \makecell{Energy, Atomic Forces,\\Vib. Freq.} & $O(N^3)$ \\
        Hartree-Fock & \makecell{Classical} & Geometry, Basis Set & Initial State & $O(N^4)$ \\
        QPE (PW) & \makecell{Quantum} & Initial State, Geometry & Energy, Ground State & $\widetilde{O}(N^3/\epsilon)$ \\
    \end{tabular}
    \end{ruledtabular}
    \caption{Computational Kernels for the computational workflows described in the designing of corrosion-resistant magnesium alloys in aqueous environment and corrosion resistant niobium alloys in temperature. $N$ is a generic parameter for number of basis functions used in the calculation (e.g., localized Gaussian orbitals, plane waves), $N_\text{iter}$ is the number of iterations of geometric optimization, and $N_\text{image}$ is the number of images (geometries) used in a nudged elastic band (NEB) calculation.}
    \label{tab: QBG Nodes}
\end{table*}

\subsection{Magnesium}

Various surface and alloy environments were selected to facilitate scalable benchmark tests based on system size and to focus on chemically relevant species for corrosion studies in industrial applications. Magnesium slabs were selected as baselines due to the breadth of experimental~\cite{Zhang2018Effect} and computational work~\cite{Gujarati2023,Polmear2017six}, as well as their widespread industrial relevance~\cite{bai2023applications,Golroudbary2022Mag}. We concentrate on three configurations of uniform magnesium as well as an alloyed magnesium-aluminum slab to assess the quantum computation capabilities for modeling surface chemistry.

We explore an \textit{hcp} Mg(0001) slab interacting with water at its surface in three different sizes to examine resource estimation. The smallest system is a \textbf{4 $\times$ 4 $\times$ 4} slab with one adsorbed water molecule (70 atoms), comparable to DFT and CASSCF studies~\cite{Gujarati2023}. The mid-sized system, \textbf{6 $\times$ 6 $\times$ 5}, includes a water monolayer of 26 molecules (257 atoms), reflecting larger DFT analyses~\cite{wurger2020first}. Lastly, a \textbf{6 $\times$ 6 $\times$ 6} slab with 99 water molecules and 36 adsorbed hydroxyl groups (586 atoms) aligns with AIMD-level simulations~\cite{Fogarty2022}.

For alloy modeling, we focus on AZ91 and EV31A, both recognized for their corrosion resistance. These alloys have a pure \textit{hcp} Mg(0001) $\alpha$-phase and a dominant $\beta$-phase consisting of cubic Mg$_{17}$Al$_{12}$ in AZ91, and Mg$_{12}$RE$_{2}$ or Mg$_{3}$RE in EV31A, where RE is a rare-earth metal such as neodymium or gadolinium. Given these alloys' complexities, it is essential to model surface interactions both at the $\beta$-phase alone and at boundary layers between primary and secondary phases.

The first proposed approach for capturing this complex environment is shown in \cref{fig: alloy_supercell}. Primary and secondary phase structures for AZ91 were obtained from the Materials Project Materials Explorer~\cite{Jain2013Commentary}. The \textit{hcp} Mg(0001) $\alpha$-phase was extended to a \textbf{12 $\times$ 6 $\times$ 12} rectangular supercell (lattice constants $a=31.7$\,{\AA}, $b=18.3$\,{\AA}, $c=42.6$\,{\AA} and angles $\theta=\varphi=\xi=\SI{90}{\degree}$) to align with the secondary Mg$_{17}$Al$_{12}$ cubic structure. A large slab was chosen to capture various boundary-layer sizes, including subsurface secondary phases and extended periodic regions dominated by the second phase. All magnesium slab supercells include a vacuum layer for consistency in surface interaction studies. Supercell structures were built using the Schr\"{o}dinger Materials Science Suite version 2024-2~\cite{matstudio}.

Modeling corrosion environments typically begins with selecting surface geometries that accurately reflect the chemical phenomena yet remain computationally tractable. This process often starts with DFT, a method known for its computational efficiency by establishing a one-to-one mapping between the electronic density and the external potential in the electronic Hamiltonian~\cite{Jones2015Density}. DFT is used to locate optimum points on the nuclear potential energy surface, revealing stable geometries and transition states. In corrosion chemistry, such optimizations commonly employ periodic boundary conditions to mimic lattice periodicity and capture surface interactions with corrosive or protective agents. Consequently, DFT provides an initial exploration of the model space and generates structures that inform higher-fidelity simulations.

In our studies of corrosion chemistry in Mg, we often care not only about the minima along the nuclear potential energy surface but also saddle points. These saddle points represent paths of least resistance during reactions, providing critical insight into reaction kinetics, and can be directly used in larger length-scale simulations~\cite{Ke2019,Li2024Roaming}. To complete transition state searches on surfaces, NEB methods interpolate images along a reaction coordinate from an initial reactant and product pair that are then iteratively optimized in search of potential transition states. NEB simulations leverage periodic DFT during this process, and scans are often performed at increasing \textbf{k}-point grid selections to systematically verify the convergence of geometries.

\subsection{Niobium}

Since we are interested in the diffusion of oxygen, we can leverage cluster expansions to fit the solution energy $E_{\text{sol}}$, as outlined in ~\cite{Samin2019ab}. For niobium alloyed with Zr, Ta, and W, $E_{\text{sol}}$, is computed as
\begin{equation}
    \begin{split}
        E_{\text{sol}} =& E(\text{Nb}_{1-(x+y+z)} \text{Zr}_x \text{Ta}_y \text{W}_z O)\\
        &- E(\text{Nb}_{1-(x+y+z)} \text{Zr}_x \text{Ta}_y \text{W}_z) - \frac{1}{2} E(O_2).
    \end{split}
\end{equation}
Let $E_{\mathrm{sol}}^{\mathrm{Oct}}$ and 
$E_{\mathrm{sol}}^{\mathrm{Tetra}}$ represent the solution energies 
computed for oxygen at fixed locations within the alloy matrix. These locations correspond, respectively, to ``octahedral''  reactant/product geometries and ``tetrahedral'' transition-state geometries obtained via CI-NEB calculations. For each type of geometry, we introduce a cluster expansion,
\[
F^{\mathrm{Oct}} : \boldsymbol{\Omega} \to \mathbb{R}
\quad\text{and}\quad
F^{\mathrm{Tetra}} : \boldsymbol{\Omega} \to \mathbb{R},
\]
which share the general form
\[
F(\boldsymbol{\sigma})
= J_{0} \;+\;\sum_{\alpha} m_{\alpha}\,J_{\alpha} \,\Theta_{\alpha}(\boldsymbol{\sigma}).
\]
Here, $\alpha$ indexes all symmetrically distinct clusters of lattice sites; we  denote by $m_{\alpha}$ the multiplicity of cluster set $\alpha$ per unit 
cell, and by $J_{\alpha}$ the effective cluster interaction (ECI) coefficients, 
which are obtained by fitting this expansion to computed energies. The functions 
$\Theta_{\alpha}(\boldsymbol{\sigma})$ take the form
\[
\Theta_{\alpha}(\boldsymbol{\sigma})
= \frac{1}{N_{\alpha}} \sum_{\beta \subset \alpha} \prod_{i=1}^{N}
\phi_{\beta_i}\bigl(\sigma_i\bigr),
\]
where $N_{\alpha}$ is the number of sub-clusters $\beta \subset \alpha$, 
and $\sigma_i$ are ``spin'' variables indicating which atom occupies site 
$i$. The basis functions $\phi_{\beta_i}(\sigma_i)$ can be, for instance, 
simple indicator functions (one-hot encoding) or alternative forms such as 
polynomial or trigonometric functions.

The configuration vector $\boldsymbol{\sigma}$ corresponds to possible choices for various elements at each position in the alloy. The configuration vector for a lattice of $M$ sites is an element of a space of configurations $\boldsymbol{\sigma} \in \Omega_1 \times \Omega_2 \times \cdots \Omega_{M} = \boldsymbol{\Omega}$, for example a 4 site lattice could be described by $\boldsymbol{\Omega} = \{0 ,1,2,3\}^4 $ and $\boldsymbol{\sigma}=(0,1,0,3)$ where $0$ is Nb, $1$ is Zr, $2$ is Ta, and $3$ is W. For the training data, we collect energies computed at the full-configuration interaction (FCI) level for $E_{\text{sol}}^{\text{Oct}}$ and  $E_{\text{sol}}^{\text{Tetra}}$ corresponding to configurations $(\boldsymbol{\sigma})$. This training is then used to find an optimal set of ECI coefficients $\boldsymbol{J}=(J_0,J_{\alpha_1},...)$ which requires us to solve the following least squares problem. 

Let $\mathbf{\Pi}(\boldsymbol{\sigma})$ be the vector of functions $\Theta_{\alpha_i} (\boldsymbol{\sigma})$ for $i=1,...,d$
\begin{equation}
    \mathbf{\Pi}(\boldsymbol{\sigma}) = (\Theta_{0} (\boldsymbol{\sigma}) \equiv 1, \Theta_{\alpha_1} (\boldsymbol{\sigma}),\Theta_{\alpha_2} (\boldsymbol{\sigma}),...  ) \in \mathbb{R}^d.
\end{equation}
\noindent We can then compute the cluster expansion with a dot product
\begin{equation}
    F(\boldsymbol{\sigma}) = \mathbf{\Pi}(\boldsymbol \sigma) \cdot \boldsymbol{J}.
\end{equation}

\noindent We then construct the matrix $\boldsymbol{\Pi}$ that contains the values for the vector $\mathbf{\Pi}(\boldsymbol{\sigma}_j)$ for each training configuration $\boldsymbol{\sigma}_j$ for $j=1,...,m$
\begin{equation}
    \boldsymbol{\Pi} = (\mathbf{\Pi}(\boldsymbol{\sigma}_1), \mathbf{\Pi}(\boldsymbol{\sigma}_2),...,\mathbf{\Pi}(\boldsymbol{\sigma}_m))^T \in \mathbb{R}^{m \times d}
\end{equation}
\noindent Let $\boldsymbol{E} = (E_1,...E_m)$ be the FCI-level energies associated with the $m$ training structures, then the least squares problem is posed as
\begin{equation}
    \boldsymbol{J}^* = \argmin_{\boldsymbol{J}} \| \boldsymbol{\Pi} \boldsymbol{J} - \boldsymbol{E}\|_2^2 + \rho(\boldsymbol{J}).
\end{equation}
\noindent where $\rho$ is a regularization term for which there are few different choices. 

Evaluating the quality of the fit typically involves computing cross-validation (CV) and related quantities. The canonical choice is the leave-one-out CV which is given by
\begin{equation}
    CV = \sqrt{\frac{1}{n}\sum_{i=1}^n (E_i - \hat{E}_i)^2}
\end{equation}
\noindent where $E_i$ are the FCI-level energies computed for structure $i$ and $\hat{E}_i$ is the predicted energy from a cluster expansion trained on other $(n-1)$ structures~\cite{Samin2019ab}. A good score for leave-one-out cross-validation is below $5$ m$e$V/\text{atom}.

Once sufficiently accurate cluster expansions are computed, we can estimate the energy barrier between octahedral sites $i$ and $j$, denoted by $E_{i \to j}$, as in Ref.~\cite{Samin2019ab}
\begin{equation}
    E_{i \to j} = F^{\text{Tetra}}(\boldsymbol{\sigma}_{i\to j}) - F^{\text{Oct}}(\boldsymbol{\sigma}).
\end{equation}
These energy barriers are then fed into KMC simulations of oxygen on a random walk through the alloy's sublattice of ``octahedral'' sites. As outlined in Ref.~\cite{Samin2019ab}, for each time step, we generate two random numbers $r_1 , r_2 \sim \text{Unif}(0,1)$, which determines the transition path and the moving time of the simulation. If a system in state $i$, then the transition rate between $i$ and another site $j$ in the transition path $P = (i_1, i_2, ..., i_{n-1}, j)$
\begin{equation}
    k_{i \to j} = \nu_0 \exp(- E_{i \to j}/ k_B T)
\end{equation}
where $T$ is temperature, $k_B$ is the Boltzmann constant, and $\nu_0$ is the attempt frequency, defined as 
\begin{equation}
    \nu_0 \approx \left(\prod_{i=1}^{3N-4}\frac{\nu_i^{\text{IS}}}{ \nu_i^{\text{TS}}}\right)\nu_{3N-3}^{\text{IS}}.
\end{equation} 
In the interest of computational efficiency, we adopt the strategy outlined in Ref.~\cite{Samin2019ab}, setting  $\nu_0$ to be the value for pure Nb and true octahedral and tetrahedral initial (IS) and transition state (TS).

The total transition rate $k_{\text{tot}}$ is the sum of all $k_{i \to j}$ for all possible moving paths $(i_1, i_2, ..., i_{n-1}, j)$. If
\[
\sum_{P_{n-1}} k_{i \to j} \leq r_1 k_{\text{tot}} \leq \sum_{P_n} k_{i \to j}
\]
where $P_{n-1} = (i_1, i_2, ..., i_{n-1})$ and $P_n = (i_1, i_2, ..., i_{n} = j^*)$, then $j^*$ is selected for $i_n$ in the path $P$. The system moves to the next state with moving time calculated according to the time step equation ${\delta t = (1/k_{\text{tot}}) \ln(1/r_2)}$. This random process is repeated in order to estimate the diffusivity at various temperatures $T$ given by the formula
\begin{equation}
    D(T) = \frac{1}{n_s} \sum_{m=1}^{n_s} \frac{\| \boldsymbol{x}(t_m) - \boldsymbol{x}(t_{m-1}) \|_2^2}{6 \delta t}
    \label{eq:diffusivity_at_T}
\end{equation}
\noindent where $n_s$ is the number of sampled trajectories, $\boldsymbol{x}(t_{m})$ is the position of an oxygen atom at time $t_m$ and ${\boldsymbol{x}(t_{m}) = \boldsymbol{x}(t_{m-1} + \delta t)}$. The KMC calculation assumes that $\delta t \gg \nu_0^{-1}$ that kinetic jump times of oxygen are large enough to include all local jump correlations and $n_s$ should also be large enough to obtain a statistically meaningful diffusion coefficient~\cite{Samin2019ab}. Alloy compositions with low diffusivities are selected as this is an indication of their oxidation resistance, however, this is not the only factor to consider. Therefore, this computational workflow helps narrow down the search for a desirable alloy formula.

\subsection{Hamiltonian Generation}
\label{sec: Hamiltonian Generation}

A central challenge in electronic-structure calculations is choosing a finite basis set to discretize the Hamiltonian. Under the Born--Oppenheimer approximation, wavefunction ``cusps'' appear at nuclear coordinates, often suggesting local Gaussian orbitals for molecular systems. However, for large periodic systems, plane-wave (PW) bases are typically more suitable. Here, we adopt a first-quantized PW basis following Ref.~\cite{Su2021Fault}, defining plane waves on a cubic reciprocal lattice:
\begin{equation}
\label{eq:G}
    \begin{split}
        k_p &= \frac{2 \pi p}{\Omega^{1/3}}, 
        \qquad \qquad p \in G,
        \\
        G &= \left[-\frac{N^{1/3}-1}{2}, \frac{N^{1/3}-1}{2}\right]^3 \subset \mathbb{Z}^3,
    \end{split}
\end{equation}
where $N$ is the total number of basis functions or grid points, and $\Omega$ is the volume of the simulation cell~\cite{Martin2004}. 
One advantage of the plane wave basis is that it enables quantum algorithms with lower asymptotic gate complexity compared to those using Gaussian orbitals.
In first quantization, the Born--Oppenheimer Hamiltonian is represented as:
\begin{equation}
    \label{eq: dual plane-wave first quant Ham}
    H_{\rm BO} = T + U + V + \frac{1}{2}\sum_{\ell \neq \kappa=1}^{L} 
    \frac{\zeta_\ell \zeta_\kappa}{\left\|R_\ell - R_\kappa\right\|},
\end{equation}
where $r_i$ represent the positions of electrons whereas $R_\ell$ represent the positions of nuclei, 
$\zeta_\ell$ are the atomic numbers of nuclei; the kinetic ($T$), nuclear potential ($U$), and electron--electron interaction ($V$) terms are defined as:
\begin{align*}
    T &=  \sum_{i=1}^{\eta}  \sum_{p\in G} \frac{\left \| k_p\right\|^2}{2} 
    \ket{p}\!\bra{p}_{i}, \\
	U & =-\frac{4\pi}{\Omega}\sum_{\ell=1}^{L}\sum_{i=1}^{\eta}\sum_{\substack{p,q\in G\\ p\neq q}}
	\Bigg(\zeta_\ell\frac{e^{ik_{q-p}\cdot R_\ell}}{\norm{k_{p-q}}^2}\Bigg)\ket{p}\!\bra{q}_i, \\
    V &=\frac{2 \pi}{\Omega} \sum_{i\neq j=1}^{\eta}\sum_{p,q\in G} 
    \sum_{\substack{\nu\in G_0\\ (p+\nu)\in G\\ (q-\nu)\in G}}
    \frac{1}{\left\| k_{\nu}\right\|^2}
    \ket{p + \nu}\!\bra{p}_i \,\ket{q-\nu}\!\bra{q}_j,
\end{align*}
with $G_0 = [-N^{1/3}, N^{1/3}]^3 \subset \mathbb{Z}^3 \backslash\{(0,0,0)\}$ being the set of valid frequency differences. The wavefunction is stored in a computational basis that encodes configurations of $\eta$ electrons among $N$ basis functions, i.e., 
$\ket{\varphi_1 \varphi_2 \cdots \varphi_{\eta}}$, 
where each $\varphi_j$ denotes the index of an occupied basis function.

The discretization error in this PW approach is asymptotically equivalent to Galerkin discretizations using other single-particle bases (including Gaussian orbitals)~\cite{Babbush2018LowDepth}. 
Moreover, the first quantization framework offers a much more space-efficient (algorithmically) representation compared to the second quantization formalism. This is particularly critical for our calculations, as we require a large number of plane waves to accurately capture the system's electronic structure. However, we also consider the second quantization framework. In Supplementary Materials~\ref{supp_mat:second_quantization}, we discuss the Hamiltonian formulation and the implementation of the quantum algorithm within this framework, following the approach of~\cite{babbush2018encoding}.

\subsection{Vacuum Space and Wavefunction Cutoff Convergence}
\label{sec: cutoff convergence}

Another key consideration is ensuring wavefunction convergence with respect to the vacuum space allocated for simulating surface environments. In surface chemistry, periodicity is typically restricted to the \textit{xy}-plane, so the \textit{z}-dimension must be chosen carefully to prevent surface molecules from interacting with the metal slab’s periodic image in the next cell. However, increasing the cell size also increases the required \textbf{k}-point sampling, adding to the overall computational expense.

\begin{figure*}[ht]
    \centering
    \includegraphics[scale=0.75]{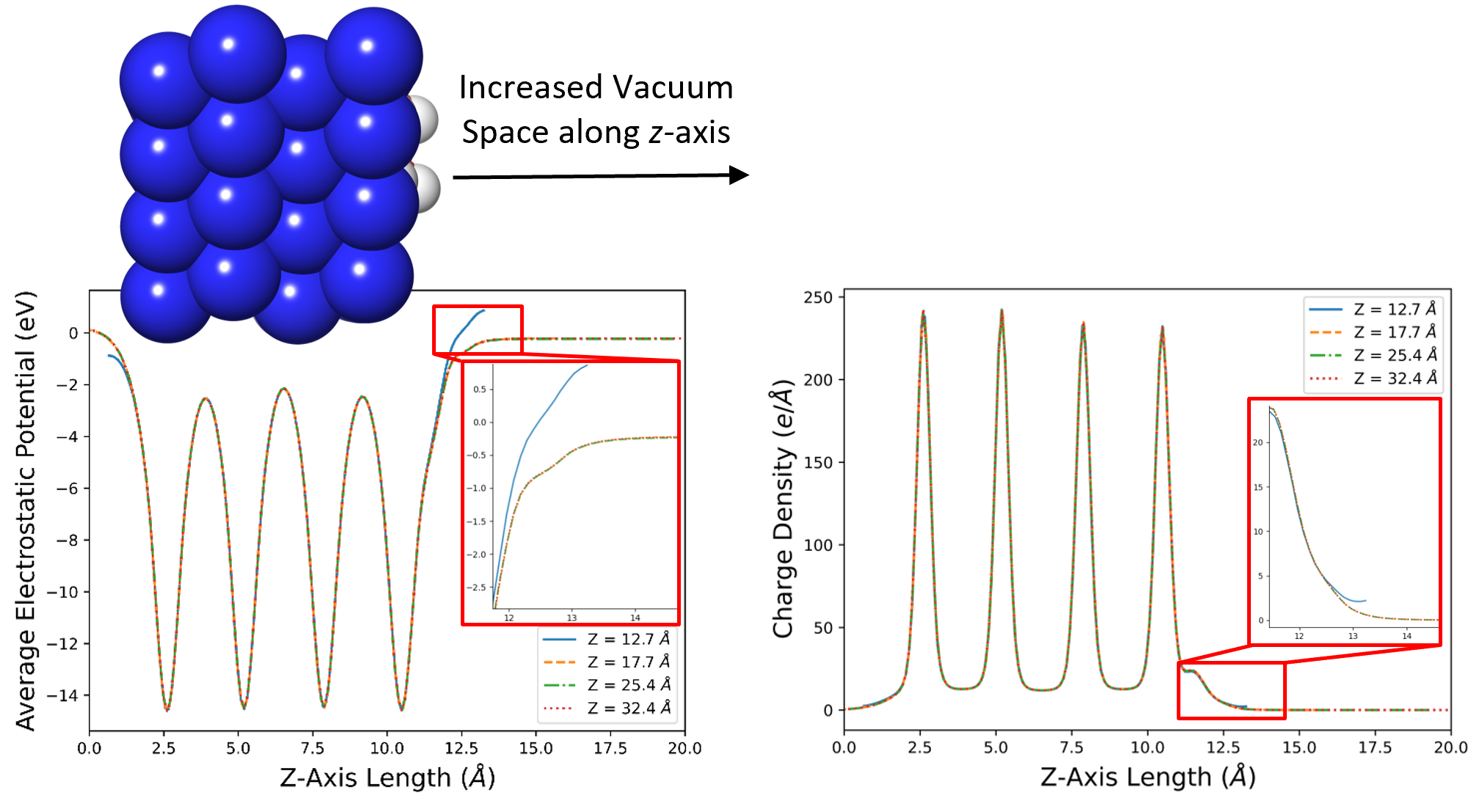}
    \includegraphics[scale=0.35]{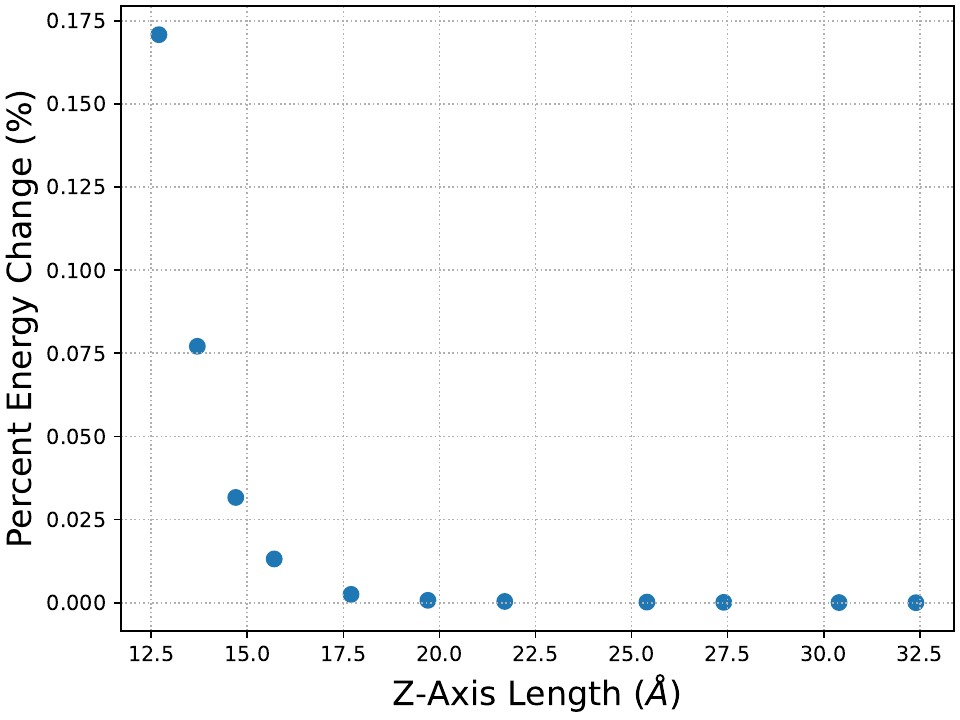}
    \caption{Determination of vacuum-space limits using electrostatic potential (top, left), charge density (top, right), and energetic convergence (bottom). Insets highlight the potential and charge density attributed to a water dimer at the surface interface.}
    \label{fig:vac_conv}
\end{figure*}

Vacuum-space convergence is assessed by examining electrostatic potentials, charge densities, and energies as the \textit{z}-axis length is varied (see \cref{fig:vac_conv}). We investigated vacuum layers between 12.7\,\AA\ and 32.4\,\AA. While the energetic analysis considered the entire range, representative cases of 12.7\,\AA, 17.7\,\AA, 25.4\,\AA, and 32.4\,\AA\ are shown for electrostatic potential and charge density. From these data, a 25.4\,\AA\ vacuum ensures that surface species do not artificially interact with their periodic images, aligning with previous recommendations~\cite{Williams2016Modeling}. Notably, this is about twice the original \textit{z}-axis dimension of the 64-atom magnesium lattice. However, more extensive vacuum layers may be required for systems featuring additional adsorbed species or realistic environment modeling.

With the cell size determined, the next step is to set the wavefunction cutoff for chemically accurate simulations. 
Furthermore, physically, the  high-frequency PW modes are used to reconstruct rapid wavefunction oscillations near atomic nuclei (the so-called ``nuclear cusp''). Fortunately, these oscillations are generally tied to (often) irrelevant core electrons. Classical calculations mitigate this by  using  so-called pseudopotentials (PPs) to represent core electrons as a classical potential and reduce computational complexity. Since plane-wave calculations are widely used for classical electronic structure (DFT, etc.), we can often adopt a validated cutoff energy.  In the context of ultrasoft pseudopotentials, we can define a low- and high-resolution bases as having kinetic cutoffs of $E_c = 30$ Ry and $E_c = 60$ Ry respectively.  This maps to a real-space lattice constant via the equation $a_0 = \gamma \sqrt{2 \pi^2 / E_c} \sim \gamma \lambda_\text{cut} $, where we assume atomic units, i.e., $a_0$ measured in Bohr radii and $E_c$ in Hartree (1~Ry~=~2~Ha), and $\lambda_\text{cut} $ is the wavelength corresponding to the highest-energy plane-wave mode and $\gamma$ is a scaling factor.  Setting the scaling factor to $\gamma = 1.0$ will place grid points at the highest PW wavelength, while $\gamma = 0.5$ will apply these every half-wavelength. An ideal value would depend on pseudopotential details, though likely this would be closer to $\gamma = 0.5$. 

We have determined the necessary wavefunction cutoffs and vacuum layer convergences to reduce computational costs with minimal energetic penalties for our specific application instances. Wavefunction convergence criteria were tested for Vanderbilt ultrasoft (USPPs), Goedecker-Teter-Hutter (GTH)~\cite{Goedecker1996Separable}, and Hartwigsen-Goedecker-Hutter (HGH)~\cite{Harwigsen1998Relativistic} pseudopotentials. Systematic evaluation of these pseudopotentials across different convergence criteria informs decisions during the resource estimation stage. See Section \ref{supp_mat:first_quantization} in the Supplementary Materials for more details.   

\subsection{Quantum Subroutines}
\label{sec: Quantum subroutines}

In this work, we employ the qubitized quantum phase estimation (QPE) algorithm, as described in~\cite{Su2021Fault}, to extract spectral information, specifically the ground state energies, for chemical systems of interest. Qubitized QPE encodes information about the Hamiltonian using a Szegedy walk via the technique of qubitization. The encoding procedure is as follows.

\begin{figure*}
    \centering
    \includegraphics[width = \textwidth]{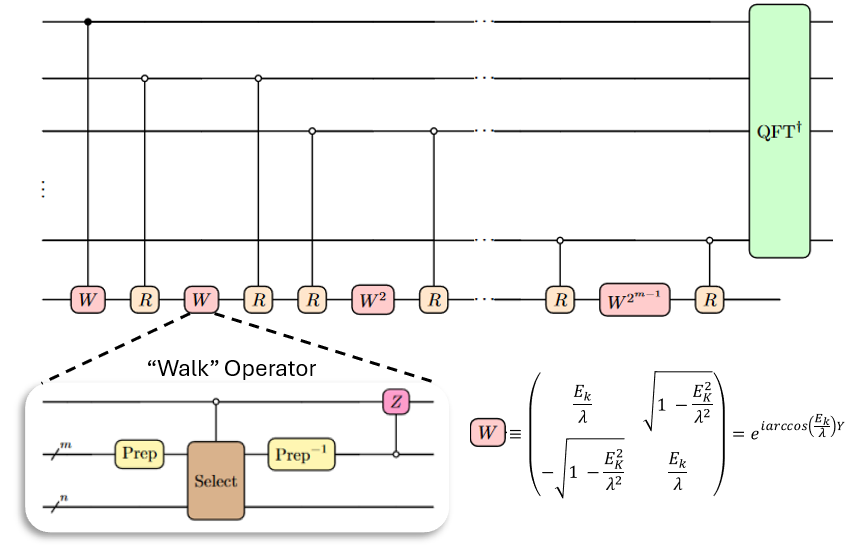}
    \caption{Quantum circuit illustration of quantum phase estimation algorithm based on qubitization. This algorithm queries powers of the qubitized walk operator $W$ instead of the standard time evolution oracle. The eigenphases $\phi_k$ of $W$ have a simple functional relation with the eigenenergies $E_k$ of $H$. Specifically, $\phi_k \approx  \text{arccos}(E_k / \lambda)$, where $\lambda$ is a normalization factor in the block encoding procedure of $H$ and can be taken as the 1-norm of $H$. This relationship allows us to extract the eigenenergies using phase estimation. }
    \label{fig: qubitized_phase_estimation}
\end{figure*}

First, we express the electronic structure Hamiltonian as a linear combination of unitaries (LCU), i.e., 
\begin{equation}
    H = \sum_{\ell=0}^{L-1} w_\ell H_\ell \hspace{1 cm} H_\ell^2 = \mathbf{1}
\end{equation}
where $w_\ell\in \mathbb{R}^+_0$ and $H_\ell$ are self-inverse operators on $n$ qubits. In particular, $H$ is the Hamiltonian arise from performing Galerkin discretization of the Hamiltonian under the plane-wave basis functions as described in Methods~\ref{sec: Hamiltonian Generation}. Next, we encode the Hamiltonian spectra in a Szegedy quantum walk $W$ defined as 
\begin{equation}
    W = R \cdot \text{SELECT}, \hspace{1 cm} R = 2 |\mathcal{L} \rangle \langle \mathcal{L} | \otimes \mathbf{1} - \mathbf{1}.
\end{equation}
where $\ket{\mathcal{L} }$ is a state that can be prepared by the ``preparation oracle'', referred to as PREPARE, and SELECT is the ``Hamiltonian selection oracle''. PREPARE is defined as 
\begin{subequations}
    \label{eq: PREPARE oracle}
    \begin{gather}
        \text{PREPARE} \equiv \sum_{\ell=0}^{L-1} \sqrt{\frac{w_\ell}{\lambda}} \ket{\ell} \bra{0},\\
        \text{PREPARE} |0\rangle^{\otimes \log L} \mapsto \sum_{\ell=0}^{L-1} \sqrt{\frac{w_\ell}{\lambda}} |\ell \rangle \equiv |\mathcal{L} \rangle
    \end{gather}
\end{subequations}
with $\lambda$ (1-norm) defined as $ \lambda= \sum_{\ell=0}^{L-1} w_{\ell} $, and SELECT is defined as 
\begin{subequations}
    \label{eq: SELECT oracle}
    \begin{gather}
        \text{SELECT} \equiv \sum_{\ell =0}^L \ket{\ell} \bra{\ell} \otimes H_{\ell}, \\
        \text{SELECT} |\ell\rangle |\psi \rangle \mapsto |\ell\rangle H_\ell |\psi \rangle.
    \end{gather}   
\end{subequations}
Note that $R$ and SELECT are both reflection operators, ensuring that \(W\) takes the form of a Szegedy walk. In particular, \(W\) exhibits a block-diagonal structure with each block corresponding to a two-dimensional subspace spanned by \( \{|0\rangle |\phi_k \rangle, |\psi^\perp \rangle \} \), where \(|\phi_k \rangle \) is an eigenstate of \(H\) with eigenvalue \(E_k\), and \(|\psi^\perp \rangle\)  is a state that is orthogonal to \(|0\rangle |\phi_k \rangle\). Furthermore, within the subspace \(\{|0\rangle |\phi_k \rangle, |\psi^\perp \rangle \}\), \(W\) acts as follows:
\begin{subequations}
    \label{eq: action of W}
    \begin{gather}
        W |0\rangle |\Phi_k\rangle = \cos(\theta_k) |0\rangle |\Phi_k\rangle - \sin(\theta_k) |\psi^\perp\rangle, \\
        W |\psi^\perp\rangle = \cos(\theta_k) |\psi^\perp\rangle + \sin(\theta_k) |0\rangle |\Phi_k\rangle.
    \end{gather}   
\end{subequations}
where \( \cos(\theta_k) = E_k / \lambda \). The qubitization procedure exploits this structure to encode spectral information in the phase of \( W \), enabling the use of QPE on \(W\) to efficiently extract eigenvalues of \(H\).
The $\lambda$ parameter has a significant impact on the computational cost of the algorithm and is therefore an important parameter in the resource estimation. Hence, we report this value for each of the instances considered in this paper as shown in Sections \ref{sec: supp_resource_estimates_1st_quantization} and \ref{sec: supp_resource_estimates_2nd_quantization}. 
The overall algorithm can be implemented with a quantum circuit as shown in Figure \ref{fig: qubitized_phase_estimation}. 
Note that this particular implementation of the QPE circuit differs from the standard approach, where one typically applies a controlled operation on each of the \(W^{2^k}\) operators. In this implementation, instead of leaving the ancilla qubit in the \( |0 \rangle \) state unchanged, we apply an inverse unitary. This adjustment effectively doubles the effectiveness of the phase estimation procedure. It can be achieved by simply removing the controls from the \(W^{2^k}\) operators and inserting controlled reflection operators \(R\) into the circuit, as shown in Figure \ref{fig: qubitized_phase_estimation}.
We discuss the implementation of each component in the first quantization framework in more detail in the Supplementary~Materials~\ref{supp_mat:first_quantization}. We then discuss the implementation of this algorithm in the second quantization framework, following the Linear-T encoding by Babbush et al.~\cite{babbush2018encoding}, in Section \ref{supp_mat:second_quantization}.  The code for our implementation is available in Ref.~\cite{Obenland2024pyliqtr}.

A major factor in the computational expense of using QPE for ground-state energy estimation lies in choosing the initial state for the algorithm. The success probability of QPE depends on how well this initial state overlaps with the true ground state. In particular, the likelihood of successfully projecting onto the ground state scales inverse polynomially with the overlap between the initial state and the true ground state.

In much of the quantum computing for quantum chemistry literature, the initial state is determined by the Hartree-Fock (HF) state—a single Slater determinant used as the starting point for QPE. Although HF works well for smaller chemical systems, there are concerns about its ability to achieve adequate overlap with the true ground state in larger systems. Since initial state preparation is beyond the purview of this work, we defer this discussion to Supplementary~Materials.

\subsection{Tools and Methodology}
The pyLIQTR workflow consists of specifying a Hamiltonian, selecting a block encoding, and building the circuit for the desired algorithm based on those choices. One feature of the pyLIQTR circuits is their hierarchical nature, where the top-level circuit is the entire qubitized algorithm which can then be decomposed in stages all the way down to one and two qubit gates. This structure allows for more efficient resource estimation, since repeated elements can be analyzed once with the result cached and used again later.

The circuits for our application instances were generated as follows. First, the Hamiltonian coefficients for the PW basis output by PEST were fed into the pyLIQTR circuit generation framework. 
Then the pyLIQTR implementation of the first quantization of the electronic structure block encoding presented in Ref.~\cite{Su2021Fault} was chosen. 
Next, the qubitized phase estimation circuit was generated for an error target of $0.001$. Finally, this circuit was passed into the pyLIQTR resource estimation protocol.

The pyLIQTR resource estimation protocol is built atop Qualtran’s T-complexity methods~\cite{Qualtran} and can be used to determine the Clifford+T cost of any circuit generated with pyLIQTR. For a given circuit, the resources reported by pyLIQTR include logical qubits, T gates, and Clifford gates. The gate counts include the additional T and Clifford gates from rotation synthesis, which are estimated using a heuristic model that depends on the user specified rotation gate precision. For the resources reported in the following sections, a gate precision of $10^{-10}$ was used. This choice of gate precision was driven by the targeted energy error value of $10^{-3}$. Consequently, the upper bound for gate precision should be on the order of $10^{-6}$. Therefore, a gate precision value of $ 10^{-10}$ was selected as a lower bound. This implies that our resource estimate values will be higher than actually needed.

\section*{Acknowledgements}
This material is based upon work supported by the Defense Advanced Research Projects Agency under Contract No. HR001122C0074. J.E., K.M., and K.O. also specifically acknowledge support by the Defense Advanced Research Projects Agency under Air Force Contract No. FA8702-15-D-0001. Any opinions, findings and conclusions or recommendations expressed in this material are those of the authors and do not necessarily reflect the views of the Defense Advanced Research Projects Agency.  

M.J.B. acknowledges the support of the ARC Centre of Excellence for Quantum Computation and Communication Technology (CQC2T), project number CE17010001.

N.N. and K.S.W. acknowledge Boeing Technical Fellow David Heck for numerous helpful discussions on magnesium alloy composition, properties and aerospace usage. B.L. and K.S.W. acknowledge Christopher Taylor from DNV and Ohio State University for providing insights on first principles modeling of HER and methods to construct realistic solvation models. B.L. thanks  Casey Brock from Schr\"{o}dinger, LLC for technical assistance with constructing the magnesium supercells.

N.N., K.S.W., and B.L. would like to thank the Boeing DC\&N organization, Jay Lowell, and Marna Kagele for creating an environment that made this research possible.  

The authors thank John Carpenter for his support in creating high-resolution figures for this paper.

\bibliographystyle{apsrev4-2}
\bibliography{references}

\onecolumngrid
\newpage

\section*{Suplementary Materials}
\setcounter{section}{0}
\setcounter{equation}{0}
\setcounter{figure}{0}
\setcounter{table}{0}
\renewcommand{\thesection}{S-\Roman{section}}
\renewcommand{\theequation}{S.\arabic{equation}}
\renewcommand{\thefigure}{S.\arabic{figure}}
\renewcommand{\thetable}{S.\Roman{table}}

\section{Designing Mg-Rich Sacrificial Coatings and Corrosion-Resistant Mg Alloys for Aqueous Environments}
\label{sec: Mg corrosion}
\subsection{Problem Overview}

Magnesium (Mg) alloys have been increasingly sought after for lightweight structural applications in industries such as automobile, aerospace, and defense~\cite{zhang2023magnesium,calado2022rare}. In addition to their inherent lightweight nature, magnesium alloys demonstrate an array of advantageous mechanical properties, including high stiffness and impressive tensile strength. Furthermore, these alloys offer notable benefits such as excellent castability and machinability, making them highly versatile and attractive for a wide range of applications. However, a major limitation of Mg alloys is their high aqueous corrosion rates~\cite{Esmaily2017Fundamentals}. 

Because of these high corrosion rates in aqueous environments, magnesium is commonly used as a sacrificial anode to protect structure in submersibles. Magnesium has also been investigated as a metal additive for sacrificial coatings. Similar to Zn-rich and Al-rich coatings, Mg-rich coatings contain magnesium alloy particles (or flakes) that protect the underlying metallic substrate through a sacrificial galvanic mechanism~\cite{mcmahon2023corrosion,Moraes2021}. In this type of application, magnesium serves as a galvanizing-like barrier, 
making its high corrosion rates beneficial, provided the magnesium particles 
are effectively dispersed in the polymer matrix. By understanding the corrosion 
mechanisms at both the particle surface and the particle, substrate interface, 
formulators can more efficiently develop Mg-rich coatings.

The corrosion of magnesium alloys is dominated by the highly exothermic hydrogen evolution reaction (HER) which constitutes the cathodic part of the electrochemical corrosion reaction~\cite{Esmaily2017Fundamentals}. This phenomenon is further complicated by the highly reactive and short-lived corrosion intermediates formed during cathodic polarization, presenting a significant challenge in both computational~\cite{yuwono2019aqueous,yuwono2016electrochemical,limmer2017first,yuwono2019understanding,Ma2017,Ma2020,Taylor2016,kwak2016atoms,wurger2020first,ng2023first} and experimental~\cite{huang2020activates,benbouzid2022new,gabbardo2020hydrogen,fajardo2018critical} investigations of the reaction mechanism. Experimental techniques that are often used to estimate magnesium corrosion rates include mass loss measurements, electrochemical characterization~\cite{korjenic2024electrochemical}, and in situ monitoring of the hydrogen gas evolved during HER~\cite{sridhar2021insitu}. At present, simulating HER kinetics with the required level of precision remains out of reach for purely classical approaches. If quantum simulation could accurately predict HER rates, formulators would be able to perform high-throughput computational screening of a wide range of coating compositions before committing to costly experimental programs.

The workflows and resource estimates provided here focus on selecting and designing magnesium alloy particles, intended for use as additives in protective coatings. In practice, sacrificial coatings are more complex chemical systems that include polymeric components and various other additives. This paper does not explicitly provide workflows and resource estimates for other aspects of coating formulation, such as coating/substrate interactions or the galvanic mechanism occurring when the magnesium particles come into contact with the underlying metal substrate. However, the workflows and resource estimation tools presented in this paper are extendable to those other mechanisms, as well as the many other electrochemical reactions that happen on a corroding magnesium surface (beyond just HER). Those other reactions will be the focus of future work.

Single-reference quantum chemistry methods have known limitations, especially where strong correlation effects are significant~\cite{Chu2017,Bulik2015,Gori-Giorgi2008}. This shortcoming can hamper the accuracy of their quantitative predictions. To capture all electronic correlations, methods like full-configuration interaction (FCI) incorporate a multiconfigurational wavefunction to capture many Slater determinants. This is critical in surface chemistry as strong correlation effects from near-degeneracies in bond-breaking and metal orbitals~\cite{Norskov2011,Wodtke2016}. Incorporating highly correlated methods aims to achieve quantitatively accurate simulations of chemical processes. Key observables—such as activation energies, binding energies, and electron transfer phenomena along relevant reaction coordinates—can inform materials design strategies. Moreover, these results can be integrated into larger-scale models to predict other chemically relevant measurements, including reaction kinetics~\cite{Taylor2016}.
    
To date, only simplified versions of the workflow outlined in this study have been implemented. These implementations have concentrated on the most basic scenarios, specifically a pure \textbf{4} $\times$ \textbf{4} $\times$ \textbf{4} magnesium slab composed of 64 magnesium atoms along with a few water molecules~\cite{Williams2016First,Williams2016Modeling,Gujarati2023}. However, as we strive to create more realistic models that accurately capture the physics of the corrosion process, the complexity of the problem rapidly increases. Larger slab models and more extensive aqueous environments become necessary, rendering classical simulations intractable due to the substantial computational resources required for high-fidelity simulations. Moreover, transitioning to magnesium alloys introduces additional complexity to the surface models, both in terms of their structural intricacies and overall size, compared to pure magnesium systems.

\subsection{Computational Workflow}
\label{sec: mg-rich workflow}

There are several potential directions in which the workflow can be constructed. To simplify matters, we focus on a particular direction that we consider as the baseline. This workflow centers around gaining a comprehensive understanding of a crucial step in the corrosion process: the cathodic reaction known as the Hydrogen Evolution Reaction (HER). It is widely acknowledged that the corrosion of magnesium and magnesium-rich alloys is primarily driven by the highly exothermic HER, constituting the cathodic aspect of the electrochemical corrosion reaction, and also occuring during anodic polarization~\cite{yuwono2019understanding}. A theoretical model for the HER process has been proposed~\cite{Williams2016Modeling} based on the Volmer-Tafel reaction mechanisms, which decomposes the reaction into three simpler, elementary steps:

\begin{subequations}
    \begin{align}
    \text{Mg} + 2 \text{H}_2\text{O} &\rightarrow \text{Mg(OH}_{\text{ads}}\text{)(H}_{\text{ads}}\text{)} + \text{H}_2\text{O}, \\
    \text{Mg(OH}_{\text{ads}}\text{)(H}_{\text{ads}}\text{)} + \text{H}_2\text{O} &\rightarrow  \text{Mg(OH}_{\text{ads}}\text{)}_2\text{(H}_{\text{ads}}\text{)}_2, \text{ and} \\
    \text{Mg(OH}_{\text{ads}}\text{)}_2\text{(H}_{\text{ads}}\text{)}_2 &\rightarrow  \text{Mg(OH}_{\text{ads}}\text{)}_2 + \text{H}_2\text{(g)},
\end{align}
\end{subequations}
where (ads) designates species adsorbed on the surface and (g) designates gaseous species that are evolved from the surface.

\begin{figure}
    \centering
    \includegraphics[width=0.7\textwidth]{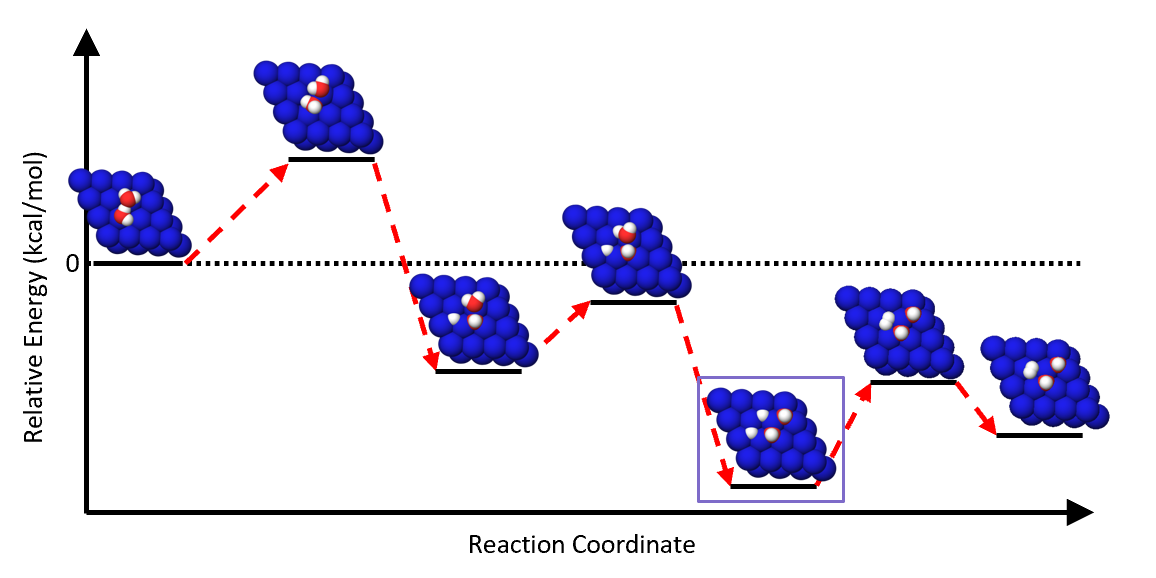}
    \caption{Volmer-Tafel reaction path for the HER on Mg surface. Energies are relative and qualitative to the reactant geometry. The purple box indicates the geometry that is used for benchmarking within this work. In principle, quantum resource estimates will be roughly equivalent across the entire reaction coordinate as during the Volmer-Tafel mechanism, number of atoms, number of electrons, and cell size are invariant. Therefore, the selected geometry will be representive of the entire scheme.}
    \label{fig:VT_schem}
\end{figure}

A schematic representation of this mechanism is listed in Figure \ref{fig:VT_schem}. In situ modeling of the rate of hydrogen gas production acts as a proxy for corrosion rate. The mechanism by which hydrogen gas is produced is the HER. The goal is then to determine computationally the reaction rate constants associated with the HER and use those rate constants to evaluate the corrosion inhibition efficiency for many prospective magnesium-rich alloys.

Along with the experimental understanding of hydrogen evolution, another aspect often addressed is characterization of magnesium-alloy environments. There has been extensive research into corrosion resistance via introduction of alloying elements including aluminum, zinc, rare earths, and many others~\cite{Somasundaram2023,Fogarty2022,Murugesan2023}. To effectively model metallic materials containing alloying elements, there should be two considerations for how these metallic phases are dispersed; either single-atom replacements within the primary Mg phase, or via phase formations (secondary-phase, beta-phase, gamma-phase, etc.) that see clusters of intruding structures into the main surface. The former case is a trivial case of atomic replacement at the \textit{ab-initio} level and is often performed in alloy case studies. The latter, however, requires detailed exploration of  phase formation and boundary interactions. Experimentally, these are readily observed via electron energy loss spectroscopy (EELS), transmission electron micrsocopy (TEM), and scanning electron microscopy (SEM)~\cite{Sudholz2011,angelini2015}. Here, models are built that capture what can be readily observed experimentally in order to garner understanding of the electronic structure of the primary-secondary phase boundary in magnesium-rich alloys. Similar procedures have been performed previously~\cite{Li2018Experimental}. This will give a more complete understanding of the electronic structure associated with boundary layers, while also pushing the limits of electronic structure simulations for corrosion models.

The computational workflow primarily consists of two stages. The first stage aims to discover the transition paths, also known as reaction pathways, associated with each of the three elementary reactions. The simplest conceptual method for identifying a transition state structure or first-order saddle point involves sampling the potential energy surface (PES), and the reaction path can be determined by tracing out the minimum energy path (MEP) that links the reactant and product structures. However, this method becomes intractable when dealing with large chemical systems with a high number of degrees of freedom. There are several alternative methods available for determining the transition path of a reaction more efficiently than sampling the PES. These include the  dimer method~\cite{Kastner2008Superlinearly}, linear synchronous transit~\cite{Halgreen1977synchronous}, conjugate peak refinement~\cite{Fischer1992Conjugate}, constrained optimization~\cite{Abashkin1995Constrained}, activation-relaxation technique~\cite{Jay2022Activation}, and nudged elastic band (NEB)~\cite{Henkelman2000climbing,Henkelman2000Improved}, among others. In this paper, we focus on the widely used NEB method due to its maturity and its status as a state-of-the-art transition state search method for surface chemistry problems. At a high level, NEB consists of constructing a series of atomic configurations (``images”) that represent intermediate atomic configurations that occur during a reaction. The resulting path is relaxed to the MEP connecting the optimized reactant and product structure. Note that the first atomic configuration (``initial system'') represents the state of the chemical reactants before the reaction has taken place, whereas the last atomic configuration (``final system'') represents the state of the chemical products after the reaction has transpired. The ``images'' interpolate between these two initial and final atomic geometries. NEB simultaneously optimizes the geometries of all intermediate images, and this can be done either serially or in parallel. Note that geometry optimization can consume a significant amount of computational resources due to the repeated calculations of ground state energies. Additionally, as the number of images increases, so does the calculation cost.

Moving to the second stage of the workflow, the focus shifts to following the reaction pathways associated with each of the three elementary HER reactions and calculating the rate constants using the Arrhenius equation:
\begin{equation}
    k = A\exp\left(-\frac{E_a }{RT} \right)
    \label{eq:Arrhenius}
\end{equation}
where $A$ is the Arrhenius factor, $R$ is the gas constant, $T$ is the temperature in Kelvin, and $E_{a}$ is standard Gibbs energy of activation. In the simplest scenario with a single transition state,  $E_a = E_{\text{TS}} - E_{\text{init}}$, where $E_{\text{TS}}$ denotes the electronic ground state energy of the transition state and $E_{\text{init}}$ represents the electronic ground state energy of the initial state. Note that this is only valid for elementary reaction steps. NEB image propagation allows for selection of geometries that are elementary steps in a larger reaction pathway. Given the exponential dependence of $E_a$ in the Arrhenius equation, an accurate determination of $E_a$ is crucial, often requiring a threshold of $10^{-3}$ to $10^{-4}$ Hartree to make quantitative statements about reaction kinetics. This underscores the importance of accurately determining transition states to achieve the required precision. However, for simplicity, if the MEP computed from NEB using DFT suffices, or if the difference between the ground state energy of the transition state found using DFT and that obtained using high-fidelity methods is negligible (within the accuracy threshold), one can consider only employing a high-fidelity electronic solver to resolve $E_a$ by recalculating the ground state energy of the reactant and transition state. Due to the extensive size of our computational models  high-fidelity classical solvers will become intractable. At this particular step within the described workflow, quantum computers can be leveraged, specifically using the quantum phase estimation algorithm. However, it's important to note that quantum computers could theoretically be deployed throughout the entire process. Doing so might result in an excessive and unnecessary use of quantum computational resources. Therefore, strategically integrating quantum computing allows us to maximize efficiency and computational resources. 

\begin{figure}
    \centering
    \includegraphics[width=1\textwidth]{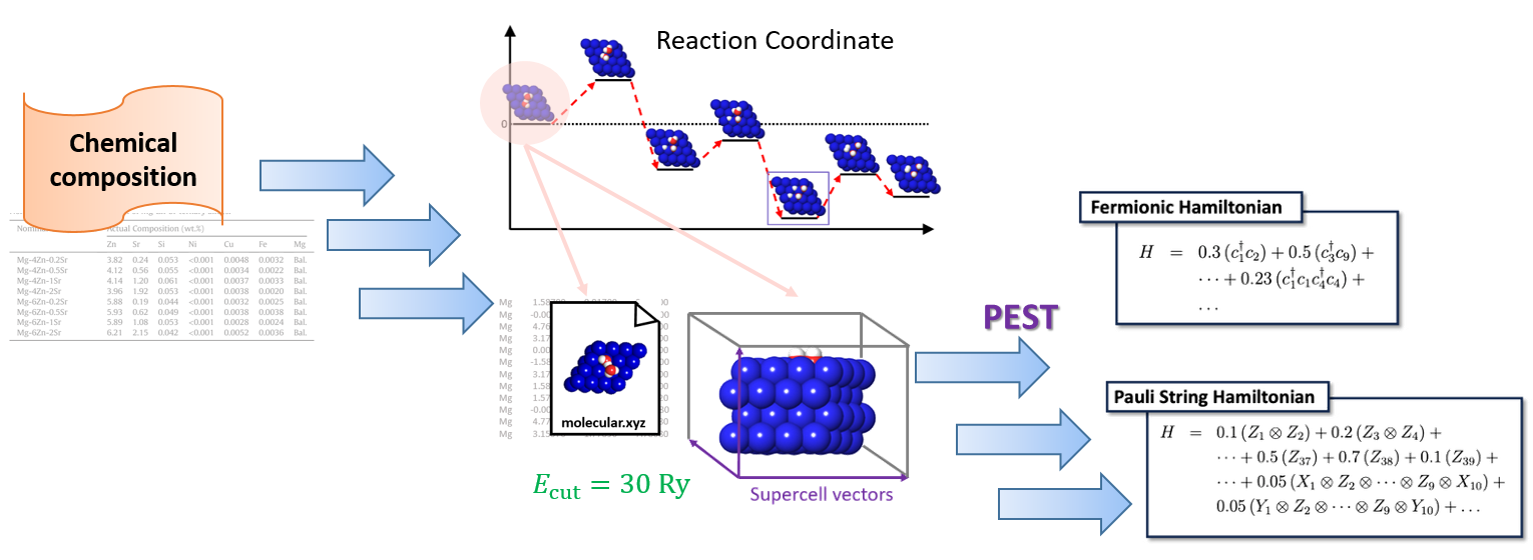}
    \caption{This figure illustrates the layered structure of input models in the outlined computational workflow. It highlights the three layers involved: the practitioner defining chemical compositions and solvent models (Layer 1), the quantum chemist creating computational models for surface reactions (Layer 2), and the quantum computing scientist generating Hamiltonians for quantum algorithms (Layer 3). Each layer represents a distinct set of inputs crucial for the subsequent computational processes. Subject matter experts assisted with specifying the inputs and outputs for Layer 1 and Layer 2. A Julia-based library called PEST was developed to take outputs from Layer 2 (inputs for Layer 3) and generate the necessary inputs to perform quantum resource estimates for the quantum algorithm. 
    }
    \label{fig: input}
\end{figure}

The computational workflow involves multiple layers, from defining the alloy composition and solvent in the first layer to constructing a quantum Hamiltonian in the third layer, with each stage building upon the previous one. These steps are visually represented in Figure \ref{fig: input}, which outlines the entire process, from material definition to quantum algorithm preparation.

In summary, the primary objective of the workflow is to uncover the most energy-efficient pathway for the proposed reaction, which will help us understand in detail how the cathodic reaction, particularly HER, occurs on magnesium-alloy surfaces. There are several methods to achieve this, depending on the selected algorithm. For example, comparing the Gibbs free energy at different transition states pinpoints the rate-limiting step of the reaction. Another approach involves directly tracking the reaction using full ab-initio molecular dynamics, which allows us to observe how atoms move and extract reaction rates. Beyond ground-state energies, chemical properties such as band structures, charge states, and population analysis are desirable for fundamental insights into charge transfer, band structure, and binding characteristics. Within this work, ground state energies along a reaction coordinate of interest are computed using a quantum computer. A simplified pseudo-code (Algorithm \ref{pseudocode: HER}) outlining the the discussed workflow. Recall that the reaction rate of the Hydrogen Evolution Reaction (HER) serves as a proxy for the corrosion rate. 

\begin{algorithm}
\DontPrintSemicolon
\KwIn{Input geometries}
\KwOut{Output activation energies}
\BlankLine

\For{each Mg-alloy stoichiometry under consideration}{
    generate a model for the alloy\;

    create an explicit model solvent by placing water molecules above the slab and relaxing at a molecular dynamics (MD) level\;
        
    run geometric optimization on MD-DFT water molecule configuration fixing slab atoms in periodic conditions \;
    
    use NEB to determine the MEP along the reaction coordinate of interest \; 
    
    determine the rate-limiting step within the reaction pathway i.e. the largest reaction barrier along the reaction coordinate.\;

    compute the ground state energy of the initial, final, and highest energy transition state, $E^{init}$ and $E^{TS} $ respectively, to high accuracy using quantum computation e.g., chemical accuracy 1.6 mHa. \;
    
    calculate the HER rate, which acts as a proxy for corrosion rate, Eq.~\eqref{eq:Arrhenius} or interface with higher length-scale models to predict reaction rates such as MKM or KMC.
}

\caption{Investigation of the HER mechanism for different magnesium alloy models.}
\label{pseudocode: HER}
\end{algorithm}

\subsection{Representative Computational Models for Resource Estimation}
\label{sec: Representative Computational Models for Mg}
Corrosion reactions are complex due to their intricate nature, making it challenging to pinpoint the precise problem size that necessitates our exploration of various parameters (e.g., the number of water molecules, the size of the slab, etc.). However, the studies conducted by Williams \textit{et al.}~\cite{Williams2016First,Williams2016Modeling} serve as an essential benchmark, representing the absolute minimum requirements for quantum hardware to offer practical utility. In the following discussion, we will enumerate several models of pure magnesium, each with an increasing level of complexity, where accurate determination of ground state energy becomes imperative to extract utility. Additionally, we will outline some magnesium alloy models of practical significance. These models provide us with a rough estimate of the computational model size necessary for our proposed workflow to deliver utility, effectively establishing lower thresholds for utility attainment.

Three distinct pure magnesium models were developed, each exhibiting increasing levels of complexity, to study various corrosion mechanisms of interest, such as the hydrogen evolution reaction (HER) mechanism. These models serve as representative systems for corrosion scientists and engineers. The smallest model consists of a four-layer slab with dimensions $4 \times 4$ magnesium atoms and two water molecules. While this simple model may not offer significant utility in designing new, effective magnesium-rich anti-corrosive coatings essential for revolutionizing industries, it serves as a valuable benchmark problem due to its simplicity.

\begin{figure}
    \centering
    \includegraphics[scale=0.8]{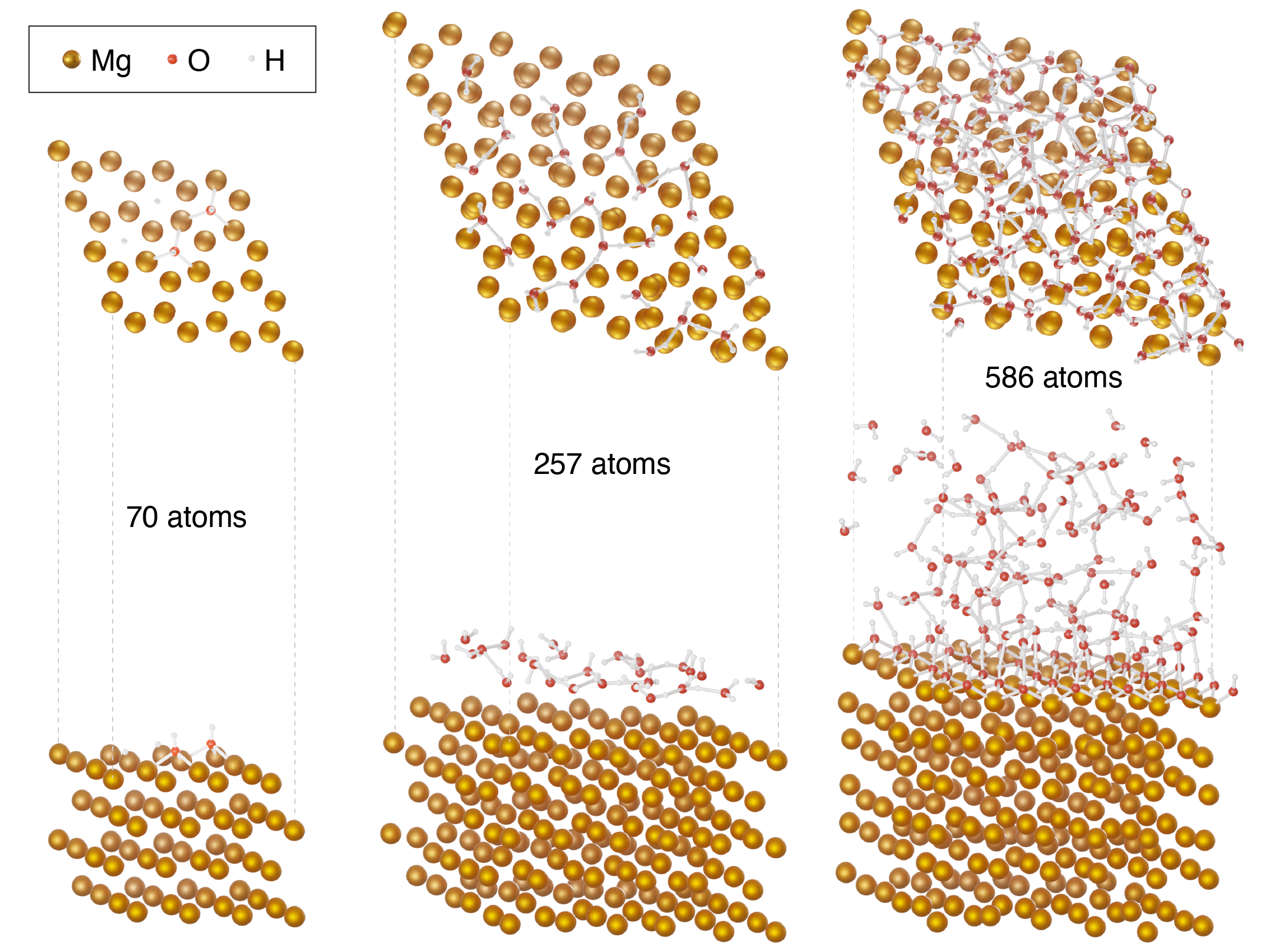}
    \caption{Various models for modeling the corrosion process on a pure magnesium surface.}
    \label{fig: PureMgModels}
\end{figure}

Our goal is to increase the realism of the model by enhancing its complexity and size. As shown in Figure \ref{fig: PureMgModels}, adding more water species on the surface and using a thicker slab ensures the model replicates the actual physical processes.

Several magnesium alloys have been approved for aircraft applications due to their high strength, improved corrosion resistance, and ability to meet flammability requirements, such as being self-extinguishing. For example, WE-43 is utilized in aircraft seat frames, and EV-31A is employed in rotorcraft gearboxes. Other industrially relevant magnesium alloys include AZ31 and AZ91E for die casting, and AMX602 for extrusion.

To construct realistic models of these alloys, the pure magnesium simulation cell described in the previous section must be modified to incorporate alloying elements (e.g., Al, Ca) and the intermetallic secondary phases that form within the magnesium matrix. These secondary phases often possess complex crystal structures, necessitating large simulation cells that must be "embedded" into a pure magnesium supercell. 

There are various methodologies for building alloy models, with one example illustrated in Figure 5 of the main text. This approach involves first expanding the 
\textbf{4} $\times$ \textbf{4} $\times\, \textbf{4}$ magnesium slab with four layers along all three axes to create a {\textbf{16}~$\times$~\textbf{16}~$\times$~\textbf{16}} slab with 16 layers, increasing the model size from 64 to 4096 atoms. Subsequently, atoms in a specific region of the top-most magnesium layer are replaced with the experimentally predicted structure of the secondary phases. In the example depicted in Figure 5, the secondary phase is Mg$_{17}$Al$_{12}$, one of the most prevalent intermetallic phases in Mg-Al alloys~\cite{ubeda2020role}.

After constructing the alloy surface model, additional layers of water molecules can be added to the simulation cell for explicit solvation, similar to the water layers shown in Figure \ref{fig: PureMgModels}. Comparing the pure magnesium models to those of a representative alloy surface (Figure 5 in the main text) highlights that incorporating intermetallic secondary phases at the water/metal interface significantly increases the complexity and size of the simulation cell. The alloy model scales from a few hundred atoms to several thousand, making quantum mechanical calculations on such large systems computationally intensive.

Table \ref{tab:computational_models} show several representative computational models. These models vary in size and complexity, incorporating different configurations of magnesium (and magnesium alloys) and water molecules. Each model is embedded in a simulation cell, and the details of these models are presented in the table. 
\begin{table}[ht]
    \centering
    \begin{tabularx}{\textwidth}{X X}
        \hline\hline
        \textbf{Model Name} & \textbf{Description} \\
        \hline
        Dimer model & 70 Mg atoms and 2 water molecules, embedded in a simulation cell of $12.7 \times 12.7 \times 19.9$ {\AA} (see Figure \ref{fig: PureMgModels} (left)). \\
        \\[-1ex]
        Monolayer model & 257 Mg atoms and 26 water molecules, embedded in a simulation cell of $19.8 \times 19.8 \times 32.3$ {\AA} (see Figure \ref{fig: PureMgModels} (middle)). \\
        \\[-1ex]
        Cluster model & 587 Mg atoms and 100 water molecules, embedded in a simulation cell of $19.8 \times 19.8 \times 58.9$ {\AA} (see Figure \ref{fig: PureMgModels} (right)). \\
        \\[-1ex]
        Single-unit Mg$_{17}$Al$_{12}$ secondary phase structure & 271 Mg atoms and 24 Al atoms, embedded in a simulation cell of $29.7 \times 9.2 \times 42.6$ {\AA}. \\
        \\[-1ex]
        Double-unit Mg$_{17}$Al$_{12}$ secondary phase structure & 258 Mg atoms and 48 Al atoms, embedded in a simulation cell of $30.1 \times 9.3 \times 43$ {\AA}. \\
        \\[-1ex]
        Quadruple-unit Mg$_{17}$Al$_{12}$ secondary phase structure & 228 Mg atoms and 96 Al atoms, embedded in a simulation cell of $30 \times 9.3 \times 42.9$ {\AA}. \\
        \\[-1ex]
        Secondary-phase supercell & An expansion of the single-unit cell, composed of 807 Mg atoms and 72 Al atoms, embedded in a simulation cell of $33 \times 31.3 \times 50.3$~{\AA}. \\
        \hline\hline
    \end{tabularx}
    \caption{Representative computational models for resource estimation.}
    \label{tab:computational_models}
\end{table}

\section{Designing High-Temperature Corrosion-Resistant Alloys}
\label{sec:Nb_workflow}
\subsection{Problem Overview}
Refractory alloys, renowned for their robustness and high-temperature resistance, have long been essential materials in aerospace, defense, and aviation applications~\cite{Philips2020New}. The pursuit of even more durable and oxidation-resistant materials has recently shifted focus toward niobium-rich (Nb-rich) refractory alloys. Despite their superior strength-to-weight ratios and higher melting points compared to nickel-based alloys~\cite{Loria1987Nb, Philips2020New}, Nb-rich alloys suffer from poor oxidation resistance, which limits their performance at elevated temperatures~\cite{Reynolds2023Niobium}.

Addressing this challenge requires not only the identification of stable Nb-rich alloy compositions but also a deep understanding of their atomic-level behavior. Specifically, investigating oxygen diffusivity within the alloy matrix is crucial for enhancing oxidation resistance and overall material performance. This investigation involves modeling oxygen diffusion using kinetic Monte Carlo (KMC) methods, which depend on accurately determined energy barriers. These energy barriers are obtained through cluster expansions trained on ab initio data.

Previous studies have successfully applied similar techniques to binary Nb alloys and Fe-Cr alloys, utilizing density functional theory (DFT) to generate the necessary ab initio training data for the cluster expansions~\cite{Chen2020Alloying, Samin2019ab, Mann2024}. However, the larger design space inherent in Nb-rich alloys with multiple alloying elements introduces additional complexity in identifying suitable multicomponent alloys. Potential alloying elements for Nb-rich systems include Ti, Ta, W, V, Y, Mg, Al, Mo, Hf, Cr, Zr, Mn, and Co.

Furthermore, it is important to consider that transition states in these systems are inherently out-of-equilibrium~\cite{Gerrits2020DensityFT}. This out-of-equilibrium nature of transition states adds another layer of complexity to accurately modeling and predicting diffusion behaviors within Nb-rich alloys.

\subsection{Computational Workflow}
At a high level, the objective for a given set of alloying elements is to accurately estimate the energy barriers associated with oxygen transitioning from one octahedral interstitial site to another. During these transitions, the oxygen atom passes through a transition state. In pure niobium, the transition state is an interstitial site with tetrahedral symmetry, making it amenable to classical computational methods~\cite{Liu2020nbti}. However, when considering disordered niobium alloys, the transition states no longer retain perfect tetrahedral symmetry. This complexity necessitates the use of a transition state search algorithm, such as the Climbing Image Nudged Elastic Band (CI-NEB) method, to accurately determine the geometry of the transition states~\cite{Chen2020Alloying}.

The classical pre-processing associated with this workflow begins with generating a set of supercells, as depicted in Fig.~3 of the main text, where niobium atoms are randomly replaced within the lattice by alloying elements in various proportions. Subsequently, oxygen atoms are randomly placed at octahedral interstitial sites, and the entire supercell is relaxed to achieve a stable configuration. Additionally, a separate set of transition state geometries must be generated, which involves running a transition state search algorithm. Both the octahedral and transition state geometries are then saved in a database for further analysis, as described in Ref.~\cite{Samin2019ab}.

Special Quasi-Random Structures (SQSs) are employed to statistically mimic the random distribution of atoms or elements in an alloy or compound while utilizing a finite, periodic unit cell, thereby making the simulations computationally manageable~\cite{Zunger1990sqs, Gao2016sqs}. 

The next step in the workflow involves fitting a cluster expansion to the total energies computed from octahedral geometries, as well as developing a separate cluster expansion for the transition state geometries. 

Let $ E_{\mathrm{sol}}^{\mathrm{Oct}}(x_i) $ and $ E_{\mathrm{sol}}^{\mathrm{Tetra}}(x_j) $ represent the solution energies computed for randomly selected oxygen positions $ x_i $ and $ x_j $ in the alloy lattice. These positions correspond, respectively, to "octahedral" reactant/product geometries and "tetrahedral" transition-state geometries obtained via CI-NEB calculations. For each type of geometry, we introduce a cluster expansion,
\[
F^{\mathrm{Oct}} : \boldsymbol{\Omega}  \to \mathbb{R} \quad \text{and} \quad F^{\mathrm{Tetra}} : \boldsymbol{\Omega} \to \mathbb{R},
\]
which share the general form
\[
F(\boldsymbol{\sigma}) = J_{0} + \sum_{\alpha} m_{\alpha}\,J_{\alpha}\,\Theta_{\alpha}(\boldsymbol{\sigma}).
\]
Here, $ \Omega_{O} $ denotes the set of possible oxygen sites, and $ \alpha $ indexes all symmetrically distinct clusters of lattice sites. In concentration-dependent local cluster expansions (CDLCE), the coefficients $ J_{\alpha}(x) $ are functions of the oxygen composition $ x \in [0,1] $, typically expanded in a polynomial series~\cite{Samin2019ab}:
\[
    J_{\alpha}(x) = J_{0,\alpha} + J_{1,\alpha}\,x + J_{2,\alpha}\,x^2 + \cdots \quad 
\]
In many cases, including up to four-body clusters in the expansion provides sufficiently accurate fits to the training data~\cite{Kleiven2021}. However, more nuanced strategies may be necessary for more complex alloy configurations~\cite{Muller2024, Zhang2019ce}.

We denote by $ m_{\alpha} $ the multiplicity of cluster set $ \alpha $ per unit cell, and by $ J_{\alpha} $ the effective cluster interaction (ECI) coefficients, which are obtained by fitting this expansion to computed energies. The functions $ \Theta_{\alpha}(\boldsymbol{\sigma}) $ take the form
\[
    \Theta_{\alpha}(\boldsymbol{\sigma}) = \frac{1}{N_{\alpha}} \sum_{\beta \subset \alpha} \prod_{i=1}^{N} \phi_{\beta_i}\bigl(\sigma_i\bigr),
\]
where $ N_{\alpha} $ is the number of sub-clusters $ \beta \subset \alpha $, and $ \sigma_i $ are ``spin" variables indicating which atom occupies site $ i $. The basis functions $ \phi_{\beta_i}(\sigma_i) $ can be, for instance, simple indicator functions (one-hot encoding) or alternative forms such as polynomial or trigonometric functions~\cite{Barroso2022Cluster}.

The configuration vector $ \boldsymbol{\sigma} $ corresponds to the possible choices for various elements at each position in the alloy. For a lattice of $ M $ sites, the configuration vector is an element of the configuration space $ \boldsymbol{\sigma} \in \Omega_1 \times \Omega_2 \times \cdots \times \Omega_{M} = \boldsymbol{\Omega} $. For example, a 4-site lattice could be described by $ \boldsymbol{\Omega} = \{0, 1, 2, 3\}^4 $ and $ \boldsymbol{\sigma} = (0, 1, 0, 3) $, where $ 0 $ is Nb, $ 1 $ is Zr, $ 2 $ is Ta, and $ 3 $ is W.

For the training data, we collect energies computed at FCI level for $ E_{\text{sol}}^{\text{Oct}}(x) $ and $ E_{\text{sol}}^{\text{Tetra}}(x) $ corresponding to configurations $ (\boldsymbol{\sigma}, x) $, where the position of the oxygen atom is tracked as an additional configuration parameter $ x $. This training data is then used to find an optimal set of ECI coefficients $ \boldsymbol{J} = (J_0, J_{\alpha_1}, \ldots) $ by solving the following least squares problem.

Let $ \mathbf{\Pi}(\boldsymbol{\sigma}) $ be the vector of functions $ \Theta_{\alpha_i}(\boldsymbol{\sigma}) $ for $ i = 1, \ldots, d $,
\begin{equation}
    \mathbf{\Pi}(\boldsymbol{\sigma}) = \left( \Theta_{0}(\boldsymbol{\sigma}) \equiv 1, \Theta_{\alpha_1}(\boldsymbol{\sigma}), \Theta_{\alpha_2}(\boldsymbol{\sigma}), \ldots \right) \in \mathbb{R}^d.
\end{equation}
We can then compute the cluster expansion using a dot product:
\begin{equation}
    F(\boldsymbol{\sigma}) = \mathbf{\Pi}(\boldsymbol{\sigma}) \cdot \boldsymbol{J}.
\end{equation}
Next, we construct the matrix $\boldsymbol{\Pi} $ that contains the values of the vector $ \mathbf{\Pi}(\boldsymbol{\sigma}_j) $ for each training configuration $ \boldsymbol{\sigma}_j $ for $ j = 1, \ldots, m $:
\begin{equation}
    \boldsymbol{\Pi} = \left( \mathbf{\Pi}(\boldsymbol{\sigma}_1), \mathbf{\Pi}(\boldsymbol{\sigma}_2), \ldots, \mathbf{\Pi}(\boldsymbol{\sigma}_m) \right)^T \in \mathbb{R}^{m \times d}.
\end{equation}
Let $ \boldsymbol{E} = (E_1, \ldots, E_m) $ be the FCI-level energies associated with the $ m $ training structures. The least squares problem is then posed as
\begin{equation}
    \boldsymbol{J}^* = \arg\min_{\boldsymbol{J}} \left\| \boldsymbol{\Pi} \boldsymbol{J} - \boldsymbol{E} \right\|_2^2 + \rho(\boldsymbol{J}),
\end{equation}
where $ \rho $ is a regularization term, with several available options~\cite{Barroso2022Cluster}.

Evaluating the quality of the fit typically involves computing cross-validation (CV) and related metrics. The canonical choice is the leave-one-out CV, defined as
\begin{equation}
    CV = \sqrt{ \frac{1}{n} \sum_{i=1}^n \left( E_i - \hat{E}_i \right)^2 },
\end{equation}
where $ E_i $ are the FCI-level energies computed for structure $ i $, and $ \hat{E}_i $ is the predicted energy from a cluster expansion trained on the other $ n - 1 $ structures~\cite{Samin2019ab, Kadkhodaei2021Cluster}. A good score for leave-one-out cross-validation is below $ 5 $ meV/atom~\cite{Kleiven2021}.

Once sufficiently accurate cluster expansions are obtained, we can estimate the energy barrier between octahedral sites $ i $ and $ j $, denoted by $ E_{i \to j} $, as described in Refs.~\cite{Chen2020Alloying, Kleiven2021}:
\begin{equation}
    E_{i \to j} = E_{\text{sol}}^{\text{Tetra}}(x_{i \to j}) - E_{\text{sol}}^{\text{Oct}}(x_i) = F^{\text{Tetra}}(\boldsymbol{\sigma}, x_{i \to j}) - F^{\text{Oct}}(\boldsymbol{\sigma}, x_i).
\end{equation}
These energy barriers are then incorporated into Kinetic Monte Carlo (KMC) simulations of oxygen undergoing a random walk through the alloy's sublattice of octahedral sites. As outlined in Ref.~\cite{Chen2020Alloying}, for each time step, two random numbers $ r_1, r_2 \sim \text{Unif}(0,1) $ are generated to determine the transition path and the moving time of the simulation. If the system is in state $ i $, the transition rate between $ i $ and another site $ j $ in the transition path $ P = (i_1, i_2, \ldots, i_{n-1}, j) $ is given by
\begin{equation}
    k_{i \to j} = \nu_0 \exp\left( -\frac{E_{i \to j}}{k_B T} \right),
\end{equation}
where $ T $ is temperature, $ k_B $ is the Boltzmann constant, and $ \nu_0 $ is the attempt frequency, defined as
\begin{equation}
    \nu_0 \approx \left( \prod_{i=1}^{3N-4} \frac{\nu_i^{\text{IS}}}{\nu_i^{\text{TS}}} \right) \nu_{3N-3}^{\text{IS}}.
\end{equation}
For computational efficiency, we adopt the strategy outlined in Ref.~\cite{Chen2020Alloying}, setting $ \nu_0 $ to the value for pure Nb and using true octahedral and tetrahedral initial (IS) and transition state (TS) frequencies.

The total transition rate $ k_{\text{tot}} $ is the sum of all $ k_{i \to j} $ for all possible moving paths $ P = (i_1, i_2, \ldots, i_{n-1}, j) $. If
\[
\sum_{P_{n-1}} k_{i \to j} \leq r_1 k_{\text{tot}} \leq \sum_{P_n} k_{i \to j},
\]
where $ P_{n-1} = (i_1, i_2, \ldots, i_{n-1}) $ and $ P_n = (i_1, i_2, \ldots, i_n = j^*) $, then $ j^* $ is selected for $ i_n $ in the path $ P $. The system transitions to the next state with a moving time calculated according to the time step equation
\[
\delta t = \frac{1}{k_{\text{tot}}} \ln\left(\frac{1}{r_2}\right).
\]
This random process is repeated to estimate the diffusivity at various temperatures $ T $ using the formula
\begin{equation}
    \label{sup_mat_eq:D(T)}
    D(T) = \frac{1}{n_s} \sum_{m=1}^{n_s} \frac{ \| \boldsymbol{x}(t_m) - \boldsymbol{x}(t_{m-1}) \|_2^2 }{6 \delta t },
\end{equation}
where $ n_s $ is the number of sampled trajectories, $ \boldsymbol{x}(t_m) $ is the position of an oxygen atom at time $ t_m $, and $ \boldsymbol{x}(t_m) = \boldsymbol{x}(t_{m-1} + \delta t) $. The KMC calculation assumes that $ \delta t \gg \nu_0^{-1} $, meaning that kinetic jump times of oxygen are large enough to include all local jump correlations. Additionally, $ n_s $ must be sufficiently large to obtain a statistically meaningful diffusion coefficient~\cite{Chen2020Alloying}. Alloy compositions with low diffusivities are selected as indicators of oxidation resistance; however, diffusivity is not the sole factor to consider. Therefore, this computational workflow aids in narrowing down the search for desirable alloy formulations.

Generic cluster expansions are challenging to apply to multi-component alloys~\cite{Zhang2019ce}. To address this complexity, either the effective pair model~\cite{Zhang2019ce} or chemical embedding techniques~\cite{Muller2024} can be employed. Additionally, the size of the supercell must be carefully considered in relation to the number of alloying elements to ensure accurate representation of the alloy's composition and structure.

The inputs for the described computational workflow aimed at identifying promising niobium-rich (Nb-rich) ultrahigh temperature alloys include a set of alloying elements $\{ \text{A}, \text{B}, \text{C}, \ldots \}$ and defined ranges for their fractional stoichiometries. For example, when considering alloys of the form $\text{Nb}_{1-(x+y+z)} \text{A}_x \text{B}_y \text{C}_z$, the focus is exclusively on Nb-rich compositions, restricting $(x, y, z)$ to the ranges $[0, a] \times [0, b] \times [0, c]$ where $a, b, c \ll 1$. The number of alloying elements and the specified stoichiometry ranges determine the necessary size of the computational supercell. This ensures that the supercell contains a sufficient number of atoms to accurately represent the proportions $(x, y, z, \ldots)$ of the various alloying elements.

\medskip

\noindent
The primary output of this computational workflow is the oxygen diffusivity, $D(T)$, at various temperatures for different alloy compositions $\text{Nb}_{1-(x+y+z)} \text{A}_x \text{B}_y \text{C}_z$. Additionally, if a cluster expansion is trained on the alloy's formation energy, the workflow can also produce a phase diagram as $(x, y, z)$ vary. This phase diagram facilitates the evaluation of alloy stability and thermodynamic properties. Oxygen diffusivity at high temperatures, $D(T)$, is the key quantity of interest because it directly influences the rate at which the niobium alloy oxidizes. The ideal scenario is to identify an alloy composition that forms a thin oxide film, similar to aluminum alloys, which provides protection against further oxidation~\cite{Perkins1990Oxidation}.

\medskip

\noindent
In the case of Nb-Al binary alloys, the critical atom fraction of aluminum, $N_{\text{Al}}^{\text{crit}}$, required to form a protective oxide layer is related to the diffusivity of oxygen by the following equation:

\begin{equation}
    N_{\text{Al}}^{\text{crit}} = \sqrt{\frac{\pi g^*}{3} \cdot N_{\text{O}}^{\text{s}} \cdot \frac{D_{\text{O}} V_{\text{m}}}{D_{\text{Al}} V_{\text{ox}}}}
\end{equation}

\noindent
where $N_{\text{O}}^{\text{s}}$ is the oxygen solubility in the alloy, $D_{\text{O}}$ and $D_{\text{Al}}$ are the diffusivities of oxygen and aluminum in the alloy, respectively, $V_{\text{m}}$ and $V_{\text{ox}}$ are the molar volumes of the alloy and the oxide, respectively; and $g^*$ represents the critical volume fraction of the oxide scale~\cite{Perkins1990Oxidation}.

\noindent
This relationship underscores the importance of accurately determining oxygen diffusivity to design Nb-rich alloys with enhanced oxidation resistance. By integrating cluster expansions and high-fidelity simulations, the workflow efficiently navigates the extensive design space of multicomponent alloys, enabling the identification of compositions that meet the desired high-temperature performance criteria. Algorithm \ref{pseudocode: Nb} presents the simplified pseudocode outlining the workflow discussed in this section.

\begin{algorithm}
\DontPrintSemicolon
\KwIn{Nb alloy chemical composition e.g. $\text{Nb}, \text{W}, \text{Ta}, \text{Ti}$ }
\KwOut{Oxygen Diffusivities $D(T)$}
\BlankLine

$N^{\text{Oct}}_{\text{samples}} \gets $ number of cluster expansion training data samples for octahedral sites \;

$N^{\text{TS}}_{\text{samples}} \gets $ number of cluster expansion training data samples for transition state sites (``tetrahedral'') \;

\For{$i = 1,...,N^{\text{Oct}}_{\text{samples}}$}{
    Generate random composition $\text{Nb}_{1-(x+y+z)} \text{W}_x \text{Ta}_y \text{Ti}_z$ \;
    Randomly select octahedral sites for oxygen to generate structures $\text{Nb}_{1-(x+y+z)} \text{W}_x \text{Ta}_y \text{Ti}_z O$ \;
    Relax structure (DFT, geometric optimization) \;
    Compute solution energy $E_{\text{sol}} = E(\text{Nb}_{1-(x+y+z)} \text{W}_x \text{Ta}_y \text{Ti}_z \text{O}) - E(\text{Nb}_{1-(x+y+z)} \text{W}_x \text{Ta}_y \text{Ti}_z) - \frac{1}{2} E(O_2)$ (quantum computer, ground state energy)\;
}

\For{$i = 1,...,N^{\text{TS}}_{\text{samples}}$}{
    Run CI-NEB to find transition state (DFT, geometric optimization) \;
    Compute solution energy $E_{\text{sol}}$ (quantum computer, ground state energy)\;
}

Train cluster expansions on octahedral and tetrahedral (transition state) datasets.\;
Perform KMC simulations of oxygen diffusion for various stoichiometries of interest $\text{Nb}_{1-(x+y+z)} \text{W}_x \text{Ta}_y \text{Ti}_z$ \;
Estimate the diffusivity at various temperatures using \cref{sup_mat_eq:D(T)} of the main text

Select stoichiometries with low diffusivity\;

\caption{Investigation of the oxygen diffusivity in Nb-rich alloys.}
\label{pseudocode: Nb}
\end{algorithm}

\subsection{Representative Computational Models for Resource Estimation}
\label{sec: Representative Computational Models for Nb}

\begin{table}[ht]
    \centering
    \begin{tabular}{ccc}
        \hline\hline
        \textbf{Nb-alloy Compound} & \textbf{Simulation Cell Size} \\
        \hline
        Nb$_{97}$Hf$_{3}$Ti$_{22}$Zr$_6$O & $13.3 \times 13.3 \times 13.3$ {\AA} \\
        \\[-1ex]
        Nb$_{97}$Ta$_{22}$Zr$_{3}$W$_6$O & $13.3 \times 13.3 \times 13.3$ {\AA} \\
        \\[-1ex]
        Nb$_{42}$Ti$_3$Hf$_3$Ta$_3$Zr$_3$O & $10.0 \times 10.0 \times 10.1$ {\AA} \\
        \\[-1ex]
        Nb$_{65}$Zr$_6$Hf$_7$Ti$_4$W$_3$ & $8.6 \times 9.8 \times 14.0$ {\AA} \\
        \hline\hline
    \end{tabular}
    \caption{Compounds and their simulation cell dimensions.}
    \label{tab:compounds}
\end{table}

\begin{figure}[htbp]
    \centering
    \begin{minipage}{0.45\textwidth}
        \centering
        \includegraphics[width=\linewidth]{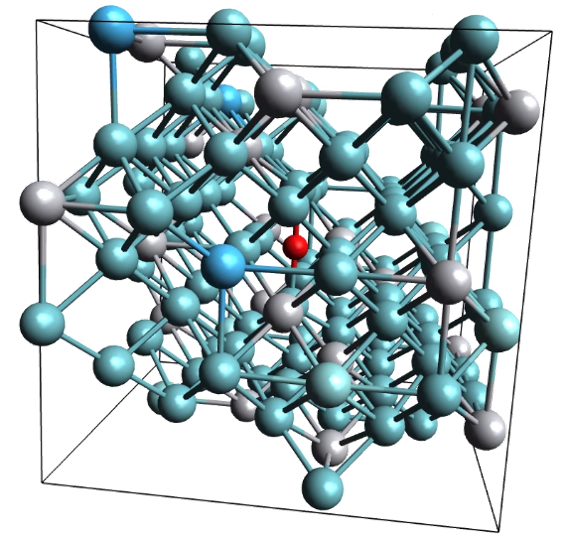}
        \subcaption{Nb$_{97}$Hf$_{3}$Ti$_{22}$Zr$_6$O} 
    \end{minipage}%
    \hfill
    \begin{minipage}{0.45\textwidth}
        \centering
        \includegraphics[width=\linewidth]{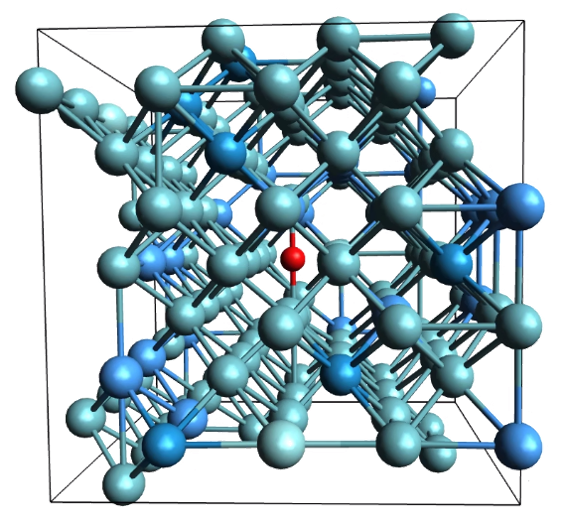}
        \subcaption{Nb$_{97}$Ta$_{22}$Zr$_{3}$W$_6$O} 
    \end{minipage}

    \vspace{0.5cm}

    \begin{minipage}{0.45\textwidth}
        \centering
        \includegraphics[width=\linewidth]{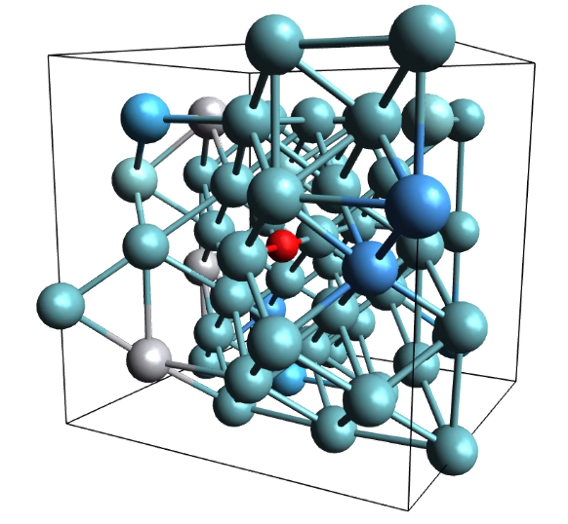}
        \subcaption{Nb$_{42}$Ti$_3$Hf$_3$Ta$_3$Zr$_3$O} 
    \end{minipage}%
    \hfill
    \begin{minipage}{0.45\textwidth}
        \centering
        \includegraphics[width=\linewidth]{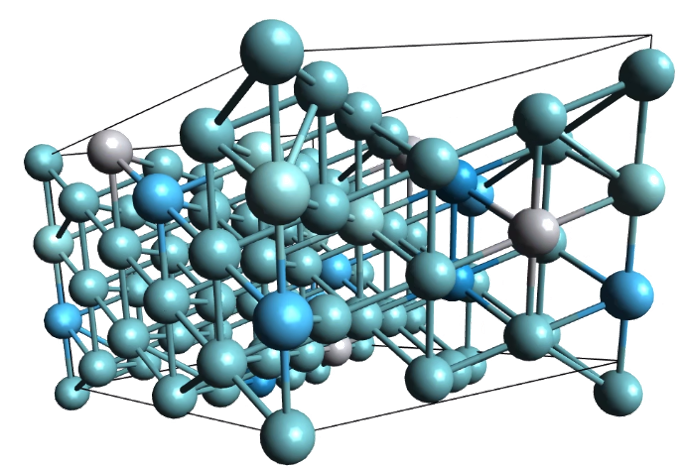}
        \subcaption{Nb$_{65}$Zr$_6$Hf$_7$Ti$_4$W$_3$} 
    \end{minipage}
    \caption{Various representative computational models of Nb alloys that are of interest to subject matter experts and are being used as example calculations in this work}
\end{figure}

\section{Comparative Analysis Between Classical and Quantum Techniques }
\label{sec: classical vs quantum}

\subsection{Limitations of Classical Methods for Estimating Ground-State Energies}
\label{sec:Limitations_of_Classical_Methods}
The solution of the electronic Schr\"{o}dinger equation to predict \emph{ab initio} properties of chemical and material systems has been a challenge for nearly a century. Within the Born-Oppenheimer approximation, the electronic Schr\"{o}dinger equation is
\begin{equation}
    \label{eq: time-independent schrodinger}
    H |\Psi(r;R)\rangle = E[\vec{r};R] |\Psi(r;R)\rangle
\end{equation}
where 
\begin{equation}
    H = -\sum_i \left( \frac{\nabla_{r_i}^2}{2} + \sum_I \frac{Z_I}{|r_i - R_I|} \right) + \sum_{i<j} \frac{1}{|r_i - r_j|} 
\end{equation}
with \(r\) being the electronic coordinate, \(R\) is the nuclear coordinate, and \(Z_I\) represents the nuclear charges. Solving this exactly is challenging due to the Coulomb repulsion between electrons. 
To convert this differential equation into a form suitable for computation, a discretization process is typically applied. One common method is the Galerkin approach, where the wavefunction is expanded in terms of a finite set of basis functions or orbitals. These basis functions are chosen to efficiently capture electronic states while minimizing the number of functions required. The Galerkin discretization provides the basis set error to be variational, which is an important feature because it guarantees that as the basis set is refined, the error decreases and the solution converges towards the exact result, making the method well-suited for numerical computations. 
The choice of basis functions introduces a parameter \( N \), the number of spatial orbitals, which directly impacts the complexity of the problem. 
It is very straightforward to write down a guaranteed exact solution for \(\eta\) electrons in \(N\) spatial orbitals, but the solution of this problem scales exponentially in both memory and floating-point operations. As a result, exact solutions are currently limited to \(\eta=26\), \(N=23\)~\cite{gao2024distributed}. 

A widely used and efficient method for electronic structure calculations is Density Functional Theory (DFT). The Hohenberg-Kohn theorem~\cite{Hohenberg1964Inhomogeneous} establishes a one-to-one correspondence between the ground-state electronic density and the defined Hamiltonian, effectively reducing the system’s degrees of freedom to just the density. This reduction allows DFT to scale significantly better than other methods, particularly as system size increases, with linear scaling $O(N)$ achieved in recent advancements~\cite{Nakata2020Large}, where $N$ is a generic parameter for number of basis functions used in the calculation. The efficiency of DFT has enabled large-scale calculations on systems with millions of atoms. For instance, CONQUEST, a linear-scaling DFT code, has been used to perform calculations on systems containing up to 2 million atoms, leveraging 705,024 physical cores on the Japanese Fujitsu-made K-computer~\cite{Nakata2020Large}. Similarly, DFT-FE, another large-scale DFT code utilizing high-order finite element discretization and adaptive spatial discretization strategies, has successfully simulated large-scale Mg dislocation systems, such as pyrIIScrewC with 6,164 Mg atoms (61,640 electrons) and pyrIIScrewC with 10,508 Mg atoms (105,080 electrons), using Summit GPU nodes~\cite{Das2019Fast}. However, despite its efficiency, DFT often falls short of delivering the high accuracy required for out-of-equilibrium chemical systems, which are the focus of our computational workflow. This is because, in practice, DFT approximates the exchange-correlation energy, a critical component for the method's accuracy. In other words, while DFT is exact in principle, it relies on approximations of this key quantity through exchange-correlation functionals, and no systematic approach exists to improve these approximations across all systems. The choice of functional depends on the specific chemical system being studied, and as a result, DFT can struggle to capture all the nuances of different bonding situations~\cite{Mardirossian2017Narbe}. 

While DFT is efficient and widely used, it is limited by its reliance on approximating the exchange-correlation functional, particularly in systems with strong electronic correlations. To overcome these limitations, many-electron wavefunction theories such as coupled-cluster (CC) theory, provide a rigorous approach to achieve high accuracy. These theories are often formulated in the second quantization framework, where the electronic Hamiltonian is written as:
\begin{equation}
    H = \sum_{p,q} h_{pq} a^\dagger_p a_q + \frac{1}{2} \sum_{p,q,r,s} h_{pqrs} a^\dagger_p a^\dagger_q a_s a_r
\end{equation}
where \( h_{pq} \) and \( h_{pqrs} \) represent one- and two-electron integrals, respectively.  In this framework, the CC wavefunction is expressed as an exponential ansatz acting on a reference state \( |\Phi_0 \rangle \) (typically a Hartree-Fock determinant):
\begin{equation} 
    \label{eq: Coupled Cluster} 
    |\psi_{\text{CC}} \rangle = e^{\hat{T}}|\Phi_0\rangle = \sum_{i} \frac{1}{i!} (\hat{T})^i |\Phi_0\rangle 
\end{equation}
where \(\hat{T}\) is the cluster operator, defined as a sum of excitation operators \(\hat{T}_n\) for \(n\)-fold electron excitations:
\begin{align} 
    \hat{T} &= \hat{T}_1 + \hat{T}_2 + \dots + \hat{T}_n \\
    &= \sum_{ai} t_i^a \hat{a}^\dagger_a \hat{a}_i
    + \frac{1}{4} \sum_{abij} t_{ij}^{ab} \hat{a}^\dagger_a \hat{a}^\dagger_b \hat{a}_j \hat{a}_i
    + \cdots + \frac{1}{(n!)^2} \sum_{i_1, \cdots, i_n, a_1, \cdots, a_n} t_{i_1 \cdots i_n}^{a_1 \cdots a_n} \hat{a}_{a_1}^\dagger \cdots \hat{a}_{a_n}^\dagger \hat{a}_{i_n} \cdots \hat{a}_{i_1} 
\end{align}
with $\hat{T}_1, \hat{T}_2, \cdots, \hat{T}_n$ being the excitation operators corresponding to single, double, and higher excitations, respectively. To make the method computationally feasible, the cluster operator is often truncated, typically including only single and double excitations, resulting in the CCSD method. The CCSD(T) method, which includes a perturbative treatment of triple excitations, is considered the gold standard in quantum chemistry and can provide highly accurate ground-state energy estimates for dynamically correlated systems. CCSD(T) scales as $O(N^7)$ and has been applied to large systems, such as a calculation on octane with 1468 basis functions using the aug-cc-pVQZ basis set with NWChem codes. The CCSD part of the calculation took 43 hours on 600 processors, and the perturbative triples took 23 hours on 1400 processors, sustaining a performance of 6.3 TFlops. Despite its somewhat favorable scaling and ability to treat dynamically correlated systems with high accuracy, CCSD(T) breaks down or yields inaccurate results for systems exhibiting static correlation (i.e., systems with multi-reference characteristics). Given the nature of the chemical systems in our computational workflow described in Section \ref{sec: Mg corrosion} and \ref{sec:Nb_workflow}, we expect that many of them may exhibit static correlations.

To address static correlation, methods such as Selected Configuration Interaction (SCI)~\cite{Olsen2011,Gagliardi2007Multiconfigurational,Szalay2012Multiconfiguration} and Density Matrix Renormalization Group (DMRG)~\cite{White1992} are often employed. SCI provides a way to systematically include important configurations in the wavefunction while avoiding the combinatorial explosion of configurations. It does this by selecting a subset of configurations based on their relevance to the system's ground state. In contrast, DMRG, considered the state-of-the-art method for treating statically correlated chemical systems, relies on an approximate wavefunction ansatz known as the matrix product state (MPS). The MPS is constructed as a product of variational objects, each corresponding to an orbital within the basis. A one-site MPS is mathematically represented as:
\begin{equation}
\label{eq: one-site MPS}
    |\psi \rangle = \sum_n A^{n_1} A^{n_2} \cdots A^{n_k} |n_1 n_2 \cdots n_k\rangle
\end{equation}
where $n_i$ represents the occupancy state of orbital $\phi_i$, which could be any of $|0 \rangle, |\uparrow \rangle , |\downarrow \rangle, |\uparrow \downarrow \rangle$. Each $A^{n_i}$ matrix within the MPS contains information regarding the wavefunction coefficients corresponding to the specific orbital occupancy of $\phi_i$. The $A^{n_i}$ matrices are typically $M \times M$ except for the first and last, which have dimensions $1 \times M$ and $M \times 1$, respectively. The parameter $M$, also referred to as the MPS bond dimension, dictates the number of normalized states, thus controlling both the accuracy and computational complexity of a DMRG calculation. When $M^2 \approx N_d$, where $N_d$ is the total number of possible determinants, the DMRG wavefunction becomes exact. Obtaining reliable results from SCI and DMRG requires converging the number $D$ of selected configurations and the bond dimension $B$ respectively, which is an algorithmic bottleneck. See \cref{tab:scaling_of_classical_methods} for a summary of the scaling for these classical approaches. 

In summary, while all of the classical methods discussed such as DFT, CCSD(T), and static correlation methods like DMRG have proven to be powerful tools in quantum chemistry, they each come with inherent limitations. DFT is fast and efficient, but it struggles with accuracy when applied to systems with strong electronic correlation. CCSD(T) can treat dynamically correlated systems but breaks down when dealing with static correlated systems.  On the other hand, methods designed to treat static correlation, such as DMRG, can provide high precision for these systems but encounter computational bottlenecks as the bond dimension must be large for complex systems, limiting scalability. In the computational workflows presented in this paper, the chemical models under consideration can exhibit both dynamically and statically correlation, thus these classical methods will encounter their respective shortcomings in these cases. Therefore, alternative approaches, such as quantum computing, are necessary to overcome these limitations and realize the proposed computational workflows.

\begin{table}[ht]
    \centering
    \begin{tabular}{ccc}
        \hline\hline
        Method & Acronym & Asymptotic Scaling \\
        \hline\\[-8pt]
        Exact Diagonalization & FCI & $\binom{N}{\eta/2}$ \\[5pt]
       \makecell{Coupled Cluster\\Singles Doubles\\Perturbative Triples} & CCSD(T) & $\eta^3(N-\eta/2)^4$ \\[15pt]
        Selected Configuration Interaction & SCI & $D \eta^2(N-\eta/2)^2$ \\[5pt]
        Density-Matrix Renormalization Group & DMRG & $B^3N^3 + B^2N^4$ \\
        \hline\hline
    \end{tabular}
    \caption{Summary of the asymptotic scaling for various classical computational methods. \(N\) refers to the number of spatial orbitals, \(\eta\) to the number of electrons,  \(D\) to the number of selected configurations in SCI, and \(B\) to the bond dimension in DMRG.}
    \label{tab:scaling_of_classical_methods}
\end{table}

\subsection{Prospective Quantum Approach}

Quantum computers are usually expected to function as hardware accelerators. Just as GPUs are specially designed to handle the bulk of the computational challenge in graphics rendering applications, so too do we expect quantum computers to target a specific portion of the workflows described in Sections \ref{sec: Mg corrosion} and \ref{sec:Nb_workflow}; namely, ground state energy estimation (GSEE). In this section, we explain the formal problem statement that underpins the belief that a quantum computer will accelerate GSEE.

GSEE is employed within the workflow after a mathematical toy model has been specified. Such a toy model is expected to incorporate several kinds of approximations, including a method of truncating the dimension of the otherwise-infinite-dimensional Hilbert space and a decision about the order at which interactions are to be neglected. These approximations are well justified within the context of simulating the chemical processes involved in corrosion.

In quantum chemistry, a Hamiltonian is typically expressed using creation and annihilation operators in single-particle and two-particle terms; the fundamental electron–nucleus and electron–electron interactions can be captured by one-body and two-body operators without requiring an infinite hierarchy of interactions. To understand why, consider the fact that the Coulomb force is intrinsically two-body, and the exponential decay of localized orbitals renders higher-order unnecessary for accurate modeling. Moreover, the Born–Oppenheimer approximation further localizes the problem by separating electronic and nuclear degrees of freedom, ensuring that most relevant physics is already contained in these comparatively simple terms. After then taking into account the fact that interaction strengths are likely to decay rapidly with geometric distance between these particles, we can invoke a notion of fixed locality, in the sense that the maximum support of the operators in the natural basis is independent of the size of the system, though it is possible that increasing the range of interactions included may increase the precision and accuracy of the results.

As such, it is reasonable to consider toy models of chemical systems in which the Hamiltonians are described by the sums of $k$-local terms. This simplification, while based on physical principles, introduces a degree of arbitrary freedom on the Hamiltonian, that introduces a level of uncertainty and error. 
While a naive application of the Jordan-Wigner transformation from the second quantized description of the system results in a set of Hamiltonian terms with high weight, recent results on fermion-to-qubit mappings have shown it is possible to develop locality-preserving mappings~\cite{Chapman2020Characterization,chapman2022free,Derby2021compact,chien2022optimizing}.

In general, a reasonable quantum-classical computation workflow requires an approximation for the nucleonic configuration ground state for the atomic configuration of the system as it evolves in time, before we can begin to consider the electronic structure Hamiltonian for a given atomic setup. Since, in general, the atoms of metallic alloys have a quasi-crystalline structure, we may approximate the nuclei-nuclei and nuclei-electron interactions by a $k$-local Hamiltonian for a lattice structure; perturbations from the regular lattice caused by misplaced atoms may then be accounted for by perturbations in the interaction strengths for each $k$-local term in the Hamiltonian. The degree of such simplifications may be informed by similar accessions made in entirely classical simulations. Additionally, we expect energy cutoffs to imply a restriction to finite-dimensional Hamiltonians that could be expressed using a total of $n$ qubits, with $n$ growing logarithmically with the dimension of the Hamiltonian.

As such the Hamiltonian may be limited to a $k$-local Hamiltonian~\cite{kempe2006complexity}.
Unfortunately, the problem of determining the ground state energy for a $k$-local Hamiltonian is known to be \textsf{QMA}-complete~\cite{kempe2006complexity,kitaev2002classical} even for general instances emerging from the electronic structure problem~\cite{ogorman2022intractability}. So finding a $k$-local estimation for a particular configuration of atoms is not, on its surface, evidence for the quantum amenability of the task of modeling the corrosion. However, recent complexity theoretic results give us reason for optimism. Many of the classical techniques currently in use give a good starting point for approximations to the ground state of a many-body Hamiltonian. These classical approximations can be used to effectively `guide' the quantum computer to the true ground state and energy. Such a process has recently been shown to be amenable for a quantum machine through the results on classifying the \emph{guided local Hamiltonian (GLH) problem}~\cite{Gharibian2022}.

It was shown in Ref.~\cite{Gharibian2022} that \emph{GSEE is classically amenable} for a $k$-local Hamiltonian if the accuracy target grows polynomially with the number of qubits on which the Hamiltonian is acting. For an accuracy target that is approximately constant, \citet{cade2023Guided} show that the problem of guiding a quantum computer to the ground state energy estimation using a specific family of semi-classical states is $\textsf{BQP}$-complete, even when the Hamiltonian is rescued to 2-local. Note that a problem being $\mathsf{BQP}$-complete implies that, insofar as the $k$-local Hamiltonian constitutes an accurate model of the system, GSEE for metallic corrosion by exposure \emph{is} quantum amenable and even when likely not classically amenable. It's also worth noting that the authors provide a specific family semi-classical guiding state as an \textit{indicative class}, rather than the only one that will work. While we do not believe that the ground sates in question are likely to be close to a Gaussian state in the traditional sense~\cite{Chapman2020Characterization}, it is more likely that the there is a reasonable guiding state to be found based on close-ness to a `disguised' free-fermion solvable model~\cite{Elman2021,*chapman2023unified}. Investigating which families of semi-classical states constitute good guiding states~\cite{Waite2025} or whether there are provable guarantees quantum speedup for guiding the chemical Hamiltonian problem~\cite{ogorman2022intractability} would be interesting pursuit.

That being said, quantum amenability is not a binary property; the extent to which the computation is amenable for a quantum device is dependent on the choices of the size of the system being modeled. The locality restriction of the terms will play a role in determining this. So too, will the total number of Hamiltonian terms, as well as the precision and accuracy to which their interaction strengths must be specified.

Results on the guidability of a quantum Hamiltonian have important consequences for our quantum amenability analysis.
The challenge is that the guided $k$-local Hamiltonian problem requires a semi-classical guiding state which has to be somehow determined. Further, the size of the overlap, $\delta$, with the true ground state will determine whether the problem is in \textsf{BQP}, as the number of repetitions of the QPE subroutine will depend on $\delta$ as well as the unscaled Hamiltonian norm. A poor overlap would yield exponential scaling, hence not quantum amenable.

\subsection{Quantum-phase-estimation-based Workflow}
\label{subsub:QPE workflow}

\begin{figure*}[t]
    \centering
    \includegraphics[width=0.6\textwidth]{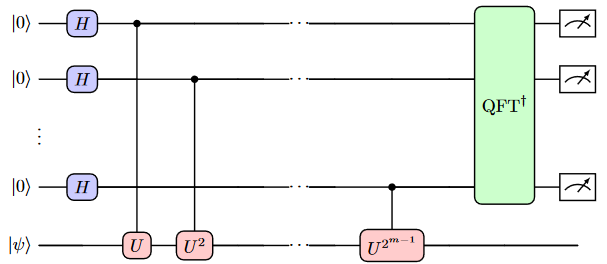}
    \caption{ QPE circuit with queries to \(U = e^{-iHt} \) for extracting the eigenspectrum of \(H\).}
    \label{fig:OriginalQPE}
\end{figure*}

Quantum phase estimation (QPE) is an important quantum subroutine that  estimates the eigenvalue $\lambda = e^{i2\pi\varphi}$, corresponding to the eigenvector $\ket{\nu}$ of the unitary operator $U$. In the context of ground state energy estimation, the unitary is defined by the a normalized Hamiltonian as $U = e^{-i t H}$, such that $\varphi \in (0,1)$. The original formulation of QPE due to Kitaev~\cite{kitaev1995quantum} accomplishes an estimate of the value of $\varphi$ to the additive error $\varepsilon$ with high probability, using $O(\log(\frac{1}{\varepsilon}))$ qubits and $O( \frac{1}{\varepsilon})$ controlled-$U$ operations. In the context of ground state energy estimation, $U = e^{-i tH}$. See Figure \ref{fig:OriginalQPE} for the circuit implementation of this algorithm.

Note that, in general, the eigenvalues of $H$ are not within the range of $(0,1)$ or even positive. Thus, when implementing QPE to perform ground state energy estimation, we are actually performing $U = e^{-i \hat{H}}$, where
\[
    \hat{H} = \frac{2\pi(H - E_{\text{min}}\mathbf{1})}{E_{\text{max}} - E_{\text{min}}},
\]
and we have replaced $t$ with
\[
    t = \frac{2\pi}{E_{\text{max}} - E_{\text{min}}}.
\]
However, this is just a constant shift and rescaling of the original Hamiltonian, where the constants can be easily bounded. Now, if we had the ability to prepare $\ket{\nu}$ to be the ground state of $H$, QPE would output the ground state energy; however, this assumption is not realistic. Instead, we assume to be able to prepare a quantum state $\ket{\tilde{g}}$ such that 
\begin{equation}
    | \braket{\tilde{g}}{g}|^2 \leq \delta \in \Omega \left(\frac{q}{\textsf{poly}(n)}\right),
\end{equation}
where $\ket{g}$ is the ground state of the Hamiltonian on $n$ qubits. The state $\ket{\tilde{g}}$ can be seen as the guiding state. Under this assumption, QPE will output an approximation of the ground state energy with probability $\delta$ and one of the excited state energies with probability $1-\delta$. Thus, using QPE for GSEE requires having an upper bound on the ground state energy that allows distinguishing it from the energy of an excited state.

\begin{figure*}
    \centering
    \includegraphics[width=.8\textwidth]{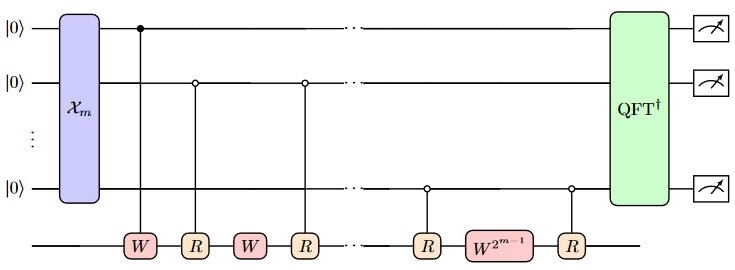}
    \caption{Quantum phase estimation on the "Walk" Operator \(W\) using double phase difference trick. }
    \label{fig:qubitized_QPE_full}
\end{figure*}

Performing Hamiltonian simulation within the phase estimation introduces the difficulties of avoiding error introduced by the Hamiltonian simulation that would then need to be taken into account in bounding the overall error. Additionally, there can be ambiguities in the phase that require simulation of the Hamiltonian over very short times to eliminate.
Hence, in this work, we employ a Heisenberg-limited QPE, where the Hamiltonian information is encoded via a Szegedy walk. Within this framework, the first step is to decompose the specified electronic structure Hamiltonian as a linear combination of unitaries (LCU). In other words, 
\begin{equation}
    H = \sum_{\ell=0}^{L-1} w_\ell H_\ell \hspace{1 cm} H_\ell^2 = \mathbf{1}
    \label{eq: HamiltonianLCU}
\end{equation}
where $w_\ell\in \mathbb{R}^+_0$ and $H_\ell$ are self-inverse operators on $n$ qubits. In particular, $H$ is the Hamiltonian that arises from performing a Galerkin discretization of the original Hamiltonian using the plane-wave basis functions in first quantization, as described in the Hamiltonian Generation section of the main text. Next, we use the technique of qubitization to block encode the Hamiltonian spectra as a \textit{Szegedy quantum walk} $W$ defined as 
\begin{equation}
    W = R \cdot \text{SELECT}, \hspace{1 cm} R = 2 |\mathcal{L} \rangle \langle \mathcal{L} | \otimes \mathbf{1} - \mathbf{1}.
\end{equation}
where $\ket{\mathcal{L} }$ is a state that can be prepared by the ``preparation oracle'', referred to as PREPARE (or PREP), and SELECT (or SEL) is the ``Hamiltonian selection oracle''. PREPARE is defined as 
\begin{equation}
    \text{PREPARE} \equiv \sum_{\ell=0}^{L-1} \sqrt{\frac{w_\ell}{\lambda}} \ket{\ell} \bra{0}, \hspace{1 cm} \text{PREPARE} |0\rangle^{\otimes \log L} \mapsto \sum_{\ell=0}^{L-1} \sqrt{\frac{w_\ell}{\lambda}} |\ell \rangle \equiv |\mathcal{L} \rangle
\end{equation}
with $\lambda$ (1-norm) defined as $ \lambda= \sum_{\ell=0}^{L-1} w_{\ell} $, and SELECT is defined as 
\begin{equation}
     \text{SELECT} \equiv \sum_{\ell =0}^L \ket{\ell} \bra{\ell} \otimes H_{\ell}, \hspace{1 cm}  \text{SELECT} |\ell\rangle |\psi \rangle \mapsto |\ell\rangle H_\ell |\psi \rangle.   
\end{equation}
First, note that both \( R \) and \text{SELECT} are reflection operators, since \( R^2 = \text{SELECT}^2 = \mathbf{1} \),  which implies that \(W\) takes the form of a Szegedy quantum walk and, as we will show, can be put in block-diagonal form with each block of size 2 -- a consequent of Jordan's lemma. Notably, \( R \) reflects about the state \( \ket{\mathcal{L}} \), while \text{SELECT} reflects about the controlled application of Hamiltonian terms. Next, observe that
\begin{equation}
    U_H = (\text{PREPARE}^\dagger \otimes \mathbf{1}) \cdot \text{SELECT} \cdot (\text{PREPARE} \otimes \mathbf{1}) = 
    \begin{bmatrix}
    H/\lambda & * \\
    * & *
    \end{bmatrix} 
\end{equation}
where \(U_H\) is unitary, is a block encoding of $H$. In other words, 
\begin{equation}
    H/\lambda = (\langle \mathcal{L}  | \otimes \mathbf{1}) \text{SELECT} (|\mathcal{L} \rangle \otimes \mathbf{1} )
    \label{eq: W}
\end{equation}
However, the eigenvalues of \(U_H\) are not directly related to the eigenvalues of \(H\). Nevertheless, the construction of \(W\), which serves as a specific block encoding of \(H\), naturally encodes the Hamiltonian's spectral properties. 
This can be seen as follows: Let \(|\phi_k \rangle \) be the \(k\)-th eigenstate of \(H\) with eigenvalue \(E_k\). Then observe that 
\begin{subequations}
    \begin{gather}
        W |0\rangle |\Phi_k\rangle =  |0\rangle \frac{E_k}{\lambda}|\Phi_k\rangle + \sqrt{1 - \frac{E_k^2}{\lambda^2}} |\psi^\perp\rangle, \\
        W |\psi^\perp\rangle = \frac{E_k}{\lambda} |\psi^\perp\rangle - \sqrt{1 - \frac{E_k^2}{\lambda^2}} |0\rangle |\Phi_k\rangle.
    \end{gather}   
\end{subequations}
where \(|\psi^\perp\rangle\) is a state that is orthogonal to \(|0\rangle |\phi_k \rangle\). Thus, the action of \(W\) leaves a set of 2-dimensional subspaces span by \(\{|0\rangle |\Phi_k\rangle, |\psi^\perp\rangle \}\) invariant. In particular, within this subspace, \(W\) can be expressed as  
\begin{equation}
    W = \begin{pmatrix}
        \frac{E_k}{\lambda} & \sqrt{1 - \frac{E_k^2}{\lambda^2}} \\
        - \sqrt{1 - \frac{E_k^2}{\lambda^2}} & \frac{E_k}{\lambda} 
    \end{pmatrix}
\end{equation}
Let \( \cos(\theta_k) = E_k / \lambda \) then we have 
\begin{subequations}
    \begin{gather}
        W |0\rangle |\Phi_k\rangle = \cos(\theta_k) |0\rangle |\Phi_k\rangle - \sin(\theta_k) |\psi^\perp\rangle, \\
        W |\psi^\perp\rangle = \cos(\theta_k) |\psi^\perp\rangle + \sin(\theta_k) |0\rangle |\Phi_k\rangle.
    \end{gather}   
\end{subequations}
Now, \(W\) can be expressed as 
\begin{equation}
    W = \begin{pmatrix}
        \cos(\theta_k) & -\sin(\theta_k) \\
        \sin(\theta_k) & cos(\theta_k) 
    \end{pmatrix}
\end{equation}
within \(\{|0\rangle |\Phi_k\rangle, |\psi^\perp\rangle \}\). The eigenvalues of \(W\) within each of this subspace is \(e^{\pm i \theta_k} = e^{\pm i arccos(E_k / \lambda )} \). More generally, \(W\) exhibits a block-diagonal structure with each block corresponding to a two-dimensional subspace spanned by \( \{|0\rangle |\phi_k \rangle, |\psi^\perp \rangle \} \).
We can then extract the eigenvalues of \(H\) by performing phase estimation on \(W\). 
The parameter $\lambda$ significantly impacts the computational cost of the algorithm and is therefore a crucial factor in resource estimation.  The overall phase estimation procedure can be implemented using the quantum circuit illustrated in Figure~\ref{fig:qubitized_QPE_full}.

Note that this particular implementation of the QPE circuit differs from the standard approach, where one typically applies a controlled operation on each of the \(W^{2^k}\) operators. In this implementation, instead of leaving the ancilla qubit in the \( |0 \rangle \) state unchanged, we apply an inverse unitary. This adjustment effectively doubles the effectiveness of the phase estimation procedure.  As discussed in~\cite{babbush2018encoding}, this can be achieved by simply removing the controls from the \(W^{2^k}\) operators and inserting controlled reflection operators \(R\) into the circuit, as shown in Figure \ref{fig:qubitized_QPE_full}, which allows for the application of either \(W^{2^k}\) or \( (W^\dagger)^{2^k}\) depending on the state of the ancilla. However, this introduces phase ambiguity. In particular, a phase modulo \(\pi\) cannot distinguish between \(\arccos(E_k / \lambda )\) and \(\arccos(E_k / \lambda ) + \pi\). To resolve this ambiguity, a controlled \(W\) is placed at the beginning of the circuit.
In contrast to the typical phase estimation, where we prepare the state of the ancilla qubits in a uniform superposition state through Hadamard gates, here we apply a unitary \(\mathcal{X}_m\) that maps the state of the ancilla qubits as follows:
\begin{equation}
\mathcal{X}_m |0\rangle^{\otimes m} = \sqrt{\frac{2}{2^m + 1}} \sum_{n=0}^{2^m-1} \sin\left(\frac{\pi (n+1)}{2^m + 1}\right) |n \rangle
\label{eq: X_m unitary}
\end{equation}
This unitary is achieved with a circuit complexity of \(\tilde{O}(m)\), which is negligible in the overall algorithm cost. Similarly, the cost of the inverse quantum Fourier transform \( \text{QFT}^\dagger \) does not significantly impact the resource estimation. Thus, the main cost of the algorithm comes from the implementation of \(\text{SELECT}, \text{PREPARE}\), and \(\text{PREPARE}^\dagger\).

We discuss the implementation of each component in the first quantization framework in more detail in the Supplementary~Materials~\ref{supp_mat:first_quantization}. We then discuss the implementation of this algorithm in the second quantization framework, following the Linear-T encoding by Babbush et al.~\cite{babbush2018encoding}, in Section \ref{supp_mat:second_quantization}.  The code for our implementation is available in Ref.~\cite{Obenland2024pyliqtr}.

\subsubsection{Initial State Preparation for Quantum Phase Estimation}
\label{subsec:init_prep}

A significant computational cost associated with using QPE for ground state energy estimation and preparing the ground state is attributed to the initial `guiding state'. This is because the success probability of QPE is directly tied to the overlap between the initial state and the ground state in question. Specifically, the success probability scales as $O(1/p_0^2)$, where $p_0 = |\langle \bar{g}| g \rangle|$, with $|\bar{g}\rangle$ is the initial state and $|g\rangle$ is the ground state of the Hamiltonian. 

A common choice for the initial state is the Hartree-Fock (HF) state, a single Slater determinant that serves as a reasonable starting point for QPE. However, for larger or strongly correlated systems, the HF state may not provide sufficient overlap with the ground state~\cite{Lee2023Evaluating}. Several strategies exist to enhance this overlap. One such approach is \textit{adiabatic state preparation}, where the HF state is gradually transformed into an approximated ground state by simulating the time evolution of a time-dependent Hamiltonian. This method leverages the adiabatic theorem~\cite{kato1950On, jordan2008quantum}, which ensures convergence to the ground state if the evolution time $T$ is sufficiently large relative to the inverse of the minimum energy gap. While natural for quantum computers, the cost can be prohibitive, as $T$ scales as $O(1/\eta^2)$, where $\eta$ is the minimum spectral gap.

Alternatively, one can leverage \textit{classical quantum chemistry methods} to generate an initial state with improved overlap. This includes approaches such as Selected Configuration Interaction (SCI), Coupled Cluster (CC), or Density Matrix Renormalization Group (DMRG). These methods construct an approximate wavefunction either as a sum of Slater determinants (SOS) or as a Matrix Product State (MPS), which can then be encoded into a quantum circuit~\cite{fomichev2024initial}. The cost of preparing an SOS state with $D$ determinants scales as $O(D\log D)$ Toffoli gates, while an MPS state requires $O(N\chi^{3/2})$ Toffoli gates, where $\chi$ is the bond dimension in the DMRG algorithm and $N$ is the number of qubits in the system. Another efficient state preparation algorithm is described in \cite[Appendix B2]{Waite2025}.

For large periodic systems, methods such as adiabatic state preparation, SCI, or DMRG might become infeasible due to their high computational cost. In such cases, the only practical options are Density Functional Theory (DFT) or Hartree-Fock (HF) methods. Given that HF may not provide sufficient state overlap, one can leverage Kohn-Sham (KS) DFT orbitals with appropriately chosen exchange-correlation functionals, which have been shown to yield better initial states for many post-Hartree-Fock wavefunction methods like CC or CI~\cite{Boyn2022Elucidating}. This can lead to a higher overlap with the true ground state and offers a computationally efficient way to obtain a better guiding state. A more drastic yet effective technique is to simplify the Hamiltonian by removing interaction terms until it describes a non-interacting fermionic system. The ground state of such a system can be prepared using $O(n^2)$ quantum gates. As shown in~\cite{Bravyi2017}, the relationship between the ground state of the simplified Hamiltonian and the full interacting Hamiltonian is polynomially dependent on the number of terms removed. This method offers a scalable alternative for efficiently preparing an initial state in QPE.

Finally, to mitigate the potentially large number of repeated phase estimation circuits caused by insufficient overlap between the initial state (e.g., KS-DFT output) and the true ground state, one can leverage the partial eigenstate projection rejection method from~\cite{berry2018improved}. This technique reduces QPE costs by restarting the phase estimation procedure as soon as the estimated energy exceeds a chosen threshold \( \tilde{E}_0 \), where \( E_0 \leq \tilde{E}_0 < E_1 \), with \(E_0\) being the true ground state energy and \(E_1\) being the first excited state energy. Specifically, the method has an overall complexity scaling of \( O\left( \frac{1}{p_0^2 (E_1 - \tilde{E}_0)} + \frac{1}{\epsilon} \right) \), in contrast to the scaling \( O\left( \frac{1}{p_0^2 \epsilon} \right) \) for phase estimation with full accuracy, where \( \epsilon\) is the final error (e.g. chemical accuracy). For many chemical systems, the gap between the first excited state and the ground state is quite large, and hence \( p_0^2 (E_1 - \tilde{E_0}) \) will often be much larger than \( \epsilon \), thus effectively eliminating the overhead from \( p_0 \).

\subsubsection{High-Level Description of Quantum Circuit Construction}
\label{supp_mat: high level circuits implementation for qubitization QPE}
In this section, we discuss the high-level implementation of the qubitized QPE framework introduced earlier and depicted in Figure \ref{fig:qubitized_QPE_full}. The algorithm consists of repeated iterations of the uncontrolled Walk operator \(W\) and one controlled version of it to eliminate the phase modulo \(\pi\) ambiguity.  In both cases, the operator can be constructed from the SELECT and PREPARE oracles~\cite{babbush2018encoding}, which together block encode the Hamiltonian in a unitary circuit as shown in \Cref{fig: Walk,fig: Controlled-Walk}. For the electronic structure Hamiltonian, there are several approaches for constructing these oracles in the literature~\cite{babbush2018encoding, Su2021Fault, Lee2021hypercontraction, Berry2019qubitization, vonBurg2021Quantum, georges2025firstquantized}. The differences between these approaches stem mainly from different choices in how to represent the Hamiltonian. Here, we provide estimates for two possible choices: the first quantization representation in a plane-wave basis~\cite{Su2021Fault}, and the second quantization representation in a dual plane-wave basis~\cite{babbush2018encoding}.

\begin{figure}[ht]
    \centering
    \begin{quantikz}
        \qw  & \qw & \qw & \qw & \octrl{1}   & \\
        \lstick[3]{$\ket{\ell}$}  
          & \gate[3, style={fill=yellow!40, rounded corners}]{\text{PREPARE}} 
          & \gate[4, style={fill=brown!60, rounded corners}]{\text{SELECT}} 
          & \gate[3, style={fill=yellow!40, rounded corners}]{\text{PREPARE}^{-1}} 
          & \octrl{1} &\qw \\
          & \qw &  &    & \octrl{-1} &\qw \\
          & \qw &  &    & \ctrl{-1} &\qw \\
        \lstick{$\ket{\psi}$}  
          & \qwbundle{n} 
          & \qw & \qw & \qw  &
    \end{quantikz}
    \caption{Circuit representation of the Walk operator ($W$).}
    \label{fig: Walk}
\end{figure}

\begin{figure}[ht]
    \centering
    \begin{quantikz}
        \qw  & \qw & \ctrl{2} & \qw & \octrl{1}   &\\
        \lstick[3]{$\ket{\ell}$}  
          & \gate[3, style={fill=yellow!40, rounded corners}]{\text{PREPARE}} 
          & \gate[4, style={fill=brown!60, rounded corners}]{\text{SELECT}} 
          & \gate[3, style={fill=yellow!40, rounded corners}]{\text{PREPARE}^{-1}} 
          & \octrl{1} &\qw \\
          & \qw &  &    & \octrl{-1} &\qw \\
          & \qw &  &    & \ctrl{-1} &\qw \\
        \lstick{$\ket{\psi}$}  
          & \qwbundle{n} 
          & \qw & \qw & \qw  & 
    \end{quantikz}
    \caption{Circuit representation of the controlled-Walk operator.}
    \label{fig: Controlled-Walk}
\end{figure}

One notable primitive used for these oracles is the unary iteration technique, which produces a circuit for iterating over the one-hot unary encoding of an index register value. The implementation in~\cite{babbush2018encoding} uses compute and uncompute AND operations to construct a circuit that executes a controlled iteration over $L$ elements using only $4L-4$ T gates. Broadly speaking, this technique is useful for constructing circuits for multiplexed unitaries of the form 
\begin{equation}
\sum_j \ket{j}\bra{j}\otimes U_j.
\end{equation}
For example, the above can represent a generic SELECT oracle, where there is a different operator $U_j$ for each index $j$ of the LCU representation of the Hamiltonian. Together with the alias sampling preparation scheme from~\cite{babbush2018encoding}, this would correspond to a general purpose, widely applicable block encoding. However, the literature has shown that it is more resource optimal to use an encoding that is specialized to the problem of interest. Thus, depending on the representation used for the Hamiltonian, the quantum circuits for implementing these oracles will look very different. 

The number of repetitions of the uncontrolled walk operator will also be affected by the choice of representation since it scales proportional to the Hamiltonian 1-norm. Thus, the resources of the overall circuit can be optimized by considering techniques for reducing the 1-norm and for reducing the cost of the oracle circuit constructions. Ideally, a representation with a lower 1-norm would also correspond to a circuit with a lower cost, but this is not necessarily the case. This is why it's helpful to have detailed comparisons of different encodings for the same problem, as presented here.

\section{First Quantization: Remarks and Resource Estimates}
\label{supp_mat:first_quantization}
In this section, we will review our setup and provide the full resource estimates within the first quantization framework. These estimates are presented in \cref{tab:first_mg,tab:first_nb}.

\subsection{Hamiltonian Formulation}
In first quantization, the Born-Oppenheimer Hamiltonian in the plane-wave basis can be written as
\begin{equation}
\begin{aligned}
H &= T + U + V + \frac{1}{2}\sum_{\ell \neq \kappa = 1}^L \frac{\zeta_\ell \zeta_\kappa}{\|R_\ell - R_\kappa\|} \\
T &= \sum_{i=1}^\eta \sum_{p \in G} \frac{\|k_p\|^2}{2} |p\rangle \langle p|_i \\
U &= -\frac{4\pi}{\Omega} \sum_{\ell=1}^{L} \sum_{i=1}^{\eta} \sum_{\substack{p,q \in G \\ p \neq q}} \left( \zeta_\ell \frac{e^{ik_{q-p}\cdot R_\ell}}{\|k_{p-q}\|^2} \right) |p\rangle \langle q|_i \\
V &= \frac{2\pi}{\Omega} \sum_{i \neq j = 1}^\eta \sum_{p, q \in G} \sum_{\substack{\nu \in G_0 \\ (p+\nu) \in G \\ (q-\nu) \in G}} \frac{1}{\|k_\nu\|^2} |p+\nu\rangle \langle p|_i |q-\nu\rangle \langle q|_j
\end{aligned}
\label{eq: first_quantized_hamiltonian}
\end{equation}
where $\eta$ is the number of electrons, $L$ is the number of atoms, $R_\ell$ is the location of each atom, $\zeta_\ell$ is the atomic number, $\Omega$ is the computational cell volume, $k_p = \frac{2\pi p}{\Omega^{1/3}}$ is a 3-dimensional reciprocal lattice vector, and $p \in G$, where
\[
G = \left[-\frac{N^{1/3}-1}{2}, \frac{N^{1/3}-1}{2}\right]^3 \subset \mathbb{Z}^3, \quad G_0 = \left[-N^{1/3}, N^{1/3}\right]^3 \subset \mathbb{Z}^3 \setminus \{(0,0,0)\},
\]
and $N$ is the number of plane waves forming the basis. Here, the kinetic term $T$ represents the momentum energy of electrons in the plane-wave basis, the potential $U$ describes the interaction between electrons and nuclei, and $V$ accounts for electron-electron Coulomb interactions.

In the first quantization framework, the wavefunction of a system with \( \eta \) electrons in \( N \) basis functions (orbitals) is expressed by explicitly assigning each electron to an orbital. This approach requires \( \eta \) registers, each capable of encoding an index corresponding to one of the \( N \) orbitals. The number of qubits per register is given by \( n = \lceil \log_2 N \rceil \), leading to a total qubit requirement of \( O(\eta \log N) \), which is significantly more compact than second-quantization-based approaches when $\eta << N$. We will go over this in more detail in Section \ref{supp_mat:second_quantization}. A general multi-electron wavefunction in first quantization is represented as a superposition of Slater determinants:
\begin{equation}
    |\psi\rangle = \sum_{i \in \binom{N}{\eta}} c_i A(|p_{i_1}, ..., p_{i_\eta} \rangle),
\end{equation}
where the coefficients $c_i$ satisfy $\sum_i |c_i|^2 = 1$, and $A$ is the antisymmetrization operator ensuring fermionic exchange symmetry:
\begin{equation}
    A : |p_{i_1}, ..., p_{i_\eta} \rangle \to \sum_{\sigma \in S_\eta} (-1)^{\pi(\sigma)} \frac{1}{\sqrt{\eta!}} |\sigma(p_{i_1}, ..., p_{i_\eta})\rangle.
\end{equation}
where \(S_\eta\) is the symmetric group on \(\eta\) elements, and \(\pi(\sigma)\) is the parity of the permutation. Note that unlike the second-quantized representation, which naturally enforces anti-symmetry via fermionic creation and annihilation operators, this formulation requires explicit symmetrization. Therefore, in the first quantization framework, we must take into account the anti-symmetry during the initial state preparation process. However,~\cite{berry2018improved} shows that an arbitrary state (e.g., the HF state) of \( \eta \) electrons in \( N \) orbitals can be antisymmetrized using an algorithm with gate complexity \( O(\eta \log \eta \log N) \). This procedure needs to be applied only once, as the particle exchange operator commutes with the Hamiltonian. Consequently, any simulation under the Hamiltonian will preserve the antisymmetry property, and hence the state remains antisymmetric when going through the QPE circuit.

\subsection{Block Encoding Circuit}
The qubitized quantum phase estimation (QPE) algorithm relies on block encoding the Hamiltonian using qubitization operators. As discussed earlier in the main text, this method begins by expressing the problem Hamiltonian as a linear combination of unitaries (LCU). To express the Hamiltonian \ref{eq: first_quantized_hamiltonian} as an LCU, we follow the approach outlined in~\cite{Su2021Fault} and rewrite the \(T\), \(U\), and \(V\) operators as follows:
\begin{equation}
\begin{aligned}
T &=\frac {\pi^2}{\Omega^{2/3}} \sum_{j=1}^{\eta}\sum_{w\in\{x,y,z\}} \sum_{r=0}^{n_p-2}\sum_{s=0}^{n_p-2} 2^{r+s} \sum_{b\in\{0,1\}}\left(\sum_{p\in G} (-1)^{b(p_{w,r} p_{w,s}\oplus 1)} \ket{p}\!\bra{p}_j\right), \\
U &= \sum_{\nu \in G_0} \sum_{\ell = 1}^{L} \frac{2\pi \zeta_\ell}{\Omega \| k_\nu \|^2} \sum_{j=1}^{\eta} \sum_{b \in \{0,1\}} \left( -e^{-\mathrm{i} k_\nu \cdot R_\ell} \sum_{q \in G} (-1)^{b[(q - \nu) \notin G]} \ket{q - \nu} \!\bra{q}_j \right), \\
V &= \sum_{\nu \in G_0} \frac{\pi}{\Omega \| k_\nu \|^2} \sum_{i \neq j = 1}^{\eta} \sum_{b \in \{0,1\}} \left( \sum_{p,q \in G} (-1)^{b \left( [(p + \nu) \notin G] \lor [(q - \nu) \notin G] \right)} \ket{p + \nu} \!\bra{p}_i \cdot \ket{q - \nu} \!\bra{q}_j \right),
\end{aligned}
\label{eq: first_quantized_Ham_LCU}
\end{equation}
where $ n_p=\left\lceil\log \left(N^{1/3}+1\right)\right\rceil \,$ represents the number of qubits required to store a signed binary representation of one component of the momentum of a single electron, $p_{w,r}$ is indicating bit $r$ of the $w$ component of $p$, $\zeta_\ell$ being the nuclear charges, and $\eta$ denotes the number of electrons. 

In this representation, it's helpful to separate the SELECT and PREPARE oracles into components for the kinetic energy $T$ and the potentials $U + V$, such that
\begin{equation}
    \langle 0 | \text{PREP}_T^{\dagger} \cdot \text{SEL}_T \cdot \text{PREP}_T |0\rangle = \frac{T}{\lambda_T}, \quad 
    \langle 0 | \text{PREP}_{U+V}^{\dagger} \cdot \text{SEL}_{U+V} \cdot \text{PREP}_{U+V} |0\rangle = \frac{U+V}{\lambda_U + \lambda_V}.
\end{equation}
The full Hamiltonian can then be block encoded by introducing an additional ancilla qubit that controls whether the $T$ or $U + V$ operators are applied, with the overall normalization factor being $\lambda = \lambda_T + \lambda_U + \lambda_V$, where
\begin{align}
    \lambda_T &= \frac{6\eta\pi^2}{\Omega^{2/3}} \left( 2^{n_p-1}\!-1 \right)^2 = {\cal O}\left(\frac{\eta N^{2/3}}{\Omega^{2/3}}\right),\qquad 
    \qquad
    \lambda_U = \frac{\eta \sum_\ell \zeta_\ell}{\pi \Omega^{1/3}}  \lambda_\nu = {\cal O}\left(\frac{\eta^2 N^{1/3}}{\Omega^{1/3}}\right)\\
    \lambda_V &= \frac {\eta(\eta-1)}{2\pi \Omega^{1/3}} \lambda_\nu = {\cal O}\left(\frac{\eta^2 N^{1/3}}{\Omega^{1/3}}\right),\qquad\lambda_\nu = \sum_{\nu\in G_0} \frac 1{\norm{\nu}^2} \leq \int_0^{2\pi}\!\! \!\! \int_{0}^\pi\! \! \!\int_0^{N^{1/3}} \!\!\!\!\!\! {\rm d}r \, {\rm d}\phi \,{\rm d}\theta \, \left(\frac{1}{r^2}\right)r^2 \sin \theta = {\cal O}\left(N^{1/3}\right) \, . \nonumber
\end{align}

For our estimates, we use the circuit implementation of these operators from Section II of~\cite{Su2021Fault} to construct the quantum circuits that realize the block encoding in Eq.~\ref{eq: W}. An important note for this implementation is that the preparation method for $U + V$ has a success probability of approximately $1/4$, i.e.,
\begin{equation}
    \widetilde{\text{PREP}}_{U+V}|0\rangle \otimes I \approx \frac{1}{2}|0\rangle\otimes \text{PREP}_{U+V} + \frac{\sqrt{3}}{2}|1\rangle \otimes\text{PREP}_{U+V}^{\perp},
\end{equation}
which means the operators are encoded with a normalization factor close to $4(\lambda_U + \lambda_V)$. So, an ancilla qubit is introduced to adjust the relative amplitudes between $T$ and $U + V$. This is preferable to using oblivious amplitude amplification to increase the success probability because it doesn't require additional repetitions of the preparation oracle. Then, assuming $\lambda_T < 3(\lambda_U +\lambda_V)$, the state preparation is given by
\begin{equation}
    (\cos \theta |0\rangle + \sin \theta |1\rangle) \otimes \text{PREP}_T |0\rangle \otimes \widetilde{\text{PREP}}_{U+V} |0,0\rangle,
\end{equation}
and the selection operator is
\begin{equation}
    I \otimes I \otimes |0\rangle \langle 0| \otimes \text{SEL}_{U+V} +
    |0\rangle \langle 0| \otimes \text{SEL}_T \otimes |1\rangle \langle 1| \otimes I +
    |1\rangle \langle 1| \otimes I \otimes |1\rangle \langle 1| \otimes I,
\end{equation}
where the selection of $T$ is conditioned on both the failure of the $U + V$ preparation and the ancilla qubit being in the state $|0\rangle$. Together, these steps result in the block encoding
\begin{equation}
    \frac{U+V}{4(\lambda_U + \lambda_V)} + \frac{3T \cos^2 \theta}{4 \lambda_T}.
\end{equation}
The angle $\theta$ is then determined based on the condition that the above expression is proportional to $T + U + V$, yielding
\begin{equation}
    \theta = \arccos \sqrt{\frac{\lambda_T}{3(\lambda_U + \lambda_V)}}.
\end{equation}
Note, this holds for the case $\lambda_T < 3(\lambda_U+\lambda_V)$ but the inverse of this can be handled by adjusting the select operator as discussed in~\cite{Su2021Fault}.

Using these operators, the circuit is designed such that the necessary states are prepared in several ancilla registers. During this preparation, various qubits are used to flag the success of subroutines. These flags are what ultimately determine which select operator will be applied. Then, controlled on these flag qubits, the ancilla registers corresponding to the appropriate select operator are swapped into the selection register, the select operator is applied, and then the ancilla are swapped back out and the inverse preparation circuit is executed. The benefit of this design is that the subroutines which are needed for more than one operator can be executed once, and the subroutines which are unique to one operator can be executed after the swap. As such, the most costly part of this block encoding implementation is the multiplexed swap, which is executed a total of 4 times and costs $12\eta n_p + 4\eta - 8$ Toffolis.

\subsection{Resource Estimates}
\label{sec: supp_resource_estimates_1st_quantization}
In this section, we present quantum computing resource estimates for two representative application models: 1) designing Mg-rich sacrificial coatings and corrosion-resistant Mg alloys for aqueous environments, and 2) designing high-temperature corrosion-resistant alloys. To derive these estimates, we will again follow the methodology outlined in~\cite{Su2021Fault}, specifically Sections II A-D, to construct the circuits for the state preparation and selection subroutines discussed earlier. These circuits will then be combined to form the complete phase estimation circuit, as shown in Figure 7 of the main text, which serves as the basis for our resource estimate results.

To construct block encoding circuits, several precision parameters must be defined as outlined in Section E of~\cite{Su2021Fault}. In particular, the following parameters must be specified: 
\begin{enumerate}
    \item \(b_r\): the number of precision bits used for rotating the ancilla qubit in the amplitude amplification subroutine of the uniform superposition,
    \item \(n_{\mathcal{M}}\): the number of bits used for inequality testing to prepare a state with amplitudes \(1/\|\nu\|\),
    \item \(n_T\): the number of bits used in the rotation on the qubit selecting between \(T\) and \(U+V\),
    \item \(n_R\): the number of bits used to store each component of the atomic position.
\end{enumerate}
We follow the default and set $b_r = 8$, which was shown to result in a preparation success probability greater than 0.999 (see Appendix J of~\cite{Su2021Fault}). As for \(n_{\mathcal{M}}\), \(n_T\), and \(n_R\), they need to be chosen such that the allowable total RMS error in the estimation of the energy, \(\epsilon\), satisfies the following condition:
\begin{equation}
    \epsilon^2 \ge \epsilon_{\rm pha}^2 + (\epsilon_{\mathcal{M}} + \epsilon_R + \epsilon_T)^2,
\end{equation}
where 
\begin{align}
    \epsilon_{\rm pha} &> 0,  \\
    \label{eq:Mer}
    \epsilon_\mathcal{M} &= \frac{2\eta}{2^{n_\mathcal{M}}\pi\Omega^{1/3}} \left( \eta - 1 + 2 \lambda_\zeta \right) \left( 7 \times 2^{n_p + 1} - 9n_p - 11 - 3 \times 2^{-n_p} \right),  \\
    \label{eq:Rer}
    \epsilon_R &= \frac{\eta \lambda_\zeta}{2^{n_R} \Omega^{1/3}} \sum_{\nu \in G_0} \frac{1}{\| \nu \|},  \\
    \epsilon_T &= \frac{\pi \lambda}{2^{n_T}}.
\end{align}
where\(\lambda_\zeta = \sum_\ell \zeta_\ell \), \( \epsilon_{\rm pha} \) is the allowable RMS error in phase estimation, and \( \epsilon_{\mathcal{M}} \), \( \epsilon_R \), and \( \epsilon_T \) represent the allowable errors used to determine \( n_M \), \( n_R \), and \( n_T \), respectively. In our application, we aim for an accuracy of \(\epsilon = 1.6 \times 10^{-3}\), which is required for the calculations to be useful within the proposed computational workflow. We vary the precision parameters, and while they do influence the computed resource estimates, the changes are not substantial. All resulting estimates fall within the same order of magnitude. For simplicity, we set \(\epsilon_\mathcal{M} = \epsilon_T = \epsilon_R = 1 \times 10^{-4}\), and all our reported resource estimates are based on this choice. 

\begin{figure*}
    \centering
    \includegraphics[width=\textwidth]{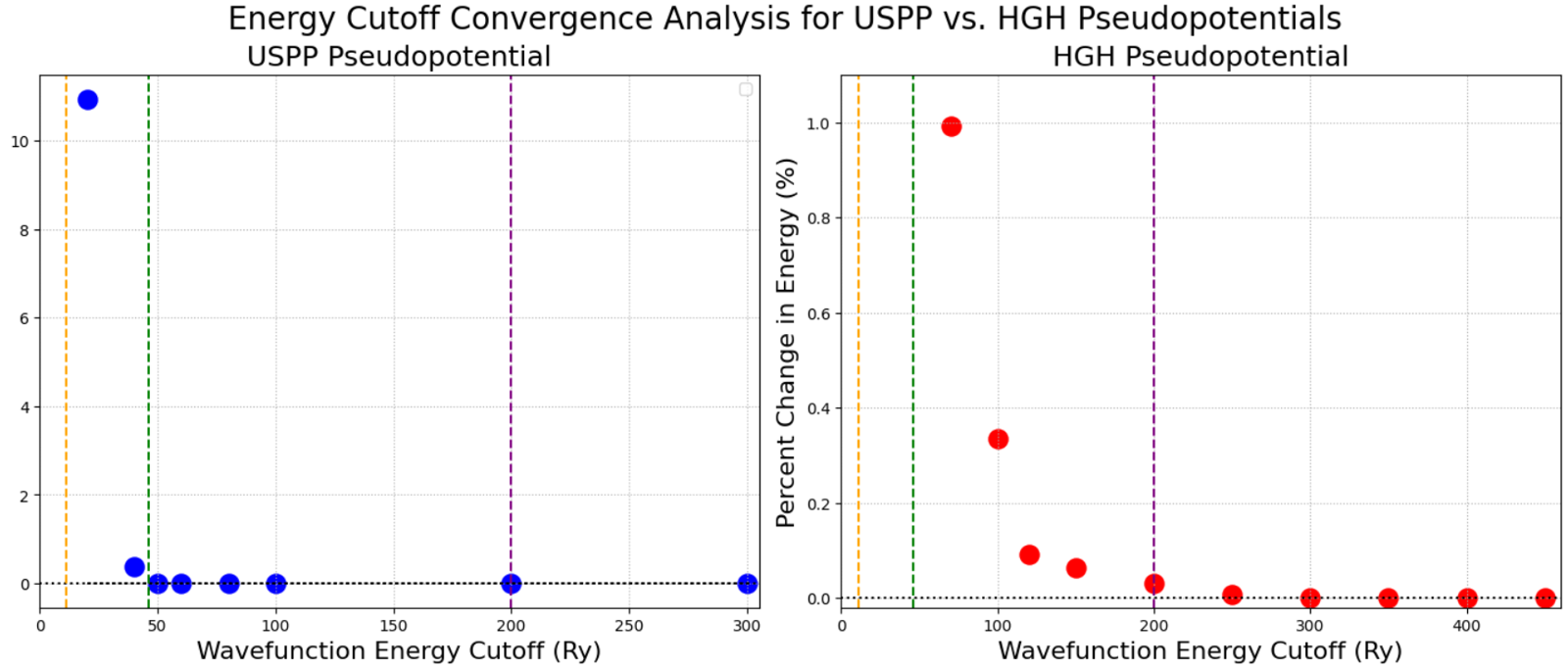}
    \caption{Energy cutoff convergence analysis in Rydberg (Ry) for the Vanderbilt ultrasoft pseudopotential (USPP) and Hartwigsen-Goedecker-Hutter (HGH) pseudopotential in the Dimer model. USPP achieves faster convergence compared to HGH. A similar convergence profile was observed for the monolayer model. Hence, a target energy cutoff value between 40 and 50 Ry is desired for these computational models. The dashed vertical lines indicate the plane-wave cutoffs where the grid resolution requires another bit. This is in contrast to the second quantization framework as discussed in Section \ref{supp_mat:second_quantization}, where additional plane waves incur additional qubits. Here, we have a range of energy cutoffs where the number of logical qubits remains the same, so one might prefer using the highest energy cutoff to achieve the best grid resolution before incurring another bit.}
    \label{fig:USPP_vs_HGH_Ecutoff}
\end{figure*}

\begin{figure*}
    \centering
    \includegraphics[width=0.75 \textwidth]{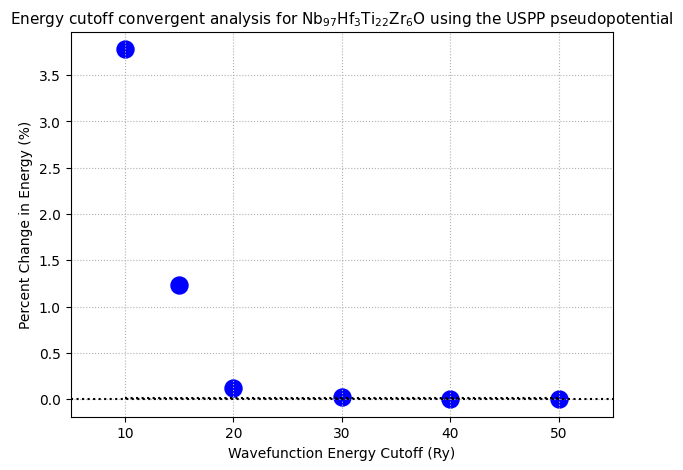}
    \caption{Energy cutoff convergence analysis in Rydberg (Ry) for the Vanderbilt ultrasoft pseudopotential (USPP) pseudopotential for  the representative computation model for Nb$_{97}$Hf$_3$Ti$_{22}$Zr$_6$O. We found that the USPP achieves faster convergence compared to HGH, similar to the previous case. A target energy cutoff value between 30 and 50 Ry is desired as indicated in the plot.}
    \label{fig: Ecut_analysis_NbAlloy1}
\end{figure*}

These resource estimates were reported at various energy cutoffs \(E_{cut}\) to give the reader a clear understanding of the computational resources required at different cutoff values. In practice, the choice of energy cutoff depends on several factors, including the chemical system of interest and the type of pseudopotential applied. In Figure \ref{fig:USPP_vs_HGH_Ecutoff}, we performed a convergence analysis for the \textit{Dimer Model} on two different pseudopotentials: the Vanderbilt ultrasoft (USPP) pseudopotential and the Hartwigsen-Goedecker-Hutter (HGH) pseudopotential. The results show that USPP converges at an energy cutoff between 40 Ry and 50 Ry, whereas HGH, a refined parameterization of the Goedecker-Teter-Hutter (GTH) potential, requires an energy cutoff of around 150 Ry to reach convergence. In Figure \ref{fig: Ecut_analysis_NbAlloy1}, a similar analysis was done for Nb$_{97}$Hf$_{3}$Ti$_{22}$Zr$_6$O, a representative computational model for niobium alloy as described in Section \ref{sec:Nb_workflow}.

Table \ref{tab:first_mg} presents the resource estimates for various representative computational models described in Section \ref{sec: Representative Computational Models for Mg} across different energy cutoffs (\(E_{\text{cut}}\)). Similarly, Table \ref{tab:first_nb} provides the resource estimates for the computational models outlined in Section \ref{sec: Representative Computational Models for Nb}. These estimates consider only the valence electrons, which is reasonable when using pseudopotentials. Pseudopotentials simplify the electron-nucleus interaction by replacing the Coulomb potential with a smoother potential, allowing for a more efficient representation of the electron wavefunction, particularly near the nucleus, without explicitly accounting for the complex behavior of core electrons. However, incorporating pseudopotentials requires modifications to the Hamiltonian, which in turn affects the circuit construction in the SELECT and PREPARE subroutines.~\cite{Berry2023QuantumSO} demonstrated a first-quantization implementation of the GTH and HGH pseudopotentials, along with instructions for constructing the corresponding quantum circuit. This implementation introduces additional computational costs, especially in the number of T gates required. In Tables \ref{tab:first_mg} and \ref{tab:first_nb}, the resource estimates do not incorporate the implementation of the pseudopotential Hamiltonian. Instead, we performed resource estimates using the original (non-pseudopotential) Hamiltonian, restricting the calculations to valence electrons. As a result, our resource estimates serve as a lower bound and should be considered "optimistic." Further optimizations in the pseudopotential implementation could bring resource requirements closer to this range.

We  also provide more conservative or "pessimistic" resource estimates, where we do not assume the use of pseudopotentials and include all electrons of the system. In this setup, the estimates will be more accurate but practically infeasible. In practice, the use of pseudopotentials is essential. Tables \ref{tab:first_mg_all_electrons} and [REF (still need to add)] show the resource estimates for the computational models in Sections \ref{sec: Representative Computational Models for Mg} and \ref{sec: Representative Computational Models for Nb}, respectively.

\begin{table}
\centering
\begin{ruledtabular}
    \begin{tabular}{l c c c c c c c}
        \makecell{\textbf{Cell size} \\ ({\AA})} & \makecell{$\boldsymbol{\eta}$ \\ (\# of e$^-$)} & \makecell{$\boldsymbol{E}_{\text{cut}}$\\ (Ry)} & \makecell{\textbf{Num. of PWs}} & \makecell{$\boldsymbol{N_{\text{Log}}}$} & \makecell{\textbf{QPE $T$ count}\\ ($\times 10^{13}$)} & \makecell{\textbf{Clifford Count}\\ ($\times 10^{13}$)} & \makecell{\textbf{1-Norm}} \\
        \hline
        \multicolumn{8}{c}{{Magnesium dimer model}} \\[5pt]
        \makecell{[12.7, 12.7, 19.9]} & \makecell{144} & \makecell{$[5, 10]$\\$[11-46]$\\$[47,200)$} & \makecell{1134 - 3211\\3380 - 27716\\ 33712 - 247860} & \makecell{1851\\2292\\2732} & \makecell{0.21\\1.04\\2.48} & \makecell{0.30\\1.42\\3.33} & \makecell{1.26 $\times 10^4$\\3.08 $\times 10^4$\\ 8.23 $\times 10^4$} \\[15pt]
        
        \multicolumn{8}{c}{{Magnesium monolayer model}} \\[5pt]
        \makecell{[19.8, 19.8, 32.3]} & \makecell{567} & \makecell{$[5, 19]$\\$[20, 79]$\\$[80, 320]$} & \makecell{4312-29068\\32076-244383\\253692-1992126} & \makecell{8648\\10358\\12067} & \makecell{14.34\\68.21\\158.10} & \makecell{20.9\\98.4\\226} & \makecell{2.20 $\times 10^5$\\4.78 $\times 10^5$\\1.09 $\times 10^6$} \\[15pt]
        
        \multicolumn{8}{c}{{Magnesium cluster model}} \\[5pt]
        \makecell{[19.8, 19.8, 32.3]} & \makecell{1477} & \makecell{$[5, 19]$\\$[20, 79]$\\$[80, 320]$} & \makecell{4312-29068\\32076-244383\\253692-1992126} & \makecell{22307\\26746\\31185} & \makecell{288\\6865\\1591} & \makecell{288\\6865\\1591} & \makecell{1.43 $\times 10^6$\\2.98 $\times 10^6$\\6.33 $\times 10^6$} \\[15pt]
        
        \multicolumn{8}{c}{Single-unit $\textrm{Mg}_{17}\textrm{Al}_{12}$ secondary phase structure} \\[5pt]
        \makecell{[29.7, 9.2, 42.6]} & \makecell{614} & \makecell{$[5, 19]$\\$[20, 83]$\\$[84, 340]$} & \makecell{4060-28392\\30160-249444\\251576-2033460} & \makecell{9354\\11204\\14021} & \makecell{30.9\\73.5\\182} & \makecell{45.1\\106\\244} & \makecell{2.64 $\times 10^5$\\5.72 $\times 10^5$\\1.30 $\times 10^6$} \\[15pt]
        
        \multicolumn{8}{c}{Double-unit $\textrm{Mg}_{17}\textrm{Al}_{12}$ secondary phase structure} \\[5pt]
        \makecell{[30.1, 9.3, 43]} & \makecell{660} & \makecell{$[5, 19]$\\$[20, 81]$\\$[81, 335]$} & \makecell{4263-29640\\30914-249444\\251576-2045784} & \makecell{10045\\12033\\17139} & \makecell{33.1\\78.8\\199} & \makecell{48.4\\113\\262} & \makecell{3.00 $\times 10^5$\\6.48 $\times 10^5$\\1.46 $\times 10^6$}\\[15pt]
        
        \multicolumn{8}{c}{Quadruple-unit $\textrm{Mg}_{17}\textrm{Al}_{12}$ secondary phase structure} \\[5pt]
        \makecell{[30.0, 9.3, 42.9]} & \makecell{744} & \makecell{$[5,19]$\\$[20,81]$\\$[82,335]$} & \makecell{4263-29640\\30914-249444\\251576-2045784} & \makecell{11305\\13545\\15785} & \makecell{37.1\\88.3\\205} & \makecell{54.3\\128\\295} & \makecell{3.79 $\times 10^5$\\8.13 $\times 10^5$\\1.82 $\times 10^6$}  \\[15pt]
        
        \multicolumn{8}{c}{$\textrm{Mg}_{17}\textrm{Al}_{12}$ secondary-phase supercell} \\[5pt]
        \makecell{[33.0, 31.3, 50.3]} & \makecell{1830} & \makecell{$[8,30]$\\$[31,126]$\\$[127,510]$} & \makecell{33669-237380\\252280-2018240\\2049264-16390350} & \makecell{33100\\38598\\44096} & \makecell{846\\1962\\4464} & \makecell{1244\\2867\\6492} & \makecell{2.78 $\times 10^6$\\5.76 $\times 10^6$\\1.22 $\times 10^7$} \\
    \end{tabular}
\end{ruledtabular}
\caption{Resource estimate summary for representative computational models related to the \textit{Design of Mg-Rich Sacrificial Coatings and Corrosion-Resistant Mg Alloys for Aqueous Environments}, as discussed in Section \ref{sec: Mg corrosion}. Here, \(\eta\) represents the number of electrons, considering only valence electrons in this scenario, which is the case when a pseudopotential is imposed in the calculation. We estimate the resources at various energy cutoff values to provide a more comprehensive picture of the required resources. In practice, the energy cutoff depends on several factors, including the choice of pseudopotential and convergence analysis, as shown in Figure \ref{fig:USPP_vs_HGH_Ecutoff}. Additionally, we report the number of plane waves (Num. of PWs), the number of logical qubits (\(N_{\text{Log}}\)), the T-gate count, the Clifford gate count, and the 1-norm (\(\lambda\)), each corresponding to a given energy cutoff value.
}
\label{tab:first_mg}
\end{table}

\begin{table}
\centering
\begin{ruledtabular}
    \begin{tabular}{l c c c c c c c}
        \makecell{\textbf{Cell size} \\ ({\AA})} & \makecell{$\boldsymbol{\eta}$ \\ (\# of e$^-$)} & \makecell{$\boldsymbol{E}_{\text{cut}}$\\ (Ry)} & \makecell{\textbf{Num. of PWs}} & \makecell{$\boldsymbol{N_{\text{Log}}}$} & \makecell{\textbf{QPE $T$ count}\\ ($\times 10^{13}$)} & \makecell{\textbf{Clifford Count}\\ ($\times 10^{13}$)} & \makecell{\textbf{1-Norm}} \\
        \hline
        \multicolumn{8}{c}{Nb$_{97}$Hf$_3$Ti$_{22}$Zr$_6$O}\\[5pt]
        \makecell{[13.3, 13.3, 13.3]} & \makecell{615} & \makecell{5-14\\15-62\\ 63-259} & \makecell{729-3375\\4096-29791\\32768-250047} & \makecell{7516\\9370\\11223} & \makecell{12.6\\62.1\\148} & \makecell{18.6\\90.4\\213} & \makecell{2.22 $\times 10^5$\\ 4.86 $\times 10^5$\\ 1.10 $\times 10^6$} \\[15pt]
        \multicolumn{8}{c}{Nb$_{97}$Ta$_{22}$Zr$_3$W$_6$O}\\[5pt]
        \makecell{[13.3, 13.3, 13.3]} & \makecell{649} & \makecell{5-14\\15-62\\ 63-259} & \makecell{729-3375\\4096-29791\\32768-250047} & \makecell{7925\\9880\\11835} & \makecell{26.5\\65.3\\155} & \makecell{39.2\\95.3\\224} & \makecell{2.46 $\times 10^5$\\ 5.38 $\times 10^5$\\ 1.21 $\times 10^6$} \\[15pt]
        \multicolumn{8}{c}{Nb$_{42}$Ti$_{3}$Hf$_{3}$Ta$_{3}$Zr$_3$O}\\[5pt]
        \makecell{[10.0, 10.0, 10.1]} & \makecell{267} & \makecell{6-24\\25-106\\107-438} & \makecell{512-3375\\4096-29791\\32768-250047} & \makecell{3333\\4142\\4951} & \makecell{2.91\\7.17\\17.0} & \makecell{2.10\\10.2\\23.8} & \makecell{6.11$\times 10^5$\\ 1.46 $\times 10^5$\\ 3.76 $\times 10^5$} \\[15pt]
        \multicolumn{8}{c}{Nb$_{65}$Zr$_6$Hf$_7$Ti$_4$W$_3$}\\[5pt]
        \makecell{[8.6, 9.8, 14.0]} & \makecell{411} & \makecell{5-21\\22-93\\93-388} & \makecell{420-3360\\3640-29725\\30914-249747} & \makecell{5064\\6305\\7546} & \makecell{8.63\\21.3\\50.5} & \makecell{12.6\\30.6\\72} & \makecell{1.28 $\times 10^5$\\ 2.92 $\times 10^5$\\ 7.04 $\times 10^5$} 
    \end{tabular}
\end{ruledtabular}
\caption{Resource estimate summary for representative computational models related to \textit{Designing High-Temperature Corrosion-Resistant Alloys}, as discussed in Section \ref{sec:Nb_workflow}. Here, $\eta$ represents the number of valence electrons, similar to Table \ref{tab:first_mg}. To provide a comprehensive picture of the required resources, we estimate them at various energy cutoff values.}
\label{tab:first_nb}
\end{table}

\begin{table}
\centering
\begin{ruledtabular}
    \begin{tabular}{l c c c c c c c}
        \makecell{\textbf{Cell size} \\ ({\AA})} & \makecell{$\boldsymbol{\eta}$ \\ (\# of e$^-$)} & \makecell{$\boldsymbol{E}_{\text{cut}}$\\ (Ry)} & \makecell{\textbf{Num. of PWs}} & \makecell{$\boldsymbol{N_{\text{Log}}}$} & \makecell{\textbf{QPE $T$ count}\\ ($\times 10^{15}$)} & \makecell{\textbf{Clifford Count}\\ ($\times 10^{15}$)} & \makecell{\textbf{1-Norm}} \\
        \hline
        \multicolumn{8}{c}{{Magnesium dimer model}} \\[5pt]
        \makecell{[12.7, 12.7, 19.9]} & \makecell{788} & \makecell{$[5, 10]$\\$[11-46]$\\$[47,199]$} & \makecell{1134 - 3211\\3380 - 27716\\ 33712 - 247860} & \makecell{9593\\11965\\14337} & \makecell{0.32\\0.79\\1.87} & \makecell{0.47\\1.15\\2.26} & \makecell{3.15 $\times 10^5$\\6.78 $\times 10^6$\\ 3.43 $\times 10^6$} \\[15pt]
        
        \multicolumn{8}{c}{{Magnesium monolayer model}} \\[5pt]
        \makecell{[19.8, 19.8, 32.3]} & \makecell{2419} & \makecell{$[5, 19]$\\$[20, 79]$\\$[80, 320]$} & \makecell{4312-29068\\32076-244383\\253692-1992126} & \makecell{36441\\43706\\50971} & \makecell{18.75\\44.6\\103} & \makecell{27.8\\65.7\\151} & \makecell{3.78 $\times 10^6$\\7.82 $\times 10^6$\\1.63 $\times 10^7$} \\[15pt]
        
        \multicolumn{8}{c}{{Magnesium cluster model}} \\[5pt]
        \makecell{[19.8, 19.8, 32.3]} & \makecell{3907} & \makecell{$[5, 19]$\\$[20, 79]$\\$[80, 320]$} & \makecell{4312-29068\\32076-244383\\253692-1992126} & \makecell{58765\\70494\\82223} & \makecell{60.2\\143\\331} & \makecell{89.6\\211\\487} & \makecell{9.80 $\times 10^6$\\2.01 $\times 10^7$\\4.15 $\times 10^7$} \\[15pt]
        
        \multicolumn{8}{c}{Single-unit $\textrm{Mg}_{17}\textrm{Al}_{12}$ secondary phase structure} \\[5pt]
        \makecell{[29.7, 9.2, 42.6]} & \makecell{3564} & \makecell{$[5, 19]$\\$[20, 83]$\\$[84, 340]$} & \makecell{4060-28392\\30160-249444\\251576-2033460} & \makecell{53618\\64319\\75019} & \makecell{54.9\\130\\303} & \makecell{81.8\\193\\444} & \makecell{8.40 $\times 10^6$\\1.73 $\times 10^7$\\3.57 $\times 10^6$}\\[15pt]
        
        \multicolumn{8}{c}{Double-unit $\textrm{Mg}_{17}\textrm{Al}_{12}$ secondary phase structure} \\[5pt]
        \makecell{[30.1, 9.3, 43]} & \makecell{3720} & \makecell{$[5, 19]$\\$[20, 81]$\\$[81, 335]$} & \makecell{4263-29640\\30914-249444\\251576-2045784} & \makecell{55959\\67127\\78295} & \makecell{57.4\\136\\316} & \makecell{85.3\\201\\464} & \makecell{9.04 $\times 10^6$\\1.86 $\times 10^7$\\3.83 $\times 10^6$}\\[15pt]
        
        \multicolumn{8}{c}{Quadruple-unit $\textrm{Mg}_{17}\textrm{Al}_{12}$ secondary phase structure} \\[5pt]
        \makecell{[30.0, 9.3, 42.9]} & \makecell{3984} & \makecell{$[5,19]$\\$[20,81]$\\$[82,335]$} & \makecell{4263-29640\\30914-249444\\251576-2045784} & \makecell{59919\\71879\\83839} & \makecell{61.4\\145\\338} & \makecell{91.3\\215\\497} & \makecell{1.07 $\times 10^7$\\2.13 $\times 10^7$\\4.39 $\times 10^7$}  \\[15pt]
        
        \multicolumn{8}{c}{$\textrm{Mg}_{17}\textrm{Al}_{12}$ secondary-phase supercell} \\[5pt]
        \makecell{[33.0, 31.3, 50.3]} & \makecell{10620} & \makecell{$[8,30]$\\$[31,126]$\\$[127,510]$} & \makecell{33669-237380\\252280-2018240\\2049264-16390350} & \makecell{191336\\223204\\255072} & \makecell{1547\\3586\\8158} & \makecell{2292\\5826\\11975} & \makecell{9.14 $\times 10^6$\\1.85 $\times 10^8$\\3.75 $\times 10^8$} \\
    \end{tabular}
\end{ruledtabular}
\caption{This table presents the "pessimistic" (worst-case scenario) resource estimates from Table \ref{tab:first_mg}. These estimates correspond to the computational models outlined in \textit{Design of Mg-Rich Sacrificial Coatings and Corrosion-Resistant Mg Alloys for Aqueous Environments}, as discussed in Section \ref{sec: Mg corrosion}. In contrast to Table \ref{tab:first_mg}, $\eta$ here represents the total number of electrons in the system, rather than just the valence electrons. As in the previous case, we estimate resources for various energy cutoff values, providing a comprehensive assessment. The table reports the number of plane waves (Num. of PWs), the number of logical qubits ($N_{\text{Log}}$), T-gate count, Clifford gate count, and the 1-norm ($\lambda$), each corresponding to a given energy cutoff value.}
\label{tab:first_mg_all_electrons}
\end{table}

\begin{table}
\centering
\begin{ruledtabular}
    \begin{tabular}{l c c c c c c c}
        \makecell{\textbf{Cell size} \\ ({\AA})} & \makecell{$\boldsymbol{\eta}$ \\ (\# of e$^-$)} & \makecell{$\boldsymbol{E}_{\text{cut}}$\\ (Ry)} & \makecell{\textbf{Num. of PWs}} & \makecell{$\boldsymbol{N_{\text{Log}}}$} & \makecell{\textbf{QPE $T$ count}\\ ($\times 10^{15}$)} & \makecell{\textbf{Clifford Count}\\ ($\times 10^{15}$)} & \makecell{\textbf{1-Norm}} \\
        \hline
        \multicolumn{8}{c}{Nb$_{97}$Hf$_3$Ti$_{22}$Zr$_6$O}\\[5pt]
        \makecell{[13.3, 13.3, 13.3]} & \makecell{4925} & \makecell{5-14\\15-62\\ 63-259} & \makecell{729-3375\\4096-29791\\32768-250047} & \makecell{59254\\74037\\88820} & \makecell{61.4\\151\\360} & \makecell{92.5\\225\\532} & \makecell{1.35 $\times 10^5$\\ 2.81 $\times 10^7$\\ 5.79 $\times 10^7$} \\[15pt]
        
        \multicolumn{8}{c}{Nb$_{97}$Ta$_{22}$Zr$_3$W$_6$O}\\[5pt]
        \makecell{[13.3, 13.3, 13.3]} & \makecell{6155} & \makecell{5-14\\15-62\\ 63-259} & \makecell{729-3375\\4096-29791\\32768-250047} & \makecell{74016\\92489\\110962} & \makecell{153\\378\\899} & \makecell{232\\563\\1330} & \makecell{2.10 $\times 10^7$\\ 4.37 $\times 10^7$\\ 8.99 $\times 10^7$}  \\[15pt]
        
        \multicolumn{8}{c}{Nb$_{42}$Ti$_{3}$Hf$_{3}$Ta$_{3}$Zr$_3$O}\\[5pt]
        \makecell{[10.0, 10.0, 10.1]} & \makecell{2351} & \makecell{6-24\\25-106\\107-438} & \makecell{512-3375\\4096-29791\\32768-250047} & \makecell{28359\\35420\\42481} & \makecell{14.8\\36.5\\86.7} & \makecell{22.2\\54.1\\127} & \makecell{3.83 $\times 10^6$\\ 8.06 $\times 10^6$\\ 1.70 $\times 10^7$}  \\[15pt]
        
        \multicolumn{8}{c}{Nb$_{65}$Zr$_6$Hf$_7$Ti$_4$W$_3$}\\[5pt]
        \makecell{[8.6, 9.8, 14.0]} & \makecell{3719} & \makecell{5-21\\22-93\\93-388} & \makecell{420-3360\\3640-29725\\30914-249747} & \makecell{44779\\55944\\67109} & \makecell{46.5\\114\\272} & \makecell{69.9\\170\\402} & \makecell{9.50 $\times 10^6$\\ 1.99 $\times 10^7$\\ 4.14 $\times 10^7$} 
    \end{tabular}
\end{ruledtabular}
\caption{This table presents the "pessimistic" (worst-case scenario) resource estimates from Table \ref{tab:first_nb}. These estimates correspond to the computational models outlined in \textit{Designing High-Temperature Corrosion-Resistant Alloys}, as discussed in Section \ref{sec:Nb_workflow}. In contrast to Table \ref{tab:first_nb}, $\eta$ here represents the total number of electrons in the system, rather than just the valence electrons. As in the previous case, we estimate resources for various energy cutoff values, providing a comprehensive assessment. The table reports the number of plane waves (Num. of PWs), the number of logical qubits ($N_{\text{Log}}$), T-gate count, Clifford gate count, and the 1-norm ($\lambda$), each corresponding to a given energy cutoff value.}
\label{tab:first_nb_all_electrons}
\end{table}

\section{Second Quantization: Remarks and Resource Estimates}
\label{supp_mat:second_quantization}

While the workflow is ansatz independent, the resource estimates will be heavily influenced by the Hamiltonian picture and the fermionic embedding. While we have presented in the main text our resource estimates for first quantization, we also performed resource estimates for the same GSEE problems using second quantization.

\subsection{Hamiltonian Formulation}

\begin{figure*}
    \centering
    \includegraphics[scale=0.31]{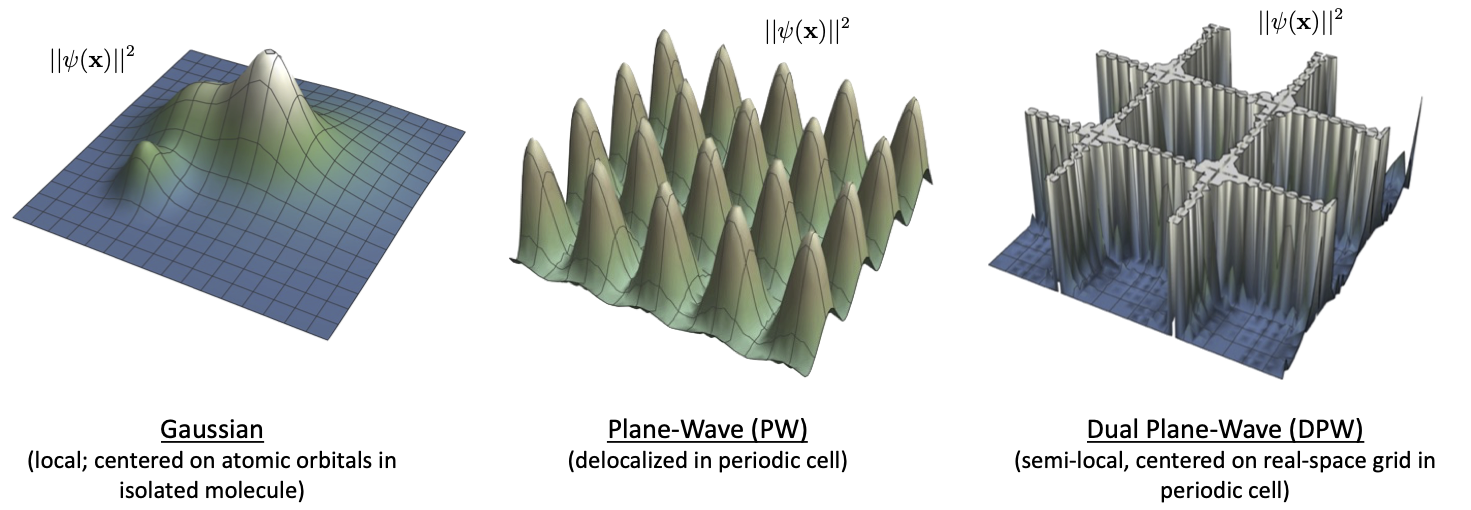}
    \caption{Comparison between Gaussian local basis sets and plane-wave (PW) and dual plane-wave (DPW).  }
    \label{fig:wavefunctions}
\end{figure*}

A central challenge in electronic-structure calculations is choosing a finite basis set to discretize the Hamiltonian. Under the Born--Oppenheimer approximation, wavefunction ``cusps'' occur at nuclei, motivating local Gaussian orbitals for molecular simulations. However, when modeling large periodic systems, nonlocal plane-wave bases are often more suitable than local Gaussian orbitals. Since plane waves lack information about the nuclei positions, achieving the same level of basis set accuracy as Gaussian orbitals requires a greater number of plane waves. Under the plane-wave basis, the second quantized electronic structure Hamiltonian can be written as 
\begin{equation}
    \label{eq: plane-wave 2nd quant Ham}
    H = \frac{1}{2} \sum_{p,q,\sigma} k_p^2 a_{p, \sigma}^\dagger a_{p, \sigma} - 
   \frac{4\pi}{\Omega} \sum_{\substack{p\neq q,j\sigma }}  \Big(\zeta_j \, \frac{e^{\imath k_{p-q} \cdot R_j}}{k_{p-q}^2} \Big)\, a^\dagger_{p,\sigma} a_{q,\sigma} 
   + \frac{2\pi}{\Omega} \sum_{\substack{(p,\sigma) \neq (q,\sigma') \\ \nu \neq 0}} \Big(\frac{1}{k_\nu^2} \Big)\,\, a^\dagger_{p,\sigma} a^\dagger_{q,\sigma} a_{q+\nu,\sigma'} a_{p-\nu,\sigma}
\end{equation}
where $a^\dagger_{p,\sigma}$ and $a_{p,\sigma}$ are fermionic creation and annihilation operators on spatial orbital $p$ with spin $\sigma \in \{\uparrow,\downarrow\}$.
Here, we employ the dual plane-wave (DPW) basis, which drastically decreases the number of terms in the electronic-structure Hamiltonian \cite{Babbush2018LowDepth}. Figure~\ref{fig:wavefunctions} contrasts local Gaussian, plane-wave, and DPW functions. This intermediate locality facilitates efficient Hamiltonian factorization and supports fault-tolerant ground-state estimation algorithms with linear \emph{T}-complexity. The plane-wave Hamiltonian (Equation \ref{eq: plane-wave 2nd quant Ham}) can be mapped to the DPW basis Hamiltonian through a discrete Fourier transform. In the DPW basis, the Hamiltonian can be written as:
\begin{equation}
    \label{eq: dual plane-wave 2nd quant Ham}
    \begin{split}
        H = &\sum_{p,q,\sigma} T(p-q) a^\dagger_{p,\sigma}a_{q,\sigma} + \sum_{p,\sigma} U(p) n_{p,\sigma} + \sum_{(p,\alpha)\neq(q,\beta)} V(p-q)n_{p,\alpha}n_{q,\beta}
    \end{split}
\end{equation}
where $n_{p,\sigma} = a^\dagger_{p,\sigma}a_{p,\sigma}$ is the number operator. The coefficients $T, U, V$ can be represented as: 
\begin{align*}
T(p-q) &= \frac{1}{2N_{\text{so}}} \sum_\nu k_\nu^2 \, \cos \big[ k_\nu \cdot r_{p-q} \big]\\
U(p) &= -\frac{4\pi}{\Omega} \sum_{j,\nu \neq 0 }  \frac{\zeta_j \, \cos \big[ k_\nu \cdot (R_j - r_p)\big]}{k_\nu^2}\\
V(p-q) &= \frac{2\pi}{\Omega} \sum_{\nu \neq 0} \frac{\cos \big[k_\nu \cdot r_{p-q} \big]}{k_\nu^2},
\end{align*}
where each spatial orbital $p$ is associated with an orbital centroid $r_p = p(2\Omega/N_{\text{so}})^{1/3}$, $\Omega$ is the computational cell volume, $R_j$ is the position of atom $j$ with atomic number $\zeta_j$, and $N_{\text{so}}$ is the number of spin-orbitals, which is twice the number of basis functions (DPW or PW). The momentum modes are defined as $k_\nu = 2\pi\nu / \Omega^{1/3}$ with $\nu \in [-(N_{\text{so}}/2)^{1/3}, (N_{\text{so}}/2)^{1/3}]^{\otimes 3}$~\cite{babbush2018encoding}. In this dual plane-wave basis, the discretization error is asymptotically equivalent to a Galerkin discretization using any other single-particle basis functions, including Gaussian orbitals~\cite{Babbush2018LowDepth}.

The second quantized  electronic Hamiltonian under DPW basis can be represented in the Pauli matrices basis under the Jordan-Wigner transformation as~\cite{Babbush2018LowDepth, babbush2018encoding}:
\begin{equation}
    \label{eq: dual plane-wave 2nd quant Ham in Pauli Rep.}
    \begin{split}
        H =& \sum_{p\neq q,\sigma} \frac{T(p-q)}{2} \left( X_{p,\sigma} \vec{Z} X_{q,\sigma} + Y_{p,\sigma} \vec{Z} Y_{q,\sigma}\right)\\
        &+ \sum_{(p,\alpha)\neq(q,\beta)} \frac{V(p-q)}{4} Z_{p,\alpha}Z_{q,\beta} \\
    &- \sum_{p,\sigma} \left( \frac{T(0)+U(p)+\sum_q V(p-q)}{2} \right) Z_{p,\sigma} \\
    &+ \sum_{p} \left( T(0)+U(p)+\sum_q \frac{V(p-q)}{2} \right)
    \end{split}
\end{equation}
where we adopt the notation $A_j \vec{Z} A_k$ introduced in Ref.~\cite{babbush2018encoding} to represent the operator $A_j Z_{j+1}\dots Z_{k-1} A_k$. Note, the index $p$ is a $d$-dimensional vector with components ${p_i \in [0,M_i-1]}$, such that the total number of spin-orbitals is ${N_{\text{so}}=2\prod_{i=0}^{d-1} M_i}$. This vector index is mapped onto an integer using the mapping function
\[
    f(p, \sigma) = \delta_{\sigma,\downarrow}\prod_{j=0}^{d-1}M_j+\sum_{i=0}^{d-1}p_i \prod_{j=0}^{i-1}M_j.
\]

\subsection{Block Encoding Circuit}
\label{supp_mat:second_quantization circuits implementation}

The second quantized Hamiltonian in the DPW basis as written in Eq.~\ref{eq: dual plane-wave 2nd quant Ham in Pauli Rep.} can be encoded using the LCU framework described above. In this representation, we use the SELECT and PREPARE oracles from Section IV of~\cite{babbush2018encoding}, where the prepared coefficients correspond to those of the Hamiltonian after the Jordan-Wigner transformation, i.e.,
\begin{align}
&\text{PREPARE} |0\rangle^{\otimes 3+2\log_2N} \rightarrow \sum_{p,\sigma} \tilde{U}(p) |\theta_p\rangle |1\rangle_U |0\rangle_V |p, \sigma, p, \sigma\rangle|\text{temp}_U\rangle  \nonumber\\
        &+ \sum_{p\neq q, \sigma} \tilde{T}(p-q) |\theta_{p-q}^{(0)}\rangle |0\rangle_U |0\rangle_V |p, \sigma, q, \sigma\rangle|\text{temp}_T\rangle 
        + \sum_{(p,\alpha)\neq(q,\beta)} \tilde{V}(p-q) |\theta_{p-q}^{(1)}\rangle |0\rangle_U |1\rangle_V  |p,\alpha,q,\beta\rangle|\text{temp}_V\rangle
\end{align}
where $|\text{temp}_i\rangle$ represents the state of ancilla qubits used for the preparation which are not used by the SELECT oracle and the coefficients are defined as
\begin{align}
&\tilde{U}(p) = \sqrt{\frac{\left|T(0) + U(p) +\sum_q V(p-q)\right|}{2\lambda}}, &\tilde{T}(p) = \sqrt{\frac{\left|T(p)\right|}{\lambda}}, &&\tilde{V}(p) = \sqrt{\frac{\left|V(p)\right|}{4\lambda}}\\
& \theta_p = \frac{1-\text{sign}\left(-T(0)-U(p)-\sum_q V(p-q)\right)}{2}, &  \theta_p^{(0)} = \frac{1-\text{sign}\left(T(p)\right)}{2}, && \theta_p^{(1)} = \frac{1-\text{sign}\left(V(p)\right)}{2},
\end{align}
with the 1-norm given by
\[
\lambda = \sum_{pq} |T(p-q)| + \sum_p |U(p)| + \sum_{p\neq q} |V(p-q)|.
\]
Note, here the selection register, denoted $\ket{\ell}$ in Fig.~\ref{fig: Walk}, is composed of several registers labeled $\ket{\theta}, \ket{U}, \ket{V}, \ket{p}, \ket{\alpha}, \ket{q},$ and $\ket{\beta}$. The $\ket{\theta}$ register is used to store information about the sign of the coefficient, the $\ket{U}$ and $\ket{V}$ registers are used to select between the $T$, $U$ and $V$ Hamiltonian coefficients, and the $\ket{p}, \ket{\alpha}, \ket{q},$ and $\ket{\beta}$ registers are used to index the Hamiltonian term acting on spin-orbitals $(p,\alpha)$ and $(q,\beta)$, where $\alpha$ and $\beta$ specify the spins $\{\uparrow,\downarrow\}$. Due to entanglement, the registers used during the preparation are not all perfectly uncomputed by a ($\text{PREPARE}^{-1} \cdot \text{SELECT} \cdot \text{PREPARE}$) series and so must be kept and passed to later iterates. 

The circuit construction for this PREPARE oracle uses a modified alias sampling technique that takes advantage of symmetries in the Hamiltonian to minimize the amount of data that's loaded. This results in the selection register prepared in a superposition state weighted by approximate Hamiltonian coefficients, where the approximation is parameterized by the error $\epsilon_{\text{prep}}$. The SELECT oracle then makes use of the unary iteration scheme to selectively apply strings of Pauli operators to the system register based on the state prepared on the selection register. The preparation approximation error can be related to the phase estimation energy error $\Delta E$ via the following set of equations~\cite{babbush2018encoding}:
\begin{align}
    \epsilon_{\text{prep}} &\leq \frac{1}{2^\mu L} \label{eq:ep_prep} \\
    \mu &= \left\lceil \log \left(\frac{2\sqrt{2}\lambda}{\Delta E}\right) + \log\left(1+\frac{\Delta E^2}{8\lambda^2}\right) - \log\left(1-\frac{\norm{H}^2}{\lambda^2}\right) \right\rceil \label{eq:mu}
\end{align}
For the estimates reported in this paper, we set $\Delta E = 0.001$ and use the above equations to calculate $\epsilon_{\text{prep}}$.

\subsection{Resource Estimates}
\label{sec: supp_resource_estimates_2nd_quantization}
This section reports quantum resource estimates for the computational models presented in \ref{sec: Representative Computational Models for Mg} and \ref{sec: Representative Computational Models for Nb} using the qubitized quantum phase estimation algorithm within the second quantization framework, with the circuit implementation detailed in Section \ref{supp_mat:second_quantization circuits implementation}. Again, resource estimates are provided in terms of the number of logical qubits and the number of $T$ gates.

\begin{figure}
    \centering
    \includegraphics[width=.65\textwidth]{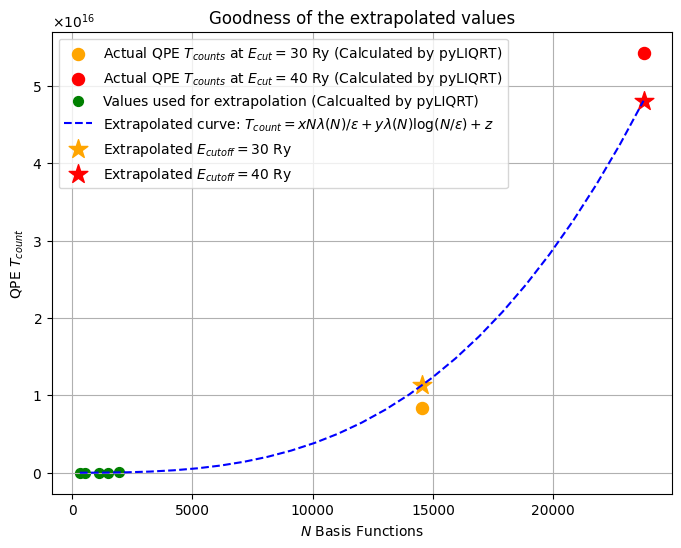}
    \caption{Reliability of extrapolated results obtained from the Dimer model, where exact resource estimates are feasible at high $E_{\text{cut}}$ thresholds. Extrapolation based on resource estimate values obtained at select low $E_{\text{cut}}$ threshold values, employing equations \ref{eq:extrapolated_equations}. Extrapolated number of $T$ gates required for QPE at energy cutoffs $E_{\text{cut}} = 30, 40$ Ry is compared with the actual computed number of $T$ gates using pyLIQRT under the same energy cutoff values.}
    \label{fig: goodness_of_fit}
\end{figure}

\begin{figure}
    \centering

    \includegraphics[width = \textwidth]{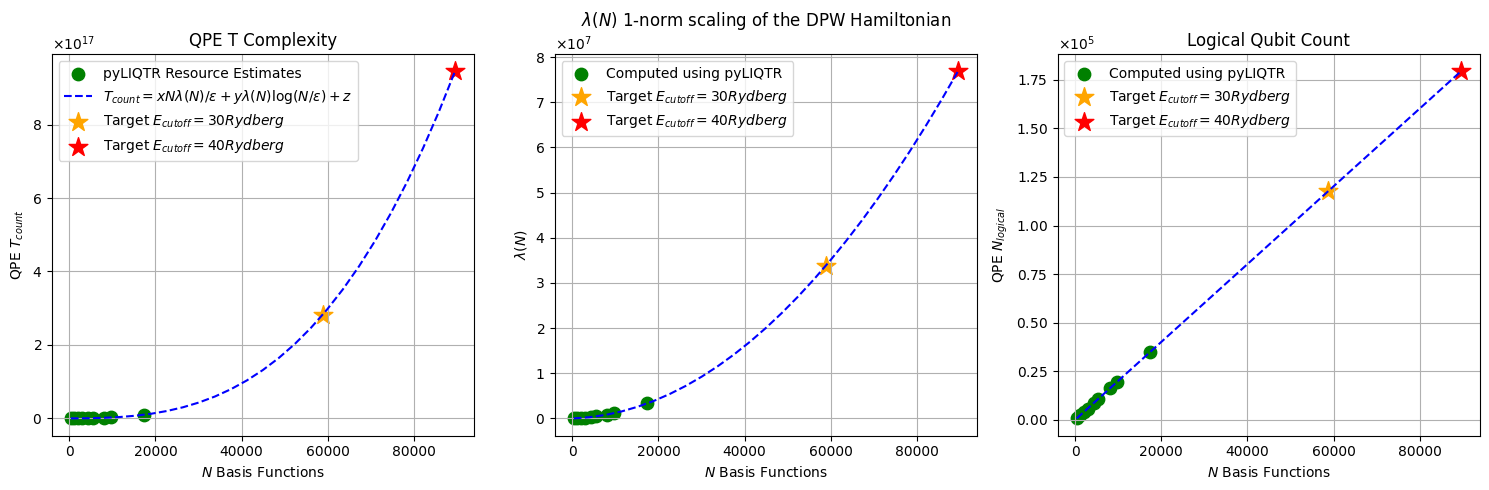}
\\
    \includegraphics[width = \textwidth]{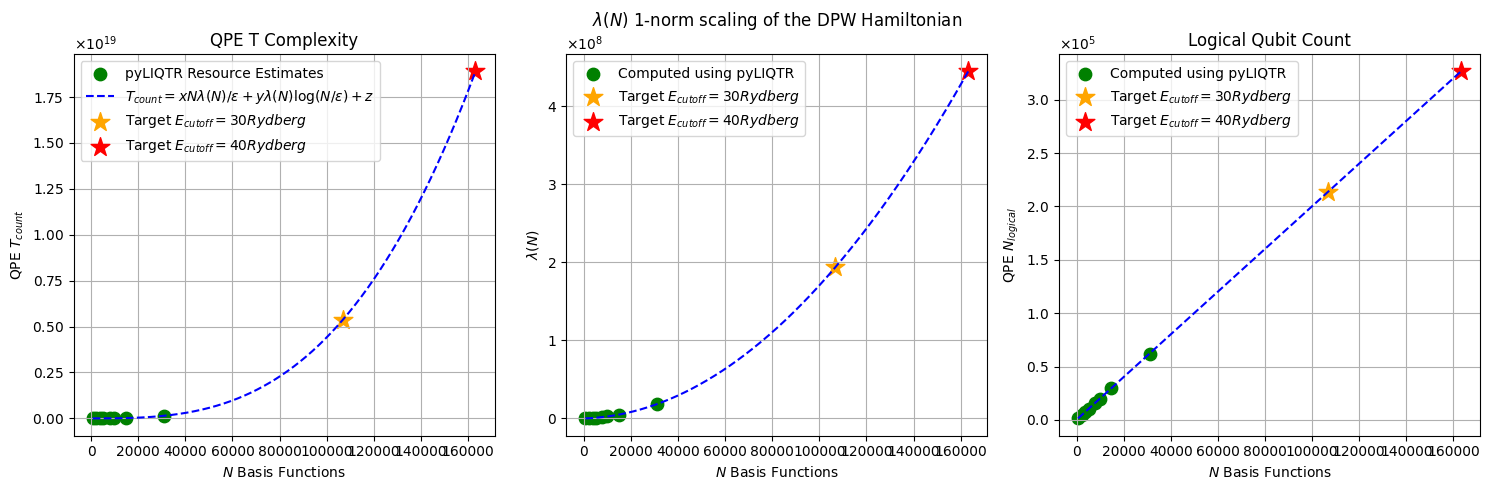}
\\
    \includegraphics[width = \textwidth]{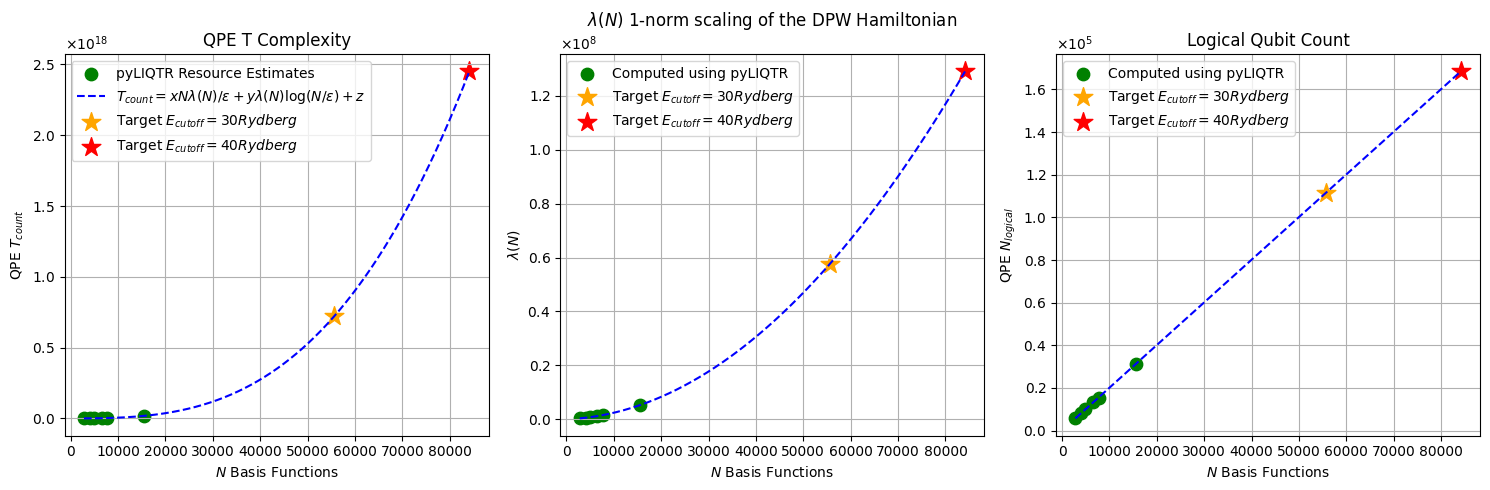}  
    \caption{Resource estimate results for the Monolayer model, Cluster model, and the single-unit Mg$_{17}$Al$_{12}$ structure. The $x$-axis represents the number of basis functions, which is directly related to the energy cutoff, $E_{\text{cut}}$, threshold. All green data points correspond to explicit resource estimate values computed using pyLIQTR. These data points were used to extrapolate to desired $E_{\text{cut}}$ values at higher thresholds, particularly at $E_{\text{cut}} = 30$ and $40$ Ry. (Top row) Monolayer model (see Figure \ref{fig: PureMgModels} (middle)), comprising 257 Mg atoms and 26 water molecules, embedded in a simulation cell of $19.8 \times 19.8 \times 32.3$ {\AA}. (Middle row) Cluster model (see Figure \ref{fig: PureMgModels} (right)), composed of 587 Mg atoms and 100 water molecules, embedded in a simulation cell of $19.8 \times 19.8 \times 58.9$ {\AA}. (Bottom row) Single-unit Mg$_{17}$Al$_{12}$ secondary phase structure embedded in a simulation cell of $29.7 \times 9.2 \times 42.6$ {\AA}.  }
    \label{fig: extrapolated_result_1}
\end{figure}

\begin{figure}
    \centering
    \includegraphics[width = \textwidth]{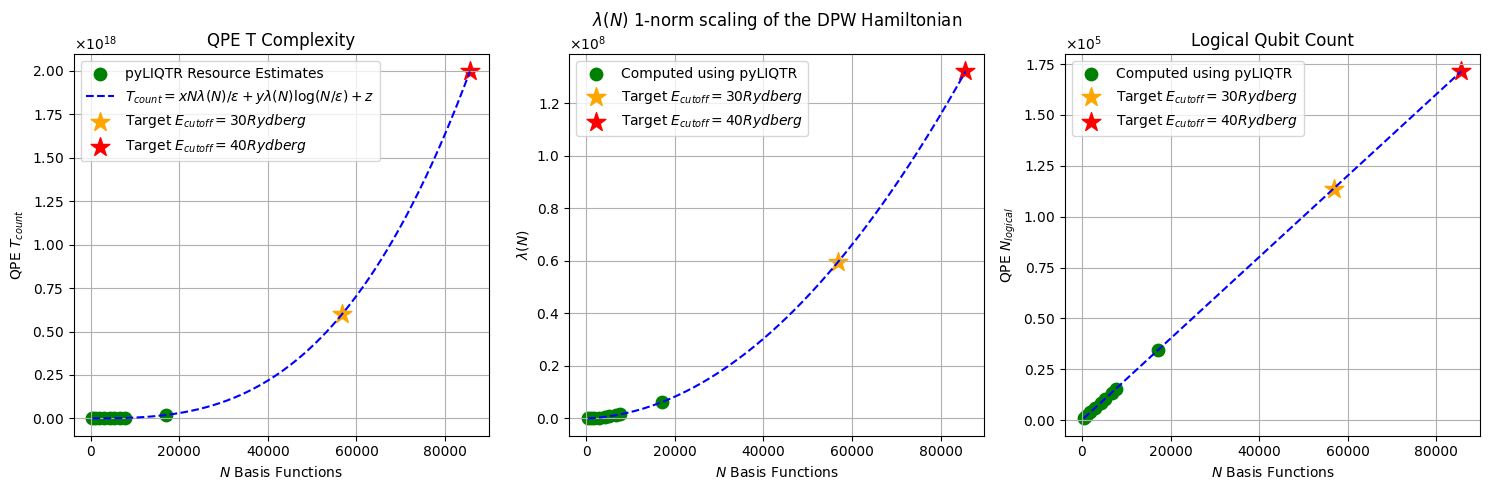}
    \\
    \includegraphics[width = \textwidth]{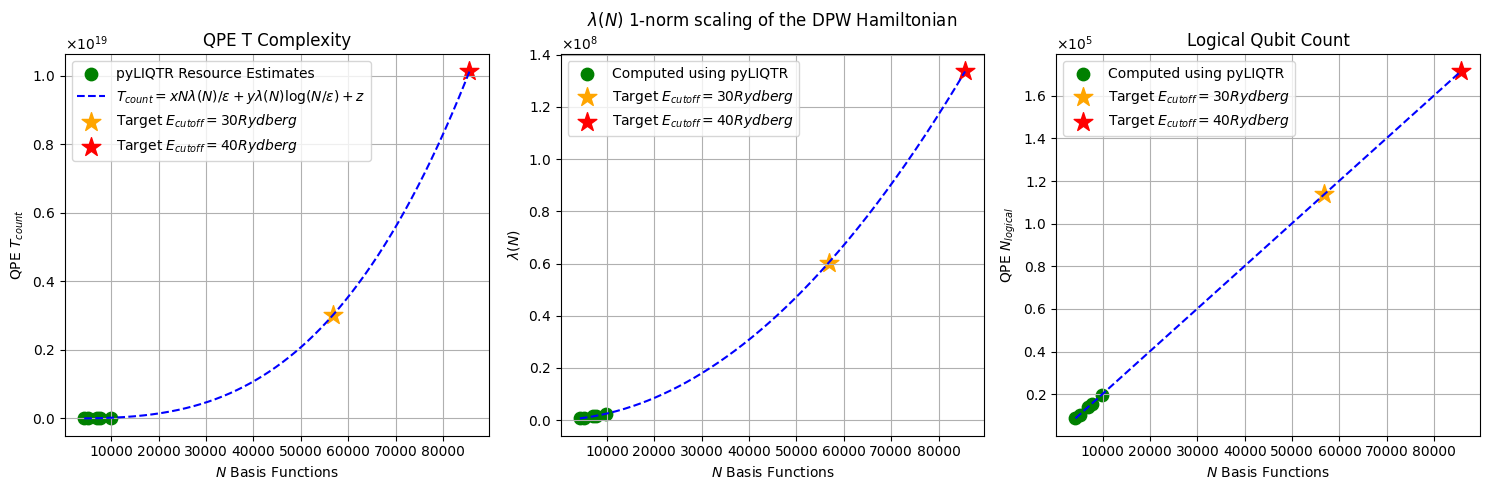}
    \\
    \includegraphics[width = \textwidth]{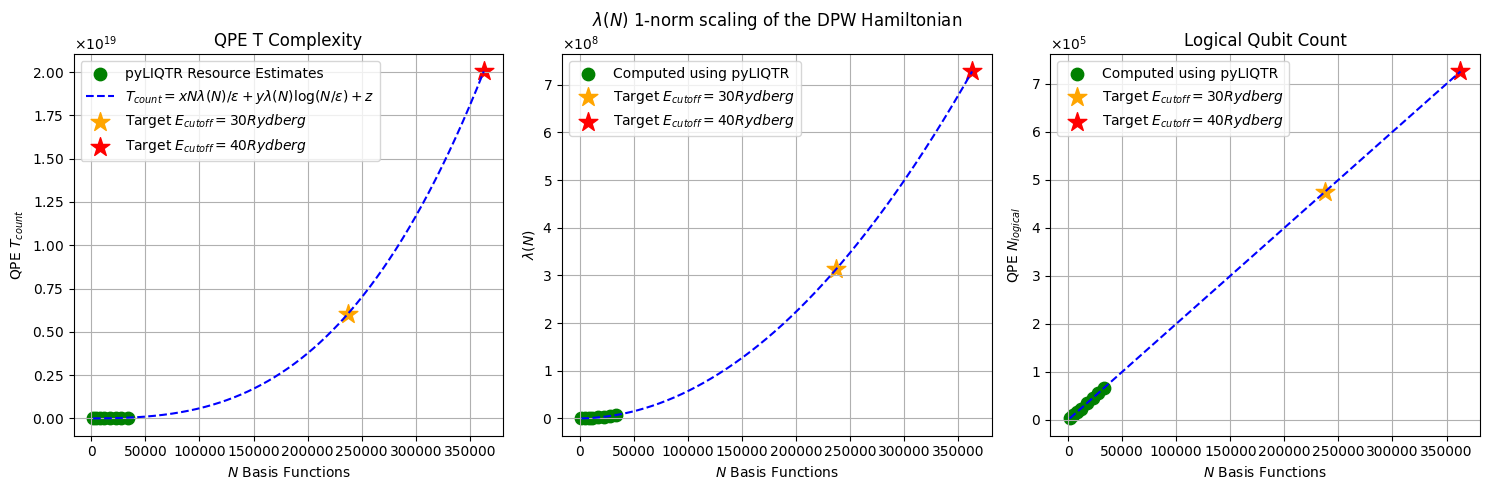}
    
    \caption{Resource estimate results for the double-unit Mg$_{17}$Al$_{12}$ secondary phase structure, quadruple-unit Mg$_{17}$Al$_{12}$  secondary phase structure, and the  Mg$_{17}$Al$_{12}$ secondary-phase supercell. The $x$-axis represents the number of basis functions, which is directly related to the energy cutoff, $E_{\text{cut}}$, threshold. All green data points correspond to explicit resource estimate values computed using pyLIQTR. These data points were used to extrapolate to desired $E_{\text{cut}}$ values at higher thresholds, particularly at $E_{\text{cut}} = 30$ and $40$ Ry. (Top row) The double-unit Mg$_{17}$Al$_{12}$ secondary phase structure, embedded in a simulation cell of $30.1 \times 9.3 \times 43$ {\AA}. (Middle row) The quadruple-unit Mg$_{17}$Al$_{12}$  secondary phase structure, embedded in a simulation cell of $30 \times 9.3 \times 42.9$ {\AA}. (Bottom row) The  Mg$_{17}$Al$_{12}$ secondary-phase supercell as an expansion of the single-unit cell with a simulation cell of $33 \times 31.3 \times 50.3$\AA.  }
    \label{fig: extrapolated_result_2}
\end{figure}

Table \ref{tab: Mg 2nd quantization resource estimation} shows the resource estimates for the representative computational models related to the design of Mg-rich sacrificial coatings and Corrosion-Resistant Mg alloys in aqueous environments. As discussed in Section \ref{sec: supp_resource_estimates_1st_quantization}, the targeted energy cutoff for these computational models is around 30 to 50 Rydberg (Ry) when using the Vanderbilt ultrasoft pseudopotential (USPP), but it can be significantly higher when employing other types of pseudopotentials, such as the Hartwigsen-Goedecker-Hutter (HGH), as shown in Figure \ref{fig:USPP_vs_HGH_Ecutoff}. In contrast to the first quantization scenario, each increment in the energy cutoff -- and consequently, the number of basis functions -- increases the number of logical qubits. Due to the immense size of these systems, performing explicit resource estimates for them at high $E_{\text{cut}}$ values was not feasible. For instance, generating the electronic structure Hamiltonian for the quantum computing framework using PEST for the Cluster model at $E_{\text{cut}} = 13$ Ry took over 85 hours using 1 node with 64 cores equipped with AMD EPYC 7543 type processors and 64 GB of memory. At $E_{\text{cut}} = 40$ Ry, the generation of the electronic structure Hamiltonian alone would take around 1700 hours using the same computing resource. Therefore, a more computationally efficient approach was adopted by extrapolating the results to larger $E_{\text{cut}}$ values through resource estimates computed at lower $E_{\text{cut}}$. The extrapolation method is as follows:
\begin{subequations}\label{eq:extrapolated_equations}
\begin{align}
    \lambda(N) &= aN^2 + bN + c \\
    T_{\text{count}} &= x N \lambda(N)/\epsilon + y (\lambda(N)/\epsilon)\log(N/\epsilon) + z \\
    N_{\text{logical}} &= 2N + \alpha \log\left(4 \sqrt{2} \pi \lambda(N)^3 N^5/\epsilon^3\right) + \beta
\end{align}
\end{subequations}
For the Monolayer, Cluster, the single-unit Mg$_{17}$Al$_{12}$ secondary phase structure, the double-unit Mg$_{17}$Al$_{12}$ secondary phase structure, and the quadruple-unit Mg$_{17}$Al$_{12}$  secondary phase structure, explicit resource estimates are performed up to an energy cutoff ($E_{\text{cut}}$) of 13 Rydberg (Ry). Values at higher $E_{\text{cut}}$ were extrapolated using Equation \eqref{eq:extrapolated_equations}. For the secondary-phase supercell model, resource estimates were performed only up to $E_{\text{cut}} = 8$ Ry due to its large size, and all estimate values above this were extrapolated. These extrapolated values are presented in Figures \ref{fig: extrapolated_result_1} and \ref{fig: extrapolated_result_2}. Refer to Table 2 for precise numerical values of logical resource estimates for these models at different targeted $E_{\text{cut}}$ thresholds. To numerically validate the reliability of these extrapolated results, extrapolation on the Dimer model was performed using estimated values at lower $E_{\text{cut}}$ thresholds, specifically at $E_{\text{cut}} = 5, 6, 7, 8$ Ry. Extrapolated results were then compared to the actual estimated results from pyLIQTR. See Figure \ref{fig: goodness_of_fit}. 

\begin{table}
    \centering
    \begin{ruledtabular}
        \begin{tabular}{l c c c c }
            \makecell{\textbf{Cell size} \\ ({\AA})} & \makecell{$\boldsymbol{E}_{\text{cut}}$\\ (Ry)}&\makecell{$\boldsymbol{N_{\text{Log}}}$} & \makecell{\textbf{QPE $T$ count}\\ ($\times 10^{16}$)} & \makecell{$\boldsymbol{1}$-\textbf{norm}\\ $\lambda$}\\
            \hline
            \multicolumn{5}{c}{{Magnesium dimer model}}\\
            \makecell{[12.7, 12.7, 19.9]} & \makecell{13\\30\\40} & \makecell{8763\\29265\\47664} &\makecell{0.003\\0.83\\5.43}
            & \makecell{3.28 $\times10^5$\\3.22 $\times10^6$\\ 8.34 $\times10^6$}
            \\[15pt]
            \multicolumn{5}{c}{{Magnesium monolayer model}}
            \\
            \makecell{[19.8, 19.8, 32.3]}
            &\makecell{13\\30$^*$\\40$^*$}& \makecell{35007\\117785\\179235} & \makecell{1.00\\28.2\\94.4}
            & \makecell{3.32 $\times10^6$\\3.37 $\times10^7$\\ 7.68 $\times10^7$}
            \\[15pt]
            \multicolumn{5}{c}{{Magnesium cluster model}}
            \\
            \makecell{[19.8, 19.8, 32.3]} &\makecell{13\\30$^*$\\40$^*$}&
            \makecell{62120\\213625\\326530} & \makecell{14.1\\535\\1890}
            & \makecell{1.83 $\times10^7$ \\ 1.93 $\times10^8$\\ 4.44 $\times10^8$}
            \\[15pt]
            \multicolumn{5}{c}{Single-unit $\textrm{Mg}_{17}\textrm{Al}_{12}$ secondary phase structure}
           \\
           \makecell{[29.7, 9.2, 42.6]} &\makecell{13\\30$^*$\\40$^*$}&  \makecell{31180\\111504\\168445} & \makecell{1.79\\72.3\\245}
           & \makecell{5.23 $\times10^6$\\5.78 $\times10^7$\\ 1.30 $\times10^8$}
           \\[15pt]
           \multicolumn{5}{c}{Double-unit $\textrm{Mg}_{17}\textrm{Al}_{12}$ secondary phase structure}
           \\
           \makecell{[30.1, 9.3, 43]} &\makecell{13\\30$^*$\\40$^*$}&  \makecell{34286\\113777\\171399} & \makecell{1.97\\60.1\\199}
           & \makecell{6.22 $\times10^6$\\5.99 $\times10^7$\\ 1.33 $\times10^8$}
           \\[15pt]
           \multicolumn{5}{c}{Quadruple-unit $\textrm{Mg}_{17}\textrm{Al}_{12}$ secondary phase structure}
           \\
           \makecell{[30.0, 9.3, 42.9]} &\makecell{13\\30$^*$\\40$^*$}&  \makecell{34286\\113777\\171399} & \makecell{1.96\\51.5\\166}
           & \makecell{6.36 $\times10^6$\\6.06 $\times10^7$\\ 1.34 $\times10^7$}
           \\[15pt]
           \multicolumn{5}{c}{$\textrm{Mg}_{17}\textrm{Al}_{12}$ secondary-phase supercell}
           \\
           \makecell{[33.0, 31.3, 50.3]} &\makecell{13$^*$\\30$^*$\\40$^*$}&  \makecell{134813\\474954\\725956} & \makecell{21.2\\604\\2000}
           & \makecell{2.66 $\times10^7$\\3.15 $\times10^8$\\ 7.28 $\times10^8$}
           \\
        \end{tabular}
    \end{ruledtabular}
    \caption{Resource estimate summary for representative computational workflows for \textit{Designing Mg-Rich Sacrificial Coatings and Corrosion-Resistant Mg Alloys for Aqueous Environments} as described in Section \ref{sec: Mg corrosion} using second quantization representation with linaer-T encoding as described in~\cite{babbush2018encoding}. Calculations were done using scaling factor $\gamma = 1$, multiply resource estimates by $8$ to get estimates for $\gamma = 0.5$ resources. For $E_{\text{cut}}$ values with an asterisk ($*$) attached to them, the resource estimates are derived from extrapolation using Equation \eqref{eq:extrapolated_equations}.}
    \label{tab: Mg 2nd quantization resource estimation}
\end{table}

In Table \ref{tab: Nb 2nd quantization resource estimation}, we report the resource estimates for various representative computational models designed for high-temperature corrosion-resistant alloys. In contrast to the previous case, no extrapolation was needed, and explicit resource estimates were computed for a range of energy cutoffs from 10 to 40 Ry.
\noindent 
\begin{table}
\centering
\begin{ruledtabular}
    \begin{tabular}{l c c c c }
        \makecell{\textbf{Cell size} \\ ({\AA})} & \makecell{$\boldsymbol{E}_{\text{cut}}$\\ (Ry)}&\makecell{$\boldsymbol{N_{\text{Log}}}$} & \makecell{\textbf{QPE $T$ count}\\ ($\times 10^{14}$)} & \makecell{$\boldsymbol{1}$-\textbf{norm}\\ $\lambda$}\\
        \hline
        \multicolumn{5}{c}{Nb$_{97}$Hf$_3$Ti$_{22}$Zr$_6$O}\\
        \makecell{[13.3, 13.3, 13.3]} & \makecell{10\\30\\40} & \makecell{4526\\21451\\35311} & \makecell{0.414\\30.4\\200}
        &\makecell{8.25 $\times 10^4$\\1.55 $\times 10^6$\\4.07 $\times 10^6$}
        \\[15pt]
        \multicolumn{5}{c}{{Nb$_{97}$Ta$_{22}$Zr$_3$W$_6$O}}
        \\
        \makecell{[13.3, 13.3, 13.3]} & \makecell{10\\30\\40}& \makecell{4526\\21451\\35311} & \makecell{0.414\\30.4\\200}
        &\makecell{8.25 $\times 10^4$\\1.55 $\times 10^6$\\4.07 $\times 10^6$}
        \\[15pt]
        \multicolumn{5}{c}{{Nb$_{42}$Ti$_{3}$Hf$_{3}$Ta$_{3}$Zr$_3$O}}
        \\
        \makecell{[10.0, 10.0, 10.1]} & \makecell{10\\30\\40}& \makecell{2128\\9973\\16151} & \makecell{0.051\\3.60\\23.2}
        &\makecell{2.61 $\times 10^4$\\4.61 $\times 10^5$\\1.18 $\times 10^6$}
        \\[15pt]
        \multicolumn{5}{c}{Nb$_{65}$Zr$_6$Hf$_7$Ti$_4$W$_3$}
        \\
        \makecell{[8.6, 9.8, 14.0]} &\makecell{10\\30\\40}& \makecell{2651\\12389\\17597} & \makecell{0.126\\8.89\\25.1}&\makecell{5.64 $\times 10^4$\\8.72 $\times 10^5$\\1.69 $\times 10^6$}
    \end{tabular}
\end{ruledtabular}
\caption{Resource estimate summary for representative computational workflows for \textit{Designing High-Temperature Corrosion-Resistant Alloys}, as described in Section \ref{sec:Nb_workflow}, using second quantization representation with linaer-T encoding as described in~\cite{babbush2018encoding}. Similar to Table \ref{tab: Mg 2nd quantization resource estimation}, calculations were done using scaling factor $\gamma = 1$, multiply resource estimates by $8$ to get estimates for $\gamma = 0.5$ resources. However, unlike Table \ref{tab: Mg 2nd quantization resource estimation}, no extrapolations was needed to obtain higher energy cutoff results.}
\label{tab: Nb 2nd quantization resource estimation}
\end{table}

\section{Remarks on First vs. Second Quantization}

\begin{figure}
    \centering
    \includegraphics[width= 1\textwidth]{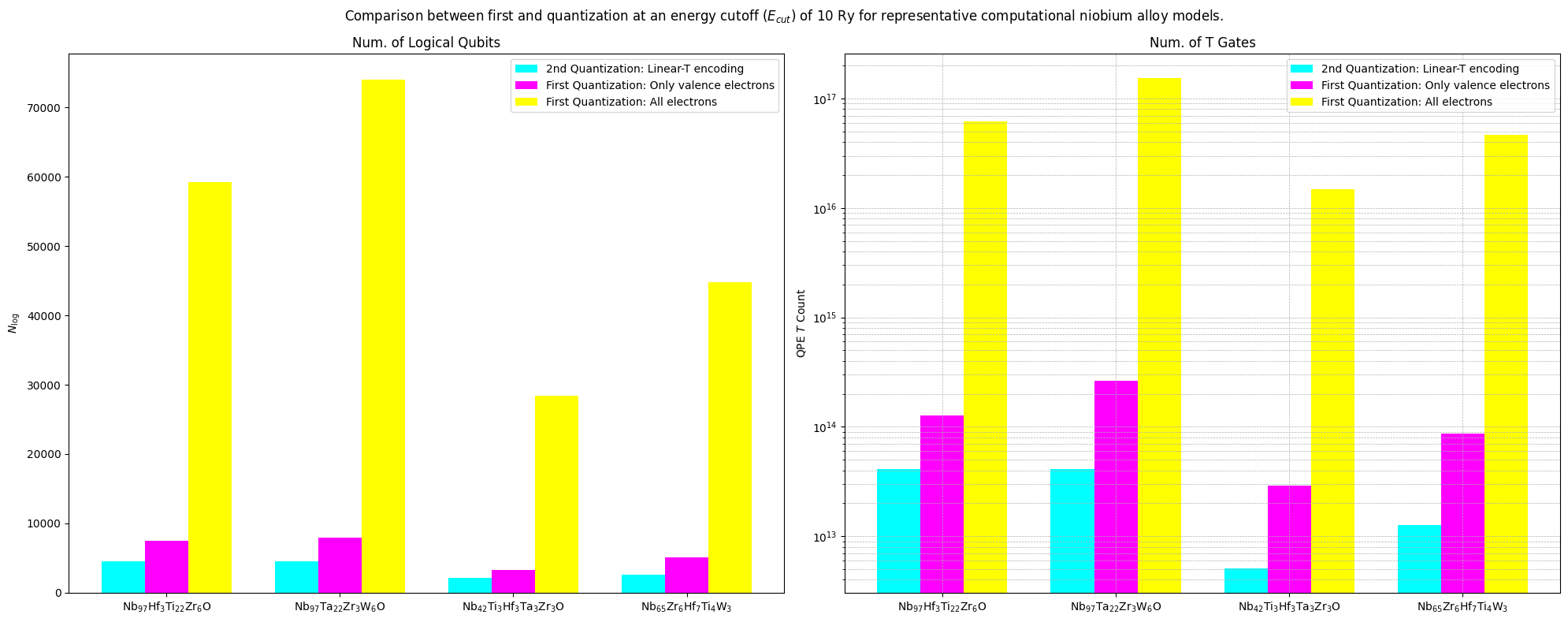}
    \caption{Comparison between first and second quantization at a low energy cutoff value (10 Ry). In the first quantization, we provide two results (as discussed in Section \ref{sec: supp_resource_estimates_1st_quantization}: 1) the optimistic scenario, where only valence electrons are considered due to the use of pseudopotentials, although the results here are not based on a pseudopotential Hamiltonian, and 2) the worst-case scenario, where all electrons are considered. This shows that there are scenarios, such as a low energy cutoff as a consequence of using the appropriate pseudopotential, where the second quantization framework with Linear-T encoding offers a better alternative than the first quantization framework, even in the best-case scenario. Therefore, one should evaluate this tradeoff for their problem of interest.}
    \label{fig: First_vs_Second_Nb_10Ry}
\end{figure}

We explicitly constructed the quantum circuit and estimated the resource requirements for the qubitized QPE algorithm in both first and second quantization representations for the representative computational models outlined in the workflows described in Section~\ref{sec: Mg corrosion} and Section~\ref{sec:Nb_workflow}. While first quantization is promising due to its space efficiency, especially when \( \eta \ll N \), and allows for an increase in the number of basis functions within a certain range without incurring additional computational costs as shown in section ~\ref{sec: supp_resource_estimates_1st_quantization}, it may not always be more efficient than second quantization for certain problems under specific regimes. This is particularly evident in the application of \textit{Designing Corrosion-Resistant High-Temperature Niobium Alloys} at lower energy cutoff values, even under the optimistic assumption of a pseudopotential Hamiltonian where core electrons are removed from the models, as shown in Figure \ref{fig: First_vs_Second_Nb_10Ry}. The reason is that Nb alloys contain heavy alloying elements (e.g., Hf, W), meaning that at lower energy cutoffs, the condition \( \eta \ll N \) does not hold. However, if a high energy cutoff is required, the first quantization framework becomes more efficient.  In contrast, for surface reactions on magnesium alloys, first quantization consistently outperforms second quantization, even at low energy cutoffs, even in the worst-case scenario.

\begin{figure}
    \centering
    \includegraphics[width= .8\textwidth]{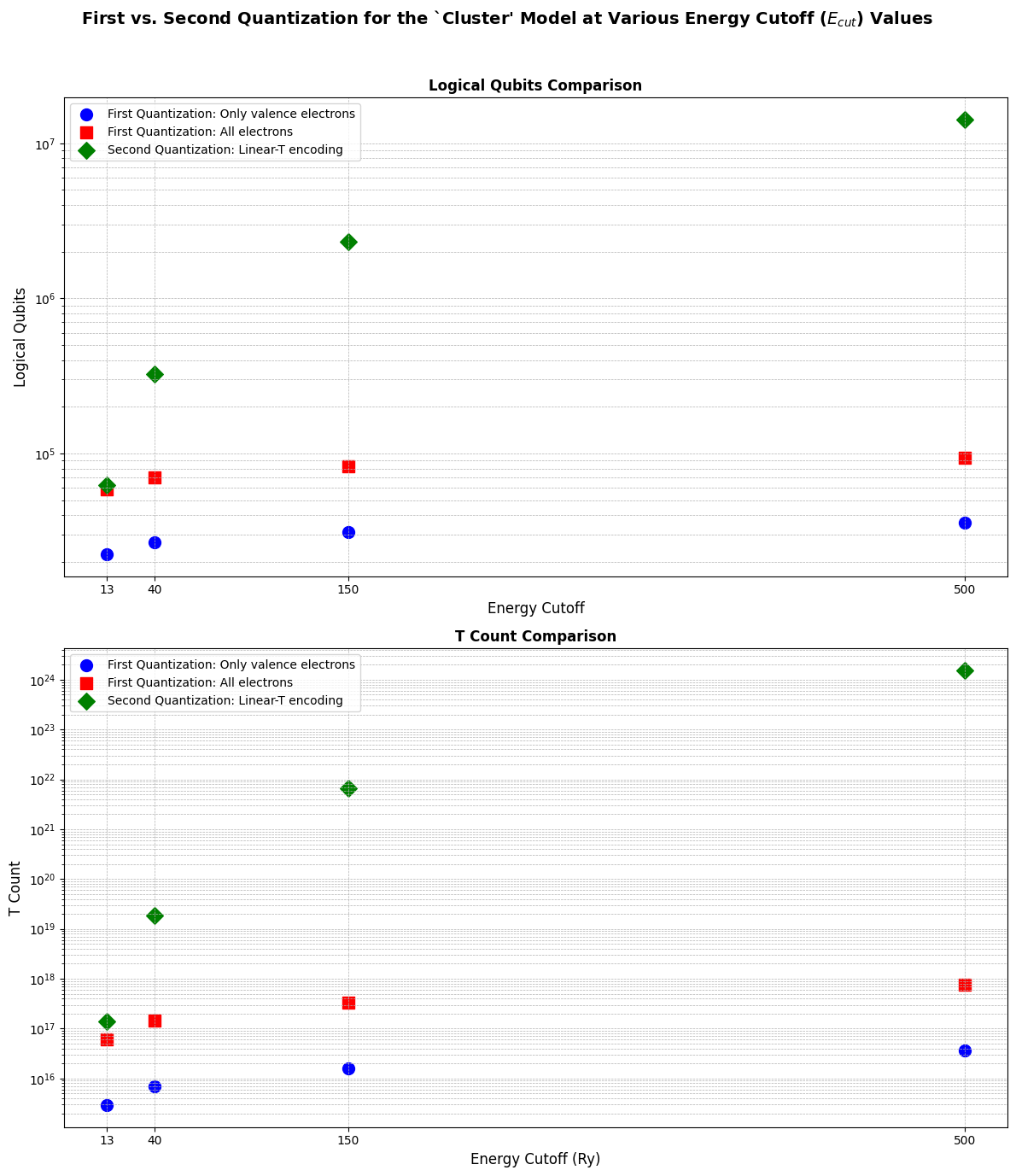}
    \caption{Comparison of resource estimation between first quantization (plane waves) and second quantization (dual-plane waves) for the 'Cluster' computational model, as described in Section \ref{sec: Representative Computational Models for Mg}, at various increasing energy cutoffs (\(E_{cutoff}\)), resulting in a corresponding increase in the number of plane waves. Significant savings are observed in the first quantization framework. }
    \label{fig: First_vs_Second_mg_cluster}
\end{figure}

\end{document}